\documentclass[twocolumn,nofootinbib,showpacs,preprintnumbers,amsmath,amssymb,floatfix]{revtex4}

\usepackage{graphicx}
\usepackage{dcolumn}
\usepackage{bm}
\usepackage{amssymb}
\usepackage{amsmath}
\usepackage{hep}
\usepackage{color}
\usepackage{multirow}

\def\bew{\begin{widetext}}    
\def\eew{\end{widetext}}      

\def\beq{\begin{eqnarray}}    
\def\eeq{\end{eqnarray}}      

\newcommand{\eeAl}{\HepProcess{\APelectron\Pelectron \HepTo \PHiggspszero \PHiggslightzero}}
\newcommand{\eeAH}{\HepProcess{\APelectron\Pelectron \HepTo \PHiggspszero \PHiggsheavyzero}}
\newcommand{\complet}{\HepProcess{\APelectron\Pelectron \HepTo \PHiggspszero \PHiggslightzero/\Azero\Hzero}}

\newcommand{\vzlA}{\PHiggspszero\,\PHiggslightzero\,\PZ^0}
\newcommand{\vzHA}{\PHiggspszero\,\PHiggsheavyzero\,\PZ^0}

\newcommand{\lag}{\ensuremath{\mathcal{L}}}

\newcommand{\rs}{\ensuremath{\hat{\Sigma}}}

\newcommand{\PG}{\ensuremath{G}}
\newcommand{\A}{\ensuremath{\PHiggspszero}}
\newcommand{\AZ}{\ensuremath{\PHiggspszero\PZ^0}}

\newcommand{\AG}{\ensuremath{\PHiggspszero \PG^0}}
\newcommand{\Hp}{\ensuremath{\PHiggs^+}}
\newcommand{\Hm}{\ensuremath{\PHiggs^-}}
\newcommand{\sw}{\ensuremath{s_W}}
\newcommand{\cw}{\ensuremath{c_W}}
\newcommand{\swd}{\ensuremath{s_W^2}}
\newcommand{\cwd}{\ensuremath{c_W^2}}

\renewcommand{\l}{\ensuremath{\PHiggslightzero}} 
\renewcommand{\H}{\ensuremath{\PHiggsheavyzero}} 
\renewcommand{\A}{\ensuremath{\PHiggspszero}} 
\newcommand{\hzero}{\ensuremath{\PHiggslightzero}} 
\newcommand{\Hzero}{\ensuremath{\PHiggsheavyzero}} 
\newcommand{\Azero}{\ensuremath{\PHiggspszero}} 

\newcommand{\Hpm}{\ensuremath{\PHiggspm}} 

\newcommand{\CP}{\ensuremath{\mathcal{C}\mathcal{P}}}
\newcommand{\jump}{\vspace{0.2cm}}
\newcommand{\plA}{\ensuremath\HepProcess{\APelectron\Pelectron\HepTo \PHiggslightzero\PHiggspszero}}
\newcommand{\pHA}{\HepProcess{\APelectron\Pelectron\HepTo \PHiggsheavyzero\PHiggspszero}}

\newcommand{\eeAlH}{\ensuremath\HepProcess{\APelectron\Pelectron\HepTo \PHiggslightzero\PHiggspszero / \PHiggsheavyzero\PHiggspszero}}

\newcommand{\founders}{Higgs:1964ia}

\newcommand{\hunter}{Hunter}

\newcommand{\order}{\ensuremath{\mathcal{O}(\alpha^3_{ew})}}
\newcommand{\orderquadrat}{\ensuremath{\mathcal{O}(\alpha^4_{ew})}}
\newcommand{\masses}{Grifols:1984xs,Garcia:1993sb}
\newcommand{\widths}{Garcia:1993rq}
\newcommand{\rself}{\ensuremath{\hat{\Sigma}}}
\newcommand{\self}{\ensuremath{\Sigma}}

\begin{document}

\hyphenation{cos-mo-lo-gi-cal sig-ni-fi-cant quin-tes-sen-ce
e-qui-ta-ble be-ha-vior fo-llo-wing en-han-cing con-si-de-ra-tion}

\title{Neutral Higgs-pair production at Linear
    Colliders within the general 2HDM: quantum effects and triple Higgs boson self-interactions}

\author{David L\'opez-Val}
 \email{dlopez@ecm.ub.es}
\author{Joan Sol\`a}%
 \email{sola@ecm.ub.es}
\affiliation{%
High Energy Physics Group, Dept. ECM, and Institut de Ci{\`e}ncies del Cosmos\\
Univ. de Barcelona, Av. Diagonal 647, E-08028 Barcelona, Catalonia, Spain}%

\date{\today}

\begin{abstract}
The pairwise production of neutral Higgs bosons (\hzero\Azero,
\Hzero\Azero) is analyzed in the context of the future linear
colliders, such as the ILC and CLIC, within the general
Two-Higgs-Doublet Model (2HDM). The corresponding cross-sections are
computed at the one-loop level, including the full set of
contributions at order $\mathcal{O}(\alpha^3_{ew})$ together with
the leading $\mathcal{O}(\alpha^4_{ew})$ terms, in full compliance
with the current phenomenological bounds and the stringent
theoretical constraints inherent to the consistency of the model. We
uncover regions across the 2HDM parameter space, mainly for low
$\tan\beta\sim 1$ and moderate -- and negative -- values of the
relevant $\lambda_5$ parameter, wherein the radiative corrections to
the Higgs-pair production cross section, $\sigma(\complet)$, can
comfortably reach $|\delta\sigma|/\sigma\sim 50 \%$. This behavior
can be traced back to the enhancement capabilities of the trilinear
Higgs self-interactions -- a trademark feature of the 2HDM, with no
counterpart in the Minimal Supersymmetric Standard Model (MSSM). The
corrections are strongly dependent on the actual value of
$\lambda_5$ as well as on the Higgs mass spectrum. Interestingly
enough, the quantum effects are positive for energies around
$\sqrt{s} \simeq 500\,\GeV$, thereby producing a significant
enhancement in the expected number of events precisely around the
fiducial startup energy of the ILC. The Higgs-pair production rates
can be substantial, typically a few tens of femtobarn, therefore
amounting to a few thousand events per $100\,\femtobarn^{-1}$ of
integrated luminosity. In contrast, the corrections are negative in
the highest energy range (viz. $\sqrt{s}\sim 1\,\TeV$ and above). We
conclude that a precise measurement of the 2H final states could
carry unambiguous footprints of an extended (non-supersymmetric)
Higgs sector. Finally, to better assess the scope of these effects,
we compare the exclusive pairwise production of Higgs bosons with
the inclusive gauge boson fusion channels leading to $2H+X$ final
states, and also with the exclusive triple Higgs boson production.
We find that these multiparticle final states can be highly
complementary in the overall Higgs bosons search strategy.
\end{abstract}

\pacs{{12.15.-x, 12.60.Fr, 12.15.Lk} 
}
\maketitle

\vskip 6mm

\section{Introduction}
\label{sect:introduction}

It was already in the 60's when the pioneering works of Higgs,
Kibble, Englert and others suggested the existence of a fundamental
spinless building-block of Nature\,\cite{\founders}, whose
non-vanishing vacuum expectation value could explain the spontaneous
breaking of the $SU(2)_L\otimes U(1)_Y$ gauge group of the
Electroweak (EW) interactions down to the $U(1)_{\rm em}$ abelian
symmetry group of the Electromagnetism. Indeed, this so-called Higgs
mechanism is the most fundamental longstanding issue that remains
experimentally unsettled in Particle Physics. For one thing it is
the only known strategy capable of building up a (perturbative)
renormalizable quantum field theoretical description of the
Electroweak Symmetry Breaking (EWSB) phenomenon. It is difficult to
overemphasize that, to a great extent, the Higgs mechanism embodies
the backbone of the SM structure and that in its absence we would
have to cope with an entirely different conception of the inner
functioning of the SM as a quantum field theory (QFT) .

The central assumption here is the existence of (at least) one
elementary spinless field -- the so-called Higgs boson. Its
relevance is particularly evident if we take into account that the
presence of one or more such fields in the structure of the model is
indispensable for the unitarity of the theory. If no Higgs bosons
exist below the TeV scale, weak interactions would actually become
strong at that scale and e.g. the $\PW^+ \PW^-$ cross-section would
violate the unitarity of the scattering matrix. Moreover, in the
absence of Higgs bosons the various particle masses could not be
generated consistently (i.e. without spoiling the ultraviolet
behavior of the theory at higher orders of perturbation theory). In
short, if no Higgs bosons are there to protect the (presumed)
perturbative structure of the weak interactions, the latter would
enter a perilous runaway regime at high energies. It is, thus,
essential to confirm experimentally the existence of one or more
Higgs boson particles through their explicit production in the
colliders.

Amazingly enough, more than 40 years after the Higgs mechanism was
naturally blended into the conventionally accepted SM landscape of
the strong and electroweak interactions\,\cite{GWS616769}, it
remains still thickly curtained and no direct evidence has been
found yet of its main testable prediction: the existence of
fundamental spinless particles. Notwithstanding the many efforts
devoted at LEP and at the Tevatron in the last decades, all
experimental searches for Higgs boson signatures have come up
empty-handed, and one has to admit that no conclusive physical
signature for the existence of these elementary scalar fields -- not
even the single one predicted by the SM -- has been found for the
time being. Thus, we do not know whether the conventional Higgs
boson exists at all, or if there are no Higgs particles of any sort.
On the contrary, it could be that there are extensions of the SM
spanning a richer and elusive spectrum of Higgs bosons which have
escaped all our dedicated experimental searches. Be that as it may,
this pressing and highly intriguing enigma will hopefully be
unveiled soon, specially with the advent of the new generation of
supercolliders, like the brand new LHC, the future International
Linear Collider (ILC)\,\cite{ILChome} and, hopefully, the Compact
Linear Collider (CLIC) too\,\cite{CLIChome}.

Therefore, we have to (and we can) be well prepared to recognize all
kind of hints from Higgs boson physics. Let us recall that apart
from the single neutral (\CP-even) Higgs boson $\PHiggs^0$ of the
SM, other opportunities arise in model building that could serve
equally well the aforementioned purposes. Perhaps the most
paradigmatic extension of the SM is the Minimal Supersymmetric
Standard Model (MSSM)\,\cite{MSSM}, whose Higgs sector involves two
doublets of complex scalar fields. The physical spectrum consists of
two charged states, $\PHiggs^{\pm}$, two neutral \CP-even states
$\hzero, \Hzero$ (with masses conventionally chosen as
$M_{\hzero}<M_{\PHiggs^0}$) and one \CP-odd state $\Azero$. We shall
not dwell here on the whys and wherefores of supersymmetry (SUSY) as
a most sought-after realization of physics beyond the SM. It
suffices to say that it provides a Higgs sector which is stable (to
all orders in perturbation theory) under embedding of the low-energy
structure into a Grand Unified Theory, and in particular it provides
the most natural link with gravity\,\cite{MSSM}.

But, how to test the structure of the Higgs sector? In the MSSM
case, such structure is highly constrained by the underlying
supersymmetry. A consequence of it, which is particularly relevant
for our considerations, is the fact that the self-interactions of
the SUSY Higgs bosons turn out to be largely immaterial from the
phenomenological point of view, in the sense that they cannot be
enhanced (at the tree-level) as compared to the ordinary gauge
interactions and, therefore, cannot provide outstanding signatures
of physics beyond the SM. The bulk of the enhancing capabilities of
the MSSM Lagrangian is concentrated, instead, in the rich structure
of Yukawa-like couplings between Higgs bosons and quarks or between
Higgs boson and squarks, or even between squarks and
chargino-neutralinos, as it was shown long ago in plentiful
phenomenological scenarios involving quantum effects on the gauge
boson masses\,\cite{\masses} and gauge boson and top quark
widths\,\cite{\widths}, as well as in many other processes and
observables (see e.g.
\,\cite{Coarasa:1996qa,Guasch:1998as,Carena:1999py,Belyaev:2001qm,Bejar:2008ub}.
For the present state of the art, cf.
\cite{Heinemeyer:2006px,Frank:2006yh}; and for recent comprehensive
reviews, see \,\cite{Djouadi:2005gj,Heinemeyer:2004gx}, for example.

The two-Higgs $SU_L(2)$-doublet structure of the Higgs sector in the
MSSM is a trademark prediction of SUSY invariance. Nonetheless, the
very same doublet structure can appear in the form of a
non-supersymmetric framework, the so-called general (unconstrained)
Two-Higgs-Doublet Model(2HDM), which does also entail a rich
phenomenology \cite{\hunter}. Again, the same spectrum of \CP-even
($\hzero, \Hzero$) and \CP-odd ($\Azero$) scalar fields arise.
However, we should keep in mind that the most distinctive aspects of
the general 2HDM Higgs bosons could well be located in sectors of
the model quite different from the MSSM case\,\cite{Bejar:2006ww}.
Indeed, even the very couplings involved in the structure of the
Higgs potential (viz. the so-called trilinear and quartic Higgs
boson self-interactions) could be the most significant ones as far
as the phenomenological implications are concerned, rather than just
the enhanced Yukawa couplings with the heavy quarks.  In such
circumstance, we should be able to detect a very different kind of
experimental signatures. Assuming, for instance, that the LHC will
find evidence of a neutral Higgs boson, a critical issue will be to
discern whether the newcomer is compatible with the SM or any of its
extensions and, in the latter case, to which of these extensions it
most likely belongs. In order to carry out this ``finer'' hunting of
the Higgs boson(s), a $\TeV$-range accessible linear collider
machine (such as the aforesaid ILC/CLIC) will be needed.
In this work, we will show how useful it could be such
collider to discriminate neutral supersymmetric Higgs bosons from
generic (non-SUSY) 2HDM ones.

The paper is organized as follows. In the next section, we remind
the reader of some relevant Higgs boson production processes in the
linear colliders. In Section~\ref{sec:2HDM}, we describe the
structure of the 2HDM and the relevant phenomenological
restrictions, including the properties of the trilinear Higgs
self-interactions and their interplay with the notion of unitarity
and vacuum stability. In Section~\ref{sec:renorm}, we develop in
detail the renormalization program for the Higgs sector of the
general 2HDM. The theoretical setup for the one-loop computation of
the Higgs boson cross-sections is elaborated in
Section~\ref{sec:general}, leaving Section~\ref{sec:results} to
present a comprehensive numerical analysis of our results. Finally,
Section~\ref{sec:conclusions} is devoted to a thorough general
discussion and to present the conclusions of our work.

%
\section{Higgs boson production at linear colliders}
\label{sec:HiggsLinear}

The leading Higgs boson production mechanisms at the LHC have been
studied thoroughly for the last twenty years, they are well under
control both in the SM\,\cite{Glover:1987nx} and in the
MSSM\,\cite{Djouadi:1991tka,Djouadi:1999rca} and there are also some
studies in the 2HDM\,\cite{Moretti:2004wa}. Hopefully, they may soon
help revealing some clues on Higgs boson physics at the LHC
--- for a review see e.g.\,\cite{Djouadi:2005gj} and references
therein. However, it is not obvious that it will be possible to
easily disentangle the nature of the potentially produced Higgs
boson(s). The next generation of TeV-class linear colliders (based
on both $\APelectron\Pelectron$ and $\Pphoton\Pphoton$ collisions),
such as the ILC and the CLIC projects \cite{ILChome,CLIChome}, will
be of paramount importance in finally settling the experimental
basis of the ``Higgs issue'' as the most fundamental theoretical
construct of the SM of electroweak interactions. Thanks to its
extremely clean environment (in contrast to the large QCD background
inherent to a hadronic machine such as the LHC), a linear collider
(\textit{linac} for short) should allow for a precise measurement of
the Higgs boson parameters, such as: i) the Higgs boson mass (or
masses, if more than one); ii) the couplings of the Higgs bosons to
quarks, leptons and gauge bosons; iii) the Higgs boson
self-couplings mentioned above, i.e. the trilinear (3H) and quartic
(4H) Higgs boson self-interactions. At the end of the day, a linac
should allow us to dig deeper than ever into the structure of the
EWSB and, hopefully, even to reconstruct the Higgs potential itself.

A detailed road map of predictions for Higgs-boson observables at
the linear colliders is called for, with a special emphasis on those
signatures which may be characteristic of the different extensions
of the SM. For example, as indicated above, triple Higgs boson
self-interactions (3H) may play a cardinal role in this endeavor
because, in favorable circumstances, they could easily distinguish
between the MSSM and the general 2HDM Higgs sectors. Such 3H
couplings can mediate a plethora of interesting processes. The key
point here is the potentially large enhancements that the 3H
couplings may accommodate in the general 2HDM case, in contrast to
the supersymmetric extensions. Actually, even for the SUSY case
there is a large number of works attempting to extract vestiges of
non-standard dynamics in these couplings, mainly through the
radiative corrections that they can undergo. For instance, the
3H-couplings have been investigated in
\cite{Djouadi:1996ah,Djouadi:1999gv,Osland:1998hv}, and in some
cases considered for possible phenomenological applications in
TeV-class linear colliders through the double-Higgs strahlung
process $\APelectron\Pelectron \to \PHiggsheavy \PHiggsheavy \PZ$ or
the double-Higgs $\PW\PW$-fusion mechanism $\APelectron \Pelectron
\to \PHiggsplus\PHiggsminus \Pnue\APnue$. These processes, which
include vertices like $\PZ\PZ\PHiggsheavy$, $\PW\PW\PHiggsheavy$,
$\PZ\PZ \PHiggsheavy \PHiggsheavy$, $\PW\PW\PHiggsheavy\PHiggsheavy$
and $\PHiggsheavy \PHiggsheavy \PHiggsheavy$, are possible both in
the SM and its extensions, such as the MSSM and the general 2HDM.
Unfortunately, the cross-section turns out to be rather small  both
in the SM and in the MSSM, being of order $10^{-3}$ pb at most, i.e.
equal or less than $1$ fb\, \cite{Djouadi:1999gv}. Even worse is the
situation regarding the triple Higgs boson production in the MSSM,
in which -- except in the case of some particular resonant
configuration -- the typical cross-sections just border the line of
$\sim 0.01\ $fb or less \,\cite{Djouadi:1999gv}. In the latter
reference, for instance, it has been shown that if the double and
triple Higgs production cross-sections would yield sufficiently high
signal rates, the system of couplings could in principle be solved
for all trilinear Higgs self-couplings up to discrete ambiguities
using only these processes. But this is perhaps a bit too optimistic
since in practice these cross-sections are manifestly too small to
be all measurable in a comfortable way.

In stark contrast to this meager panorama within the SM and the
MSSM, 3H couplings have been tested for tree-level processes in the
context of the unconstrained 2HDM. For example, in
Ref.\,\cite{Ferrera:2007sp} the tree-level production of triple
Higgs boson final states was considered in the ILC. There are three
classes of processes of this kind compatible with
$\CP$-conservation, namely
\begin{eqnarray}
&& 1)\ \APelectron\Pelectron \to \PHiggsplus\PHiggsminus
\PHiggslight \, ,\ \ \ 2)\ \APelectron\Pelectron \to \PHiggslight
\PHiggslight \Azero\,\nonumber ,\ \ \
\\
&&  3)\ \APelectron\Pelectron \to \hzero \Hzero \Azero\,, \ \ \
(\PHiggslight=\hzero,\Hzero,\Azero)\, \label{3H}
\end{eqnarray}
where, in class 2), we assume that the two neutral Higgs bosons
$\PHiggslight$ must be the same, i.e. the allowed final states are
$(\PHiggslight\, \PHiggslight \Azero)=(\hzero \hzero \Azero)$,
$(\Hzero \Hzero \Azero)$ and $(\Azero \Azero \Azero)$.  The
cross-sections in the 2HDM were shown to reach up to ${\cal
O}(0.1)\,\picobarn$\,\cite{Ferrera:2007sp}, i.e. several orders of
magnitude over the corresponding MSSM
predictions\,\cite{Djouadi:1992pu}. Besides the exclusive production
of three Higgs bosons in the final state, also the inclusive
pairwise production of Higgs bosons may be critically sensitive to
the 3H self-interactions. Sizable production rates, again in the
range of $0.1 - 1\picobarn$ have recently been reported in
Ref.\cite{Hodgkinson:2009uj}, whose focus was on the inclusive Higgs
boson-pair production at order $\mathcal{O}(\alpha^4_{ew})$ through
the mechanism of gauge-boson fusion
\begin{eqnarray}\label{2HX}
&&\APelectron\Pelectron \to V^*V^*\to h\,h + X \nonumber\\
&&(V=W^{\pm},Z\,,\ \ \ h =
\PHiggslightzero,\PHiggsheavyzero,\PHiggspszero, \PHiggs^\pm)\,.
\end{eqnarray}
Similar effects have also been recently computed on 2H-strahlung
processes of the guise $\APelectron\Pelectron \HepTo \PZ^0\, h\,h$
\cite{Arhrib:2008jp}. In a complementary way, 3H couplings can also
be probed in loop-induced processes. For instance,
Ref.~\cite{Bernal:2009rk} presents a computation of the single Higgs
boson production rate through the scattering process of i) two real
photons, using the $\Pphoton\Pphoton$ mode of a linear collider,
i.e.  $\Pphoton\Pphoton \to h$; and ii) the more traditional
mechanism of virtual photon fusion in $\APelectron\Pelectron$
colliders, $\APelectron\Pelectron \to \Pphoton^*\Pphoton^* \to h +
X$. In either case, the obtained cross sections within the general
2HDM (up to $1\,\picobarn$ for the first mechanism, and
$0.01\,\picobarn$ for the latter) are 1 to 2 orders of magnitude
larger than the expected SM yields and also well above the MSSM
results (see ~\cite{Bernal:2009rk} and references therein).
Promising signatures have also been reported within the general 2HDM
using loop-induced production of two neutral Higgs bosons through
real $\Pphoton\Pphoton$ collisions \cite{Cornet:2008nq}, although in
this case the cross-sections are smaller than in the primary
mechanism $\Pphoton\Pphoton \to h$ mentioned before.

A crucial observation concerning our aim here is the following: all
the processes described above are directly sensitive to the 3H
self-couplings already at the leading order. In this paper, we
continue exploiting the properties of these couplings in the general
2HDM, but we concentrate now on their impact at the level of
indirect effects through radiative corrections. The latter may
significantly affect processes that are kinematically more favored
(e.g. because they have a smaller number of particles in the final
state), but which are nevertheless totally insensitive to the
trilinear Higgs boson couplings at the lowest order. Specifically,
we wish to concentrate on identifying the largest quantum effects
that the 3H couplings can stamp on the cross-sections for two-body
neutral Higgs boson final states:
\begin{eqnarray}
  \APelectron\Pelectron \to 2h\,\ \ \ \ \ (2h \equiv \hzero\,\Azero; \Hzero\,\Azero)\,. \label{2h}
\end{eqnarray}
Surprisingly enough, very little attention has been paid to these
basic production processes and to the calculation of the
corresponding radiative corrections within the 2HDM. In contrast, a
lot of work has been invested on them from the point of view of the
MSSM\,\cite{Chankowski:1992er,DriesenHollik9596,Djouadi:1999gv,Djouadi:1996ah,Feng:1996xv,Heinemeyer:2001iy,Driesen:1996jd,Coniavitis:2007me}
-- see also \cite{Weiglein:2004hn} and references
therein\,\footnote{For the analysis of the tree-level double Higgs
production processes in the MSSM, see e.g. the exhaustive
overview\,\cite{Muhlleitner:2000jj}. For a detailed `anatomy' of the
MSSM Higgs sector and an updated account of its phenomenological
consequences, see e.g. \cite{Djouadi:2005gj}.}. To the best of our
knowledge, only Refs.~\cite{Arhrib:1998gr,kraft99,Guasch:2001hk}
have addressed this topic in the general 2HDM, although they are
restricted to the production of charged Higgs pairs
$\PHiggsplus\PHiggsminus$. In the present paper, we shall be
concerned exclusively on computing the quantum effects involved on
the neutral Higgs boson channels (\ref{2h}) at order $\order$ (and
eventually also the leading $\orderquadrat$ terms). As we will see,
very large quantum effects (of order $50\%$) may arise on the
cross-sections of the two-body Higgs boson channels (\ref{2h}) as a
result of the enhancement capabilities of the triple (and, to a
lesser extent, also the quartic) Higgs boson self-interactions.
These effects are completely unmatched within the MSSM and should
therefore be highly characteristic of non-supersymmetric Higgs boson
physics.

%
\section{The Two-Higgs-Doublet Model: general settings and relevant restrictions}
\label{sec:2HDM}

The Two-Higgs-Doublet Model (2HDM) is defined upon the canonical
extension of the SM Higgs sector with a second $SU(2)_L$ doublet
with weak hypercharge $Y\!=\!1$, so that it contains $4$ complex
scalar fields arranged as follows:
\begin{equation}
\Phi_1=\left(\begin{array}{c} \Phi_1^{+} \\ \Phi_1^0
\end{array} \right)
\ \ \ (Y=+1)\,,\ \ \
 \ \Phi_2=\left(\begin{array}{c} \Phi_2^{+} \\
\Phi_2^0
\end{array} \right)\ \ \ (Y=+1) \,\,.
\label{eq:H1H2}
\end{equation}
With the help of these two weak-isospin doublets, we can write down
the most general structure of the Higgs potential fulfilling the
conditions of $\mathcal{C}\mathcal{P}$-conservation, gauge
invariance and renormalizability. Moreover, a discrete $Z_2$
symmetry $\Phi_i\to (-1)^i\,\Phi_i\ (i=1,2)$ -- which will be exact
up to soft-breaking terms of dimension 2 --
is usually
imposed as a sufficient condition to guarantee the proper
suppression of the Flavor-Changing Neutral-Current (FCNC) effects
that would otherwise arise within the quark Yukawa sector \cite{GW}
{\footnote{For alternative strategies of building up
realizations of the 2HDM with no explicit $Z_2$ symmetry, see e.g.
\cite{mahmoudi} and references therein.}} (we shall return to this point
later)\,\footnote{This symmetry is automatic in the MSSM case,
although it is again violated after introducing the dimension 2
soft-SUSY breaking terms. These are essential for the EWSB, which
otherwise would not occur, see Eq.~(\ref{VMSSM}) below.}. All in
all, one arrives at the following expression for the tree-level
potential:
\begin{eqnarray}
V(\Phi_1,\Phi_2) =&& \lambda_1\,\left(\Phi_1^\dagger\,\Phi_1 -
\frac{v_1^2}{2}\right)^2 +
\lambda_2\,\left(\Phi_2^\dagger \Phi_2 - \frac{v_2^2}{2}\right)^2 \nonumber \\
&&+\lambda_3\,\left[\left(\Phi_1^\dagger\,\Phi_1 -
\frac{v_1^2}{2}\right)
+ \left(\Phi_2^\dagger\,\Phi_2 - \frac{v_2^2}{2}\right) \right]^2  \nonumber \\
&& +\lambda_4\,\left[(\Phi_1^\dagger \Phi_1)(\Phi^\dagger_2\Phi_2) -
(\Phi^\dagger_1\Phi_2)(\Phi^\dagger_2\Phi_1)\right]
\nonumber \\
&&+\lambda_5\, \left[\Re\,e\,(\Phi_1^\dagger \Phi_2) -\,
\frac{v_1\,v_2}{2}\right]^2\nonumber\\
 &&+  \lambda_6\,
\left[\Im\,m\,(\Phi_1^\dagger \Phi_2)
 \right]^2\,,
\label{eq:potential}
\end{eqnarray}
where $\lambda_i \,(i=1,\dots\,6)$ are dimensionless real parameters
and $v_{i}\, (i=1,2)$ stand for the non-vanishing VEV's that the
neutral component of each doublet acquires, normalized as follows:
$\langle\Phi^0_i\rangle=v_i/\sqrt{2}$.

The complex degrees of freedom encoded within each Higgs doublet in
Eq.(\ref{eq:H1H2}) are conveniently split into real ones in the
following way:
\begin{eqnarray}
\Phi_1 &=& \left(\begin{array}{c} \Phi_1^+ \\ \Phi_1^0 \end{array}
\right)
=\left(\begin{array}{c} \phi_1^+ \\ \frac{v_1 + \phi_1^0 +
i\chi_1^0}{\sqrt{2}}
\end{array} \right) \,,\ \ \ \ \ \nonumber \\
\Phi_2 &=& \left(\begin{array}{c} \Phi_2^+ \\ \Phi_2^0 \end{array}
\right)
=\left(\begin{array}{c} \phi_2^+ \\ \frac{v_2 + \phi_2^0 +
i\chi_2^0}{\sqrt{2}}
\end{array} \right)\,.
\label{eq:doublet2}
\end{eqnarray}
Upon diagonalization of the Higgs potential~(\ref{eq:potential}) we
may obtain the physical Higgs eigenstates in terms of the gauge
(weak-eigenstate) basis:
\begin{eqnarray}
&&\left( \begin{array}{c} \PHiggsheavyzero \\ \PHiggslightzero
\end{array} \right) = \left( \begin{array}{cc} \cos\alpha & \sin
\alpha \\ -\sin\alpha & \cos\alpha \end{array}
 \right)\,\left( \begin{array}{c} \phi_1^0 \\ \phi_2^0
 \end{array}\right)\nonumber\\
&&\left( \begin{array}{c} G^0 \\  \PHiggspszero \end{array} \right)
= \left( \begin{array}{cc} \cos\beta & \sin \beta \\ -\sin\beta &
\cos\beta \end{array} \right)\,\left( \begin{array}{c} \chi_1^0 \\
\chi_2^0 \end{array}\right)\nonumber\\
&&\left( \begin{array}{c} G^\pm \\  \PHpm \end{array} \right) =
\left( \begin{array}{cc} \cos\beta & \sin \beta \\ -\sin\beta &
\cos\beta \end{array} \right)\,\left( \begin{array}{c} \phi_1^{\pm} \\
\phi_2^{\pm} \end{array}\right) \label{eq:tl}
\end{eqnarray}
The parameter $\tan\beta$ is given by the ratio of the two VEV's
giving masses to the up- and down-like quarks:
\begin{equation}
\tan \beta \equiv \frac{\braket{\Phi_2^0}} {\braket{\Phi_1^0}}
=\frac{v_2}{v_1}\,. \label{tb}
\end{equation}
Incidentally, let us note that this parameter could be ideally
measured in $e^+e^-$ colliders\,\cite{Feng:1996xv} since the charged
Higgs event rates with mixed decays, $\APelectron\Pelectron \to
\PHiggs^+\PHiggs^- \to
\Ptop\APbottom\phantom{}{\Ptau\bar{\nu}_{\tau}},
\APtop\Pbottom\phantom{}{\nu_{\tau}\bar{\tau}}$, do not involve the
mixing angle $\alpha$. The physical content of
the 2HDM embodies one pair of $\mathcal{C}\mathcal{P}$-even Higgs
bosons ($\l$, $\H$); a single $\mathcal{C}\mathcal{P}$-odd Higgs
boson ($\A$); and a couple of charged Higgs bosons ($\Hpm$). The
free parameters of the general 2HDM are usually chosen to be: the
masses of the physical Higgs particles; the aforementioned parameter
$\tan\beta$; the mixing angle $\alpha$ between the two \CP-even
states; and finally the coupling $\lambda_5$, which cannot be
absorbed in the previous quantities and becomes tied to the
structure of the Higgs self-couplings. To summarize, the vector of
free inputs reads
\begin{equation}\label{freep}
  (M_{\hzero},M_{\Hzero},M_{\Azero},M_{\Hpm},\alpha,\tan\beta, \lambda_5)\,.
\end{equation}
Therefore, we are left with $7$ free parameters, which indeed
correspond to the original $6$ couplings $\lambda_i$ and the two
VEV's $v_1,v_2$ -- the latter being submitted to the constraint
$v^2\equiv v_1^2+v_2^2= 4\,M_W^2/g^2$, which is valid order by order
in perturbation theory, where $M_W$ is the $W^{\pm}$ mass and $g$ is
the $SU_L(2)$ gauge coupling constant. The dimension 2 term that
softly breaks the $Z_2$ symmetry can be written as
$-m_{12}^2\Phi_1^\dagger \Phi_2+\text{h.c.}$, with
\begin{equation}\label{m12}
m_{12}^2=\frac12\,\lambda_5\,v^2\,\sin\beta\cos\beta=\frac{G_F^{-1}}{2\sqrt{2}}\,\frac{\tan\beta}{1+\tan^2\beta}\,\lambda_5\,,
\end{equation}
the second equality being valid at the tree-level. Given $\tan\beta$
one may trade the parameter $\lambda_5$ for $m_{12}$ through this
formula, if desired.

Since we are going to adopt the on-shell renormalization
scheme\,\cite{Aoki82,Bohm:1986rj,Sirlin:1980nh} for the one-loop
calculation of the cross-sections (cf. section \ref{sec:renorm}), it
is convenient to introduce the electromagnetic fine structure
constant $\alpha_{em}=e^2/4\pi$, which is one of the fundamental
inputs in this scheme (together with the physical $\PZ^0$ mass,
$M_Z$, and the Higgs and fermion particle masses). The electron
charge $-e$ is connected to the original $SU_L(2)$ gauge coupling
and weak mixing angle $\theta_W$ through the well-known relation
$e=g\sin\theta_W$, which is also preserved order by order in
perturbation theory. Then the $W^{\pm}$ mass can be related to
$\alpha_{em}$ and the Fermi constant $G_F$ in the standard way in
the on-shell scheme, namely \jump
\begin{equation}\label{Deltar}
\frac{G_F}{\sqrt{2}}=\frac{\pi\,\alpha_{em}}{2\,M_W^2\,\sin^2\theta_W}\,(1-\Delta
r)\,,
\end{equation}\jump
where the parameter $\Delta r$\,\cite{Sirlin:1980nh} vanishes at the
tree-level, but is affected by the radiative corrections to
$\mu$-decay both from standard as well as from new physics, for
instance from the MSSM\,\cite{Garcia:1993rq,Heinemeyer:2006px} and
also from the 2HDM (see below). Notice that, in the above formula,
$M_W$ also enters implicitly through the relation
$\sin^2\theta_W=1-M_W^2/M_Z^2$, which is valid order by order in the
on-shell scheme. Since $G_F$ can be accurately determined from the
$\mu$-decay, and $M_Z$ has been measured with high precision at LEP,
it is natural to take them both as experimental inputs. Then, with
the help of the non-trivial relation (\ref{Deltar}), the $W^{\pm}$
mass can be accurately predicted in a modified on-shell scheme where
$G_F$, rather than $M_W$, enters as a physical input.

We shall not dwell here on the detailed structure of the Yukawa
couplings of the Higgs boson to fermions (see \cite{\hunter} for an
exhaustive treatment), since their contribution to the processes
$\APelectron\Pelectron \to \hzero\Azero/\Hzero\Azero$ under study at the one-loop level is largely subdominant
in the main regions of parameter space. However, it is precisely the
coupling pattern to fermions that motivates the distinction between
the different types of 2HDM's, so let us briefly describe them.
Assuming natural flavor conservation\,\cite{GW}, we must impose that
at most one Higgs doublet can couple to any particular fermion type.
As a result, the coupling of the 2HDM Higgs bosons with fermions
(say the quarks) can be implemented in essentially two manners which
insure the absence of potentially dangerous (tree-level) FCNC's, to
wit (cf. Table \ref{typeIandII})\,\footnote{Strictly speaking, in
the absence of \CP-violation there are four manners to have natural
flavor conservation, giving rise to four different kinds of allowed
2HDM's that do not lead to tree-level FCNC's, see
\cite{Barger:1989fj}, although we will restrict ourselves here to
the two canonical ones\,\cite{Hunter}.}: i) type-I 2HDM, wherein the
Higgs doublet ($\Phi_2$) couples to all of quarks, whereas the other
one ($\Phi_1$) does not couple to them at all; or ii) type-II 2HDM,
in which the doublet $\Phi_1$(resp. $\Phi_2$) couples only to
down-like (resp. up-like) right-handed quarks. In the latter case,
an additional discrete symmetry involving the chiral components of
the fermion sector must be imposed in order to banish the tree-level
FCNC processes, e.g. $D_R^i \to -\,D_R^i\,,\, U_R^i \to \,U_R^i\ $
for the down and up right-handed quarks in the three flavor families
$(i=1,2,3)$.

In turn, the Higgs self-couplings $\lambda_i$ in the Higgs potential
can be rewritten in terms of $\tan\beta$ and the physical parameters
of the on-shell scheme such as the masses and the electromagnetic
fine structure constant $\alpha_{em}$. At the tree-level, we have
\begin{eqnarray}
\lambda_1 &&= \frac{\lambda_5\,(1-\tan^2\beta)}{4}\, +
 \frac{\alpha_{em}\,\pi\,}{2\,M_W^2\,\swd\,\cos^2\beta}
\nonumber \\
&& \times\left[M_{H^0}^2\,\cos^2\alpha + M^2_{h^0}\,\sin^2\alpha\nonumber \right. \\
&& \left. -\frac12(M^2_{H^0}-M^2_{h^0})\,\sin 2\alpha\,\cot\beta\right]\nonumber\,, \\
\lambda_2 &=& \frac{\lambda_5\,(1-1/\tan^2\beta)}{4}\, + \frac{\alpha_{em}\,\pi\,}{2\,M^2_W\,\swd\,\sin^2\beta}\nonumber \\
&& \times\left[M_{h^0}^2\,\cos^2\alpha +
M_{H^0}^2\,\sin^2\alpha\nonumber  \right. \\
&& \left. -\frac12(M^2_{H^0}-M^2_{h^0})\sin
2\alpha\,\tan\beta\right]\nonumber
\,,  \\ \nonumber \\
\lambda_3 &=& -\,\frac{\lambda_5}{4}\,+\frac{\alpha_{em}\,\pi\,\sin
2\alpha} {2\,M^2_W\,\swd\,\sin 2\beta}\,(M^2_{H^0} -
M^2_{h^0})\nonumber\,,
\,\\
\lambda_4 &=& \frac{2\,\alpha_{em}\,\pi}{M^2_W\,\swd}\,M^2_{H^\pm}\nonumber\,,  \\ \nonumber  \\
\lambda_6 &=& \frac{2\,\alpha_{em}\,\pi}{M^2_W\,\swd}\,M^2_{A^0}\,,
\label{eq:param2hdm}
\end{eqnarray}
where $\sw=\sin\theta_W$, $\cw=\cos\theta_W$. From these Lagrangian
couplings in the Higgs potential, we can derive the ``physical
couplings'', namely those affecting the physical Higgs bosons in the
mass-eigenstate basis. We call these couplings the triple (3H) and
quartic (4H) Higgs couplings. Their behavior and enhancement
capabilities are at the very core of our discussion. Indeed, from
our analysis it will become clear that they furnish the dominant
source of quantum corrections to the Higgs-pair production processes
(\ref{2h}) within the general 2HDM.

The physical 3H and 4H couplings are not explicitly present in the
2HDM potential (\ref{eq:potential}).  They are derived from it after
spontaneous breaking of the EW symmetry and corresponding
diagonalization of the Higgs boson mass matrix using the rotation
angles $\alpha$ and $\beta$ in (\ref{eq:tl}). In the particular case
of the SM, the trilinear and quartic Higgs couplings have fixed
values, which depend uniquely on the actual mass of the Higgs boson.
In the MSSM, however, and due to the SUSY invariance, the Higgs
boson self-couplings are of pure gauge nature, as we shall revise
briefly below. This is in fact the primary reason for the tiny
production rates obtained for the triple Higgs boson processes
(\ref{3H}) within the framework of the MSSM\,\cite{Djouadi:1999gv}.
In contrast, the general 2HDM accommodates trilinear Higgs couplings
with great potential enhancement. The full list is displayed in
Table~\ref{tab:trilinear}.

As can be seen, the couplings in this table depend on the $7$ free
parameters (\ref{freep}). In the particular case where $\lambda_5 =
\lambda_6$, this table reduces to Table 1 of
Ref.~\cite{Ferrera:2007sp}. The equality of these couplings takes
place automatically e.g. in the MSSM, where in addition other
simplifications occur, as we discuss below. Let us note, for
example, that in the limit $\alpha = \beta - \pi/2$ the
$\hzero\hzero\hzero$-trilinear coupling  in
Table~\ref{tab:trilinear} reduces exactly to the SM form
$i\lambda_{HHH}^{\rm SM}=-3ig\,M_H^2/(2M_W)$ -- where we denote by
$H$ the SM Higgs boson. This situation would correspond to the
so-called decoupling limit in the MSSM\,\cite{Gunion:2002zf} since
it is correlated with $M_{\Azero}\to\infty$, although there is no
such correlation in the general 2HDM.
\begin{table}[t]
\centering{
\begin{tabular} {|c|c|c|} \hline
& type~I & type~II  \\ \hline $\hzero t\overline t$ & $\
\cos\alpha/\sin\beta$\ &  $\ \cos\alpha/\sin\beta\ $
\\ \hline $\hzero b\overline b$ & $\ \cos\alpha/\sin\beta\ $
&$\ -\sin\alpha/\cos\beta\ $ \\
\hline $\Hzero t\overline t$ & $\ \sin\alpha/\sin\beta\ $ & $\
\sin\alpha/\sin\beta\ $
\\ \hline $\Hzero b\overline b$ & $\ \sin\alpha/\sin\beta\ $
&$\ \cos\alpha/\cos\beta\ $ \\
\hline $\Azero t\overline t$ & $\ \cot\beta$ & $\cot\beta\ $
\\ \hline $\Azero b\overline b$ & $\ -\cot\beta\ $
&$\ \tan\beta\ $
\\ \hline
\end{tabular}
\caption{Neutral Higgs boson couplings to fermions in type-I and
type-II 2HDM, using third family notation. For example, the SM Higgs
boson coupling to top and bottom quarks, $-g\,m_f/2M_W\, (f=t,b)$,
multiplied by the corresponding factor in the table provides the
2HDM couplings of the $\hzero$ boson to these quarks.
Worth noticing are the characteristic enhancement
factors arising at large (and low) values of $\tan\beta$.}}
\label{typeIandII}
\end{table}
\jump
\begin{table*}
         \begin{tabular}{|c|c|} \hline & \\
$\hzero\hzero\hzero$ & $
 -\frac{3 i e}{2 M_W
   \sin2\beta\, \sw} \left[M_{h^0}^2 (2 \cos(\alpha+\beta)+\sin2\alpha
   \sin(\beta-\alpha)) \right.$ \\
& $ \phantom{-\frac{3 i e}{2 M_W
   s_{2\beta}\, \sw}} \left. -\cos(\alpha+\beta) \cos^2(\beta-\alpha)\,\ \frac{4 \lambda_5
   M_W^2\, \sw^2}{e^2}\right] $
\\ & \\ \hline & \\
$\hzero\hzero\Hzero $& $  -\frac{i e \cos(\beta-\alpha)}{2 M_W
\sin2\beta\,\sw} \left[\left(2
   M_{h^0}^2+M_{H^0}^2\right) \sin 2\alpha
  - (3 \sin\,2\alpha-\sin2\beta)\,\ \frac{2
   \lambda_5 M_W^2\,\swd}{e^2}\right] $ \\ & \\ \hline & \\ $\hzero\Hzero\Hzero$ & $
 \frac{i e \sin(\beta-\alpha)}{2 M_W \sin2\beta\,\sw}\left[\left(M_{h^0}^2+2
   M_{H^0}^2\right) \sin2\alpha
- (3 \sin2\alpha+\sin2\beta)\,\swd\,\,\frac{2 \lambda_5\,
   M_W^2}{e^2}\right] $ \\ & \\ \hline & \\
$\Hzero\Hzero\Hzero$ & $ -\frac{3 i e}{2 M_W
   \sin2\beta\,\sw} \left[M_{H^0}^2 (2 \sin(\alpha+\beta)-\cos(\beta-\alpha)
   \sin 2\alpha) \right. $ \\
& $\left. \phantom{-\frac{3 i e}{2 M_W
   \sin2\beta\,\sw}}- \sin(\alpha+\beta)
   \sin^2(\beta-\alpha)\,\swd\ \frac{4\,\lambda_5 M_W^2}{e^2}\right]
$ \\ & \\ \hline & \\ $\hzero\Azero\Azero$ & $
 -\frac{i e}{2 M_W\,\sw}\left[\frac{\cos(\alpha+\beta)}{\sin2\beta} \left(2
   M_{h^0}^2 -\,\frac{4\,\lambda_5\,
M_W^2\,\swd}{e^2}\right)-\sin(\beta-\alpha)\,\left(M_{h^0}^2
   -2 M_{A^0}^2\right) \right]
$
 \\ & \\ \hline & \\
$\hzero\Azero G^0$ & $ \frac{i e}{2 M_W\,\sw}
   \left(M_{A^0}^2-M_{h^0}^2\right)\,\cos(\beta-\alpha)
$
\\ &  \\ \hline & \\
$\hzero G^0 G^0$ & $ -\frac{i e}{2 M_W\,\sw} M_{h^0}^2
\sin(\beta-\alpha)$ \\ & \\ \hline & \\ $\Hzero \Azero \Azero $ & $
-\frac{i e}{2 M_W\,\sw} \left[\frac{\sin(\alpha+\beta)}{\sin2\beta}
\left(2
   M_{H^0}^2-\,\frac{4\,\lambda_5\, M_W^2\,\swd}{e^2}\right)-\cos(\beta-\alpha)
   \left(M_{H^0}^2-2 M_{A^0}^2\right)\right]
$ \\ & \\ \hline & \\ $ \Hzero \Azero G^0$ & $
 -\frac{i e}{2 M_W\,\sw} \left(M_{A^0}^2-M_{H^0}^2\right)
   \sin(\beta-\alpha)$ \\ & \\ \hline & \\
$ \Hzero G^0 G^0$ & $-\frac{ie}{2 M_W
\sw}\,M^2_{\Hzero}\cos(\beta-\alpha)$\\ &
\\ \hline & \\
$\hzero \Hp \Hm$ & $ -\frac{i e}{2 M_W\,\sw}
\left[\frac{\cos(\alpha+\beta)}{\sin2\beta} \left(2
   M_{h^0}^2-\,\frac{4 \lambda_5\, M_W^2\,\swd}{e^2}\right)-\left(M_{h^0}^2
   -2 M_{H^-}^2\right) \sin(\beta-\alpha)\right]
$ \\ & \\ \hline & \\ $\Hzero \Hp \Hm$ & $
 -\frac{i e}{2 M_W\,\sw} \left[\frac{\sin(\alpha+\beta)}{\sin2\beta} \left(2
   M_{H^0}^2-\,\frac{4 \lambda_5\, M_W^2\,\swd}{e^2}\right)-\cos(\beta-\alpha)
   \left(M_{H^0}^2-2 M_{H^-}^2\right)\right]
$ \\ & \\ \hline
         \end{tabular}
         \caption{Trilinear Higgs boson self-interactions ($i\lambda_{3H}$) in the Feynman
           gauge within the 2HDM. Here $G^0$ is the neutral Goldstone boson. These vertices are involved
           in the radiative corrections to the $\vzlA(\vzHA)$ bare couplings.}
         \label{tab:trilinear}
\end{table*}
\jump

Let us recall that the MSSM Higgs sector is a type-II one of a very
restricted sort: it is enforced to be SUSY invariant. This is a very
demanding requirement as, for example, it requires that the
superpotential has to be a holomorphic function of the chiral
superfields \,\cite{MSSM,\hunter}). This implies that we cannot
construct the MSSM with two $Y=+1$ Higgs superfields. Then, in order
to be able to generate masses for both the top and bottom quarks
through EWSB, the $\Phi_2$ doublet can be kept as it is with $Y=+1$,
although we must replace $\Phi_1$ with the conjugate ($Y=-1$)
$SU_L(2)$ doublet\jump
\begin{equation}\label{eq:Yminus}
H_1= \left(\begin{array}{c} H_1^0 \\
H_1^{-}\end{array} \right)\equiv\epsilon\,\Phi_1^*=\left(\begin{array}{c} \Phi_1^{0*} \\
-\Phi_1^{-}
\end{array} \right)
\ \ (Y=-1)\,,
\end{equation}\jump
where $\epsilon=i\,\sigma_2$, with $\sigma_2$ the second Pauli
matrix. Thus, for the first doublet the correspondence with the MSSM
case reads $\Phi_1=-\epsilon\,H_1^*$, and the second doublet (the
one with no change) is just relabeled $H_2$. Moreover, the
tree-level relationships imposed by Supersymmetry  between the
$\lambda_i$ couplings of the potential (\ref{eq:potential}) are the
following: \jump
\begin{eqnarray}\label{eq:lambdasMSSM}
\lambda_1 &=&\lambda_2\,,\nonumber\\
\lambda_3 &=&\frac{\pi\,\alpha_{em}}{2\,\swd\,\cwd}
-\lambda_1\,,\nonumber\\
\lambda_4 &=& 2\lambda_1 -
\frac{2\,\pi\,\alpha_{em}}{\cwd}\,,\nonumber\\
\lambda_5 &=&\lambda_6= 2\lambda_1-
\frac{2\pi\,\alpha_{em}}{\swd\,\cwd}\,.
\end{eqnarray}\jump
Substituting (\ref{eq:Yminus}) and (\ref{eq:lambdasMSSM}) in
(\ref{eq:potential}) one obtains the usual MSSM potential, which in
practice must be supplemented with soft SUSY-breaking scalar mass
terms $m_i^2\,H_i^2$, including a bilinear mixing term
$m_{12}^2\,H_1\,H_2$ for the two Higgs doublets.  This term is the
analog of (\ref{m12}) for the 2HDM, but in the soft-SUSY breaking
context is arbitrary. The result can be cast as follows: \jump
\begin{eqnarray}\label{VMSSM}
&&V(H_1,H_2) = \left(|\mu|^2+m_1^2\right)\,|H_1|^2 +
\left(|\mu|^2+m_2^2\right)\,|H_2|^2\nonumber\\
&&+\frac{\pi\alpha_{em}}{2\,\sw^2\cw^2}\left(|H_1|^2-|H_2|^2\right)^2+
\frac{2\pi\alpha_{em}}{\sw^2}\,|H_1^{\dagger}H_2|^2\nonumber\\
&&-m_{12}^2\left(\epsilon_{ij}H_1^i\,H_2^j+h.c.\right)\,,
\end{eqnarray}
\jump where $\mu$ is the higgsino mass term in the
superpotential\,\cite{MSSM}. The obtained potential is one where all
quartic couplings become proportional to $\alpha_{em}$, which is
tantamount to say that in the SUSY case these self-couplings are
purely gauge. Thus, after EWSB also the trilinear self-couplings
will be purely gauge. As warned before, this is the primary reason
for their phenomenological inconspicuousness. Under the
supersymmetric constraints, the entries of Table~\ref{tab:trilinear}
boil down to the MSSM form listed e.g. in \cite{\hunter}. Moreover,
the five constraints (\ref{eq:lambdasMSSM}) reduce the number of
free parameters from $7$ to $2$ in the supersymmetric context,
typically chosen to be $\tan\beta$ and $M_{A^0}$. For the general
2HDM case, however, we shall stick to the form presented in
Table~\ref{tab:trilinear}, where the $7$ free inputs are chosen as
in (\ref{freep}).

In another vein, it is of paramount importance when studying
possible sources of unconventional physics to make sure that the SM
behavior is sufficiently well reproduced up to the energies explored
so far. Such a requirement translates into a number of constraints
over the parameter space of the given model. In particular, these
constraints severely limit the a priori enhancement possibilities of
the Higgs boson self-interactions in the 2HDM.

\begin{itemize}

\item{
To begin with, we obviously need to keep track of the exclusion
bounds from direct searches at LEP and the Tevatron. These amount to
$M_{\l} \gtrsim 114 \GeV$ for a SM-like Higgs boson -- although the
bound is weakened down to $92.8 \,\GeV$ for models with more than
one Higgs doublet\,\cite{pdg}. In the 2HDM all the mass
bounds can be easily satisfied upon a proper choice of the different
Higgs masses -- which, unlike the SUSY case, are not related with
each other.}

\item{Second, an important requirement to be enforced is related to
the (approximate) $SU_{L+R}(2)$ custodial symmetry satisfied by
models with an arbitrary number of Higgs doublets
\cite{Einhorn:1981cy}. In practice, this restriction is implemented
in terms of the parameter $\rho$, which defines the ratio of the
neutral-to-charged current Fermi constants. In general it takes the
form $\rho=\rho_0+\delta\rho$, where $\rho_0$ is the tree-level
value. In any model containing an arbitrary number of doublets (in
particular the 2HDM), we have $\rho_0={M_W^2}/{M_Z^2\cwd}=1$, and
then $\delta\rho$ represents the deviations from $1$ induced by pure
quantum corrections. From the known SM contribution and the
experimental constraints\,\cite{pdg} we must demand that the
additional quantum effects coming from 2HDM dynamics satisfy
\phantom{}{the approximate condition} $|\delta\rho_{2HDM}|\lesssim
10^{-3}$. It is thus important to stay in a region of parameter
space where this bound is respected. In our calculation we include
the dominant part of these corrections, which involves one-loop
contributions to $\delta\rho$ mediated by the Higgs-boson and yields
\,\cite{Barbieri:1983wy}
\begin{eqnarray}\label{drho}
\delta\rho_{{\rm
2HDM}}&=&\frac{G_F}{8\sqrt{2}\,\pi^2}\Big\{M_{H^{\pm}}^2
\left[1-\frac{M_{A^0}^2}{M_{H^{\pm}}^2-M_{{A^0}}^2}\,
\ln\frac{M_{H^{\pm}}^2}{M_{A^0}^2}\right]
\nonumber \\
&+&\cos^2(\beta-\alpha)\,M_{h^0}^2\left[\frac{M_{A^0}^2}{M_{A^0}^2-M_{h^0}^2}\,
\ln\frac{M_{A^0}^2}{M_{h^0}^2} \right. \nonumber \\
&-&\left.\frac{M_{H^{\pm}}^2}{M_{H^{\pm}}^2-M_{h^0}^2}\,
\ln\frac{M_{H^{\pm}}^2}{M_{h^0}^2}\right]\nonumber\\
&+&\sin^2(\beta-\alpha)\,M_{H^0}^2\left[\frac{M_{A^0}^2}{M_{A^0}^2-M_{H^0}^2}\,
\ln\frac{M_{A^0}^2}{M_{H^0}^2} \right. \nonumber \\
&-& \left.\frac{M_{H^{\pm}}^2}{M_{H^{\pm}}^2-M_{H^0}^2}\,
\ln\frac{M_{H^{\pm}}^2}{M_{H^0}^2}\right]\Big\}\, \nonumber \\
\end{eqnarray}
From this expression, it is clear that arbitrary mass splittings
between the Higgs bosons could easily overshoot the limits on
$\delta\rho$. However, we note that if $M_{\Azero}\to M_{H^{\pm}}$
then $\delta\rho_{{\rm 2HDM}}\to 0$, and hence if the mass splitting
between $M_{A^0}$ and $M_{H^{\pm}}$ is not significant
$\delta\rho_{{\rm 2HDM}}$ can be kept under control
\footnote{For more refined analyses on $\delta\rho$ constraints, see e.g. \cite{Grimus:2007if}.}.
Let us also recall in passing
that the $\delta\rho$ correction translates into a contribution to
the parameter $\Delta r$ of (\ref{Deltar}) given by
$-(\cwd/\swd)\,\delta\rho$. Therefore, the tight bounds on $\Delta
r$, together with the direct bounds on the ratio of charged and
neutral weak interactions, restrict this contribution as indicated
above.}
\end{itemize}

\begin{itemize}
\item{Also remarkable are the restrictions over the charged Higgs masses
coming from FCNC radiative $B$-meson decays, whose branching ratio
$\mathcal{B}(b \to s \gamma)\simeq 3 \times 10^{-4}$\,\cite{pdg} is
measured with sufficient precision to be sensitive to new physics.
Current lower bounds render $M_{\Hpm}\gtrsim 295\,\GeV$ for
$\tan\beta\gtrsim 1$ \cite{Misiak:2006ab,ElKaffas:2007rq}. It must
be recalled that these bounds apply for both Type-I and Type-II models
in the region of $\tan\beta < 1$. In contrast, they are only relevant
for Type-I in the large $\tan\beta$ regime, since
for them the charged Higgs couplings to fermions are proportional to
$\cot\beta$ and hence the loop contributions are highly suppressed
at large $\tan\beta$ -- cf. Ref.\cite{Borzumati:1998tg} for detailed analytical expressions
of the contributions to $\mathcal{B}(b \to s \gamma)$ from
the 2HDM. Besides, data from $B \to l\nu_l$
may supply additional restrictions on the charged Higgs mass at
large $\tan\beta$ for the type-II 2HDM \cite{Akeroyd:2007eh}. For a
very recent updated analysis of the different $B$-physics
constraints on the 2HDM parameter space, see Ref.~\cite{mahmoudi}}.

\item{Further constraints apply to $\tan\beta$ coming from the following two sources:
i) The $\PZ^0 \to \APbottom\Pbottom$ and $\PB - \APB$ mixing
processes strongly disfavor $\tan\beta$ below 1
\cite{ElKaffas:2007rq}; ii) The requirement that the Higgs couplings
to heavy quarks remain perturbative translates into an (approximate)
allowed range of ${\cal O}(0.1) < \tan\beta < 60$
\cite{Barger:1989fj}.}
\item{ Besides the available experimental data, an additional set
of requirements ensues from the theoretical consistency of the
model. In particular, we shall introduce the following set of
conditions ensuring the stability of the vacuum
\cite{Deshpande:1977rw,Nie:1998yn,Kanemura:1999xf}:
\begin{eqnarray}
&& \lambda_1 + \lambda_3 > 0\, ; \quad \lambda_2 + \lambda_3 > 0 \nonumber \\
&& 2\sqrt{(\lambda_1 + \lambda_3)(\lambda_2 + \lambda_3)} +
2\lambda_3 + \lambda_4
\nonumber \\
&&+ \mbox{min}[0, \lambda_5 - \lambda_4, \lambda_6 - \lambda_4] > 0
\label{eq:vacuum_conditions}.
\end{eqnarray}
 }
\end{itemize}
%
The unitarity bounds deserve a separate discussion. Indeed, they
turn out to be a fundamental ingredient of our computation. As we
have already been mentioning, the trademark behavior that we aim to
explore shall depend critically on the enhancement capabilities of
the Higgs-boson self-interactions which, in turn, will be sharply
constrained by unitarity. The basic idea here is that, within
perturbative QFT, the scattering amplitudes are ``asymptotically
flat'', meaning that they cannot grow indefinitely with the energy.
This is tantamount to say that the unitarity of the S-matrix must be
guaranteed at the perturbative level. In a pioneering work by Lee,
Quigg and Thacker (LQT)\,\cite{Lee:1977eg}, the above arguments were
first applied to the SM. As a very relevant outcome, an upper bound
for the Higgs boson was obtained:
\begin{eqnarray}
M^2_{\PHiggs} < 8\pi\,v^2 = \frac{8\pi}{\sqrt{2}\,G_F}\equiv
M^2_{\mbox{LQT}}\simeq (1.2\,\TeV)^2\,. \label{eq:lqt}
\end{eqnarray}
The basic idea beneath the LQT argument is to compute the scattering
amplitudes in a variety of processes (namely scalar-scalar,
scalar-vector and vector-vector scattering; most particularly, the
longitudinal components of the vector bosons) and demand all of them
to be in accordance with the generic tree-level unitarity condition.
The latter is a pure quantum-mechanical requirement, which can be
expressed in terms of the s-wave scattering lenght in the
high-energy limit:
\begin{eqnarray}
|a_0| \leq 1/2\,. \label{eq_uncond}
\end{eqnarray}
Taking into account the equivalence between the longitudinal
components of the gauge and the Goldstone bosons (which is exact to
order $\mathcal{O}(\sqrt{M_W^2/s})$, s being the center-of-mass
energy squared), we can convince ourselves that only the
scalar-scalar processes will be relevant to this concern. Moreover,
it can also be shown that the bulk of the contribution is brought
about by the quartic scalar self-couplings -- the triple ones are
suppressed as 1/s. Thereby the quartic Higgs self-couplings (4H)
turn out to be the most strongly constrained by the bounds of
Eq.~(\ref{eq_uncond}).


Several authors have exported the above ideas to the 2HDM
\cite{Huffel:1980sk,Casalbuoni:1986hy,Casalbuoni:1987cz,Maalampi:1991fb,Kanemura:1993hm,Arhrib:2000is,Akeroyd:2000wc,Kanemura:2002vm,Kanemura:2002cc,Ginzburg:2003fe,Kanemura:2004mg,Ginzburg:2004vp,Horejsi:2005da,Ginzburg:2005dt}
and also to its complex ($\CP$-violating) extensions
\cite{Osland:2008aw}. The strategy here is to compute the S-matrix
elements $S_{ij}$ in a set of $2 \to 2$ scattering processes
involving Higgs and Goldstone bosons. In this case there are more
scalar-scalar channels than in the SM, which are coupled among
themselves. At this point, we could perform a unitary transformation
$U$ relating the S-matrix expressed in the original weak-interaction
eigenstate basis ($S_w$) with the S-matrix written in the physical
mass-eigenstate basis ($S_m$).
Since, however, the former is much simpler than the latter and both
carry exactly the same information, we may generalize the original
LQT procedure\,\cite{Lee:1977eg} by focusing on $S_w $ and requiring
that its eigenvalues satisfy the tree-level unitarity condition
$|{\alpha}_i|<1\ (\forall i)$. In practice, inspired by the
scattering length analysis (\ref{eq_uncond}), we require
$|{\alpha}_i|<1/2\ (\forall i)$. This leads to a set of conditions
over several linear combinations of the quartic couplings in the
original Higgs potential
\cite{Arhrib:2000is,Akeroyd:2000wc,Kanemura:2002vm,Kanemura:2002cc,Ginzburg:2003fe,Kanemura:2004mg,Ginzburg:2004vp,Horejsi:2005da,Ginzburg:2005dt},
as can be seen from the explicit form of the various eigenvalues
$\alpha_i$ at the tree-level:

\begin{eqnarray}
a_\pm & = & \frac{1}{16 \pi} \Big\{  3 (\lambda_1 + \lambda_2 + 2 \lambda_3) \nonumber \\
       &&\pm ( \sqrt{9 (\lambda_1-\lambda_2)^2
        + (4 \lambda_3+\lambda_4+\frac{\lambda_5}{2}+\frac{\lambda_6}{2})^2}
          \Big\}\;, \nonumber\\
b_\pm & = &\frac{1}{16 \pi} \Big\{ \lambda_1+ \lambda_2+ 2 \lambda_3 \nonumber \\
       &&\pm \sqrt{(\lambda_1-\lambda_2)^2
    + \frac{(-2 \lambda_4+\lambda_5+\lambda_6)^2}{4}}
          \Big\}\;, \nonumber \\
c_\pm & = &d_\pm = \frac{1}{16 \pi}   \Big\{ \lambda_1+ \lambda_2+ 2 \lambda_3 \nonumber \\
       &&\pm \sqrt{(\lambda_1-\lambda_2)^2 + \frac{(\lambda_5-\lambda_6)^2}{4}}
          \Big\}\;, \nonumber \\
e_1 & =& \frac{1}{16 \pi} \Big\{2 \lambda_3 - \lambda_4 -\frac{\lambda_5}{2} + 5 \frac{\lambda_6}{2} \Big\}\;, \nonumber \\
e_2 & =& \frac{1}{16 \pi} \Big\{2 \lambda_3 + \lambda_4 -
\frac{\lambda_5}{2} + \frac{\lambda_6}{2}\Big\}\;,
\nonumber \\
f_+ & = & \frac{1}{16 \pi} \Big\{2 \lambda_3 - \lambda_4 + 5 \frac{\lambda_5}{2} - \frac{\lambda_6}{2}\Big\}\;, \nonumber \\
f_- & = & \frac{1}{16 \pi} \Big\{2 \lambda_3 + \lambda_4 +
\frac{\lambda_5}{2} - \frac{\lambda_6}{2}\Big\}\;, \nonumber
\end{eqnarray}
\begin{eqnarray}
f_1 & = &f_2 = \frac{1}{16 \pi} \Big\{2 \lambda_3 +
\frac{\lambda_5}{2} + \frac{\lambda_6}{2}\Big\}\;,
\nonumber \\
p_1 &=& \frac{1}{16 \pi}\, \Big\{2(\lambda_3 + \lambda_4)
-\frac{\lambda_5+\lambda_6}{2}\Big\}\; .
\nonumber \\
\label{eq:arbounds}
\end{eqnarray}


\section{Renormalization of the 2HDM Higgs sector}
\label{sec:renorm}

\begin{figure}[htb]
\includegraphics[scale=0.8]{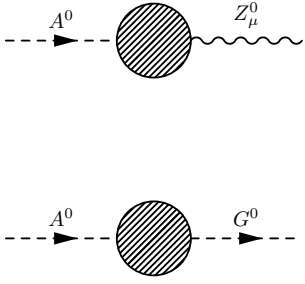}
\caption{Renormalized scalar-vector and scalar-scalar self-energies
at all orders in perturbation theory.} \label{fig:blobs}
\end{figure}

In this Section, we discuss in detail the renormalization of the
Higgs sector. The renormalization of the SM fields and parameters is
performed in the conventional on-shell scheme in the Feynman gauge,
see e.g. Ref. ~\cite{Aoki82,Bohm:1986rj,Sirlin:1980nh,Hollik95,
Denner:1991kt}. At present, highly automatized procedures are
available for loop calculations, especially in the MSSM, see e.g.
\cite{Heinemeyer:2004gx}
and\,\cite{feynarts,feynHiggs,AutomatRen,Baro:2008bg}. However, in
our case the calculation of the cross-sections for the processes
(\ref{2h}) is performed within the general (non-supersymmetric) 2HDM
and we must deal with the renormalization of the Higgs sector in
this class of generic models. To this end, we attach a
multiplicative wave-function (WF) renormalization constant to each
of the $SU_L(2)$ Higgs doublets in the 2HDM,
\begin{equation}
\left(\begin{array}{c} \Phi_1^+ \\ \Phi_1^{0}
\end{array} \right)\rightarrow Z_{\Phi_1}^{1/2}
\left(\begin{array}{c} \Phi_1^+ \\ \Phi_1^{0}
\end{array} \right)\,,\ \ \ \ \
\left(\begin{array}{c} \Phi_2^{+} \\ \Phi_2^0
\end{array} \right)\rightarrow Z_{\Phi_2}^{1/2}
\left(\begin{array}{c} \Phi_2^{+} \\ \Phi_2^0
\end{array} \right)\,,
\label{eq:doublets}
\end{equation}
The renormalized fields are those on the \textit{r.h.s.} of these
expressions. At one-loop we decompose $Z_{\Phi_{1,2}} = 1 + \delta
Z_{\Phi_{1,2}} + \mathcal{O}(\alpha_{ew})$.   These WF
renormalization constants in the weak-eigenstate basis can be used
to construct the WF renormalization constants $Z_{h_ih_j} = 1 +
\delta Z_{h_ih_j}$ in the mass-eigenstate basis by means of the set
of linear transformations (\ref{eq:tl}). We are interested only in
the production of neutral Higgs bosons in this paper, and therefore
we shall only quote the relations referring to them. The
corresponding $\delta\,Z_{h_ih_j}$ in the neutral Higgs sector
(using the notation $\delta\,Z_{h_ih_i}\equiv\delta\,Z_{h_i}$) are
determined as follows:
\begin{eqnarray}
\delta\,Z_{\PHiggslightzero} &=& \sin^2\alpha \,\delta\,Z_{\Phi_1} +
\cos^2\alpha \,\delta\,Z_{\Phi_2}\nonumber
\\
\delta\,Z_{\PHiggsheavyzero} &=&
\cos^2\alpha \,\delta\,Z_{\Phi_1} + \sin^2\alpha \,\delta\,Z_{\Phi_2}\nonumber \\
\delta\,Z_{\PHiggspszero} &=&
\sin^2\beta\,\delta\,Z_{\Phi_1} + \cos^2\beta\,\delta\,Z_{\Phi_2}\nonumber \\
\delta\,Z_{G^0} &=& \cos^2\beta \,\delta\,Z_{\Phi_1} +
\sin^2\beta\delta\,Z_{\Phi_2}\nonumber\\
\delta\,Z_{\PHiggslightzero\PHiggsheavyzero} &=&
\sin\alpha\,\cos\alpha \,(\delta\,Z_{\Phi_2} - \delta\,Z_{\Phi_1})\nonumber\\
\delta\,Z_{\PHiggspszero\, G^0} &=& \sin\beta\,\cos\beta
\,(\delta\,Z_{\Phi_2} - \delta\,Z_{\Phi_1}) \,, \label{eq:litran}
\end{eqnarray}
where in the last two equations we have also included the mixing of
the two ${\cal CP}$-even Higgs bosons and that of the ${\cal
CP}$-odd state with the neutral Goldstone boson in the Feynman
gauge. The different counterterms $\delta\,Z_{h_ih_j}$ must be
anchored by specifying a set of subtraction conditions on a finite
number of 2-point functions. To that purpose, we need to specify the
renormalized self-energies corresponding to the gauge bosons and
each of the physical Higgs-boson fields in the model. First of all
let us define the relation between the polarization tensor of a
gauge boson and the transverse self-energies. We define it as
follows,
\begin{equation}\label{poltensor}
\Sigma^{\mu\nu}_{VV'}(q)=\left(-g^{\mu\nu}+\frac{q^{\mu}q^{\nu}}{q^2}\right)\,\Sigma_{VV'}(q^2)+...\,,
\end{equation}
where the dots indicate that we neglect the longitudinal part
(proportional to $q^{\mu}q^{\nu}$) because our calculations for a
linear collider involve extremely light external fermions (electrons
and positrons) and as a result the longitudinal contributions are
suppressed as $m_e^2/M_V^2$. The same applies to the
$q^{\mu}q^{\nu}$ terms in the transverse part, of course.  We
identify $+i\Sigma_{\mu\nu}(q)$ with the Feynman diagram associated
to the structure of the polarization tensor. As a result, the
unrenormalized self-energy part is defined in such a way that
$-i\,\Sigma_{VV'}(q^2)$ (notice the minus sign) is equated to the
blob of external lines $V$ and $V'$. For simplicity, we define
$\Sigma_{V}(q^2)\equiv \Sigma_{VV}(q^2)$ and similarly for other
fields. Thus e.g. the free propagator for the gauge boson $V$
becomes ``dressed'' by the quantum effects as follows:
\begin{eqnarray}\label{renpropV}
\frac{-i\,g_{\mu\nu}}{q^2-M^{(0)2}_V}&\to&
\frac{-i\,g_{\mu\nu}}{q^2-M^{(0)2}_V+\Sigma_{V}(q^2)}\nonumber\\
&=&\frac{-i\,g_{\mu\nu}\,Z_{V}}{q^2-M^2_V+\hat{\Sigma}_{V}(q^2)}\,,
\end{eqnarray}
$M^{(0)}_V$ being the bare mass of $V$ and where in the second
equality we have introduced the renormalized form of the propagator.
Similarly, for self-energies involving two scalars (cf.
Fig.\,\ref{fig:blobs}) we identify $+i\,\Sigma_{\phi_i\phi_j}(q^2)$
with the Feynman blob with external legs $\phi_i$ and $\phi_j$. For
instance, the dressed propagator of the scalar $\phi$ with bare mass
$m^{(0)}_{\phi}$ is given by
\begin{equation}\label{renpropS}
\frac{i}{q^2-m^{(0)2}_{\phi}+\Sigma_{\phi}(q^2)}=\frac{i\,Z_{\phi}}{q^2-m^{2}_{\phi}+\hat{\Sigma}_{\phi}(q^2)}\,.
\end{equation}
The last equality refers to the corresponding renormalized scalar
field propagator, $m_{\phi}$ being the renormalized mass and
$\hat{\Sigma}_{\phi}$ the renormalized self-energy. Without wanting
to appear too pedagogical, we remind the reader that the
renormalization transformation at one-loop is performed by
canonically introducing renormalization constants relating every
bare field $\phi_i^{(0)}$ to the renormalized ones as follows
$\phi_i^{(0)}= Z^{1/2}_i\,\phi_i=(1+\frac12\delta Z_i)\,\phi_i$ --
which in this case takes on the particular form (\ref{eq:doublets})
-- and at the same time decomposing the bare masses and couplings as
the sum of the renormalized quantity plus a counterterm
($m_i^{(0)}=m_i+\delta m_i;\ g_i^{(0)}=g_i+\delta g_i$). With these
definitions, we can check that the relations (\ref{renpropV}) and
(\ref{renpropS}) are consistent at one-loop if we define the
renormalized self-energy of $V$ and $\phi$ as
\begin{eqnarray}\label{renomrphiSV}
&&\hat{\Sigma}_{V}(q^2) = \Sigma_{V}(q^2) + \delta\,Z_{V}\,(q^2 -
M^2_V) -
\delta\,M^2_V\nonumber\\
&&\hat{\Sigma}_{\phi}(q^2) = \Sigma_{\phi}(q^2) +
\delta\,Z_{\phi}\,(q^2 - m^2_{\phi}) - \delta\,m^2_{\phi}\,.
\end{eqnarray}
Finally, for blobs involving a vector ($V$) and a scalar ($\phi$)
field (cf. Fig.\,\ref{fig:blobs}), there is a free vector index
$\mu$ and  we define $\Sigma^{\mu}_{\phi\,V}(q)\equiv (\mp
iq^{\mu})\,(i\Sigma_{\phi\,V}(q^2))= \pm
q^{\mu}\,\Sigma_{\phi\,V}(q^2)$, where $\pm q$ is the external
momentum of $\phi$ flowing into (out of) the blob. As with the
tensor $i\Sigma^{\mu\nu}(q)$, we equate the vector
$\Sigma^{\mu}_{\phi\,V}(q)$ (in this case, with no $i$-factor) to
the corresponding blob with external lines $\phi$ and
$V$\,\footnote{These conventions are slightly different from
previous comprehensive calculations of Higgs boson physics presented
by some of us in the past\,\cite{Coarasa:1996qa} (basically they
differ on the sign of the self-energies) and tend to be more in
agreement with the conventions of the standard packages
\emph{FormCalc} and \emph{FeynHiggs}\,\cite{feynarts,feynHiggs}. But
there are some small differences; for instance, the standard
function \textit{SelfEnergy}$\left[S \to V,M_S\right]$ of
\textit{FormCalc} corresponds precisely to our
$-i\,\Sigma_{\phi\,V}$.}.
With these conventions, we may write the various scalar-scalar
self-energies needed for this calculation involving both Higgs
bosons and Goldstone bosons in diagonal or mixing form. For the
\CP-even sector:
\begin{eqnarray}
&&\hat{\Sigma}_{\Hzero}(q^2) = \Sigma_{\Hzero}(q^2) +
\delta\,Z_{\Hzero}\,(q^2 - m_{\Hzero}^2) - \delta\,m_{\Hzero}^2
\nonumber \\
&&\hat{\Sigma}_{\hzero\Hzero}(q^2) = \Sigma_{\hzero\Hzero}(q^2) + \frac{1}{2}\delta\,Z_{\hzero\Hzero}\,(q^2 - m_{\Hzero}^2) \nonumber \\
&&+ \frac{1}{2}\delta\,Z_{\hzero\Hzero}\,(q^2 - m_{\hzero}^2) - \delta\,m_{\hzero\Hzero}^2 \nonumber \\
&&\hat{\Sigma}_{\hzero}(q^2) = \Sigma_{\hzero}(q^2) +
\delta\,Z_{\hzero}\,(q^2 - m_{\hzero}^2) -
\delta\,m_{\hzero}^2\label{eq:sigmacp1}.
\end{eqnarray}
For the \CP-odd sector:
\begin{eqnarray}\label{eq:self}
&&\hat{\Sigma}_{\PHiggspszero}(q^2) = {\Sigma}_{\PHiggspszero}(q^2)
+ \delta\,Z_{\PHiggspszero}(q^2 - m^2_{\PHiggspszero}) -
\delta\,m^2_{\PHiggspszero}
\nonumber \\
&&\hat{\Sigma}_{G^0}(q^2) = {\Sigma}_{G^0}(q^2) +
\delta\,Z_{G^0}\,q^2 - \delta\,m_{G^0}^2\nonumber\\
&&\hat{\Sigma}_{\PHiggspszero\,G^0}(q^2) =
\Sigma_{\PHiggspszero\,G^0}(q^2)\nonumber\\
&& + \delta\,Z_{\PHiggspszero\,G^0}\,\left(q^2 -
\frac{M_{\PHiggspszero}}{2}\right) -
\delta\,m^2_{\PHiggspszero\,G^0}\,.
\end{eqnarray}
The structure of the Higgs sector beyond the leading order is
somewhat more involved. Similarly as in the SM, scalar-vector mixing
terms arise in the Lagrangian of the general 2HDM. Such
contributions originate from the gauged kinetic terms of the Higgs
doublets,
\begin{eqnarray}
\lag_K = (D_\mu\,\Phi_1)^\dagger\,(D^\mu\,\Phi_1) +
(D_\mu\,\Phi_2)^\dagger\,(D^\mu\,\Phi_2) \label{eq:kindoublets},
\end{eqnarray}
where, in our conventions, the $SU(2)_L\times U(1)_Y$ covariant
derivative reads (using pretty standard notations),
\begin{equation}\label{eq:covD}
D_\mu = \partial_\mu + i\frac{e}{2\,\sw}\,{W^a_{\mu}}{\tau_a}
+i\frac{e}{2\,\cw}\,Y\,B_{\mu}\,,
\end{equation}
with $-e$ being the electron charge. Upon spontaneous EW symmetry
breaking, the following scalar-vector mixing terms are generated:
\begin{eqnarray}
\lag^{SV}_0 &=& \frac{ie}{2\,\sw}\,(v_1\partial^\mu\phi_1^- +
v_2\partial^\mu\phi_2^-)\,W^+_\mu+ h.c.
\nonumber \\
&&- \frac{e}{2\,\sw\cw}\,(v_1\partial^\mu\chi_1^0 +
v_2\partial^\mu\chi_2^0)\,\PZ_\mu^0 \label{eq:sv1}.
\end{eqnarray}
The subindex in $\lag^{SV}_0$ means that this is a bare Lagrangian,
and therefore all fields in it are bare fields $\phi_i^{(0)}$  and
the parameters are bare parameters (couplings $g_i^{(0)}$ and masses
$m_i^{(0)}$) -- although we did not bother to label them as such in
order to avoid an increasing notational snarl-up.

Similarly for the VEV's $v_i$ of the two Higgs doublets. Here we
have to include the corresponding  WF renormalization constants from
(\ref{eq:doublets}). Therefore, the decomposition of the bare VEV's
into the renormalized ones plus counterterms reads as follows:
\begin{eqnarray}
 v_i^{(0)} &=& Z_{\Phi_i}^{\frac{1}{2}}\, (v_i + \delta\, v_i) \nonumber \\
 &=& v_i\,\left(1+\frac{1}{2}\,\delta\,Z_{\Phi_i} +
 \frac{\delta\,v_i}{v_i}\right)\,.
\label{eq:vacuumvis}
\end{eqnarray}
As we shall explain below in more detail, we will adopt the
following renormalization condition for the parameter
$\tan\beta=v_2/v_1$\,\cite{Chankowski:1992er,Dabelstein:1994hb}:
\begin{eqnarray}
 \frac{\delta\,v_1}{v_1} = \frac{\delta\,v_2}{v_2}
\label{eq:vacuumv1v2}.
\end{eqnarray}
This condition insures that the ratio ${v_2}/{v_1}$ is always
expressed in terms of the \emph{true} vacua (that is, the VEV's
resulting from carrying out the renormalization of the Higgs
potential). The corresponding counterterm $\tan\beta \rightarrow
\tan\beta + \delta\tan\beta$ associated to this definition therefore
reads
\begin{eqnarray}
 \frac{\delta\,\tan\beta}{\tan\beta} &=& \frac{\delta\,v_2}{v_2} - \frac{\delta\,v_1}{v_1}+\frac{1}{2}\,\left(\delta\,Z_{\Phi_2} - \delta\,Z_{\Phi_1}
 \right)\nonumber\\
&&=\frac{1}{2}\,\left(\delta\,Z_{\Phi_2} -
\delta\,Z_{\Phi_1}\right)\,. \label{eq:tb}
\end{eqnarray}
Proceeding in this way, we may rewrite Eq.~(\ref{eq:sv1}) as
$\lag^{SV}_{0}= \lag^{SV} + \delta\lag^{SV}$, with
\begin{eqnarray}
\lag^{SV} &=& iM_W\,\PW^+_\mu\,\partial^\mu\PG^-+h.c. -
M_Z\,\PZ^0_\mu\,\partial^\mu\,\PG^0 \label{eq:sv2}
\end{eqnarray}
and where all the fields and parameters here are the renormalized
ones. Similarly, the overall counterterm Lagrangian reads
\begin{eqnarray}
&&\delta\lag^{SV} = \nonumber\\
&&\frac{1}{2}\,\left(\delta\,Z_{G^\pm} + \delta\,Z_{\PW}
+\frac{\delta
M^2_{W}}{M^2_{W}}\right)\,i\,M_W\,W^+_\mu\,\partial^\mu\,\PG^- +
h.c.
\nonumber \\
&+& \frac{1}{2}\,\left(\delta Z_{\Hpm\PG^\pm} + \sin 2\beta\,
\frac{\delta\tan\beta}{\tan\beta}\,\right)\,i\,M_W\,W^+_\mu\,\partial^\mu\,\PHiggs^- + h.c. \nonumber \\
&-& \frac{1}{2}\,\left(\delta\,Z_{\PG^0} + \delta\,Z_{Z} + \frac{\delta\,M^2_{\PZ^0}}{M_Z^2} \right) M_Z\,\PZ_\mu^0\,\partial^\mu\,\PG^0\, \nonumber \\
&-&\frac{1}{2}\,\left(\delta\,Z_{\PG^0\PHiggspszero} + \sin
2\beta\frac{\delta\tan\beta}{\tan\beta}\right)
M_Z\,\PZ_\mu^0\,\partial^\mu\,\PHiggspszero\,, \label{eq:sv3}
\end{eqnarray}
where we have reabsorbed the parameter $v = \sqrt{v_1^2 + v_2^2}$
into the gauge boson masses, $M_W = g v/2,\, M_Z = M_W/\cw$ and
their associated counterterms.

The specific structure of the various mass and WF counterterms
becomes fixed after we choose the appropriate renormalization
conditions in a given subtraction scheme. In particular, a set of
prescriptions for the determination of the Higgs mass counterterms
arising in Eq.~(\ref{eq:self}) is called for. Following the on-shell
prescription, we arrange that the four renormalized Higgs boson
masses coincide with the four input (physical) Higgs masses, and
hence we enforce them to be the pole masses of the corresponding
renormalized propagators (\ref{renpropS}). Equivalently, we declare
that the real part of the corresponding renormalized self-energies
vanishes for on-shell Higgs bosons. In a nutshell:
\begin{eqnarray}
 \Re e\rself_{\hzero}\, (M^2_{\PHiggslightzero}) = 0\nonumber\\
 \Re e\rself_{\Hzero}\, (M^2_{\PHiggsheavyzero}) = 0\nonumber\\
 \Re e\rself_{\Azero}\, (M^2_{\PHiggspszero}) = 0\nonumber\\
 \Re e\rself_{\PHiggs^\pm}\, (M^2_{\PHiggs^\pm}) = 0 \label{eq:massc}.
\end{eqnarray}
Incidentally, let us point out that this is a different approach
with respect to the MSSM (see e.g. \cite{Heinemeyer:2004gx}), in
which case and owing to the supersymmetric invariance, less freedom
is left over to settle independent renormalization conditions.

However, additional conditions are needed due to the extended
structure of the Higgs sector of the 2HDM with respect to the
simpler SM case. In particular, the condition (\ref{eq:vacuumv1v2})
has been applied, but we still have to fix the structure of the
counterterm $\delta\tan\beta$ in (\ref{eq:sv3}), which we will
unravel later on below.

The scalar-vector mixing contributions at the tree-level in
(\ref{eq:sv3}) are canceled by the addition of gauge-fixing terms.
Using the 't Hooft-Feynman gauge choice $\xi = 1$ in the class of
linear $R_\xi$ gauges, we have:
\begin{eqnarray}
\lag^{GF} &=& -\mathcal{F}^+\mathcal{F}^-
-\frac{1}{2}|\mathcal{F}^0|^2\, ,
\nonumber \\
\mathcal{F}^\pm &=& \partial^\mu\PW^{\pm}_\mu \pm iM_W\PG^\pm \nonumber \\
 \mathcal{F}^0 &=& \partial^\mu\PZ^0_\mu + M_Z \PG^0\,.
\label{eq:sv4}
\end{eqnarray}
Let us, however, point out that the scalar-vector mixing can always
take place via quantum corrections -- see the diagrams of
Fig.~\ref{fig:blobs}, and thereby a related set of counterterms must
be introduced in order to renormalize the corresponding mixed
self-energies. In particular, from Eq.~(\ref{eq:sv3}) it follows
that the $\A - \PZ^0$ mixing counterterm reads
\begin{eqnarray}
\delta\,Z_{\PHiggspszero\PZ^0} &=&
-\frac12\delta\,Z_{\PG^0\PHiggspszero} -
\frac12\sin2\beta\,\frac{\delta\tan\beta}{\tan\beta}\nonumber\\
&=& -\frac12\,\sin 2\beta\,\left(\delta Z_{\Phi_2}-\delta
Z_{\Phi_1}\right)\,, \label{eq:cAZ}
\end{eqnarray}
where in the last equality we have taken $\delta\,Z_{\PHiggspszero\,
G^0}$ from Eq.\,(\ref{eq:litran}) and used (\ref{eq:tb}).
Let us also derive the explicit expression for the renormalized $\A
- \PZ^0$ mixing self-energy, as displayed in Fig.~\ref{fig:blobs}a.
By plugging the corresponding Feynman rules, we may cast
conveniently the amplitude in the form
\begin{eqnarray}
\Delta^{\AZ}_{\mu}\equiv \frac{i}{q^2-M_{\A}^2}\,\left[ q^\nu
\hat\Sigma_{\AZ}(q^2)\right]\, \frac{-i g_{\mu\nu} }{q^2-M_Z^2}\,,
\label{eq:greenAZ}
\end{eqnarray}
where the structure of the renormalized $\A -\PZ^0$ self-energy
encapsulates both the corresponding bare self-energy contribution
and the related counterterm as follows:
\begin{equation}
\hat\Sigma_{\AZ}(q^2)=\Sigma_{\AZ}(q^2) +
\,M_Z\,\delta\,Z_{\PHiggspszero\PZ^0}\,,
\label{eq:renAZ}\end{equation}
with $\delta\,Z_{\PHiggspszero\PZ^0}$ given by Eq.\,(\ref{eq:cAZ}).
In the same vein, a similar expression can be derived for the mixing
with the Goldstone boson, $G^0$:
\begin{equation}
\Delta^{\AG}\equiv \frac{i}{q^2-M^2_{\PHiggspszero}}
\left[i\hat\Sigma_{\AG} (q^2)\right] \frac{i}{q^2-M_Z^2}\,,
\label{eq:greenAG}
\end{equation}
with $\hat\Sigma_{\AG} $ defined in (\ref{eq:self}). We are now
ready to tackle the renormalization of the Higgs-boson fields. Two
independent conditions are needed to fix $\delta\,Z_{\Phi_{1}},
\,\delta\,Z_{\Phi_{2}}$. Insofar as we will be dealing with the
neutral Higgs sector, it is more convenient for us to impose such
conditions on the $\PHiggspszero$ boson:
\begin{eqnarray}
 \Re e\rself '_{\PHiggspszero}(q^2)\Bigg]_{q^2=M^2_{\PHiggspszero}} &=& 0\,, \label{eq:cond1} \\
 \Re e\rself_{\PHiggspszero\PZ^0}(q^2)\Bigg]_{q^2=M^2_{\PHiggspszero}} &=&
 0\,,
\label{eq:cond2}
\end{eqnarray}
\noindent where the short-hand notation $f'(q^2) \equiv
d\,f(q^2)/dq^2$ has been employed. The above conditions, together
with (\ref{eq:massc}), ensure that the outgoing $\PHiggspszero$
particle is on its mass shell and that the residue of the
renormalized propagator (\ref{renpropS}) is precisely equal to
$1/[1+\rself '_{\PHiggspszero}(M^2_{\PHiggspszero})]=1$ -- see
section \ref{sec:general}. Besides, the second condition guarantees
that no mixing between the $\PHiggspszero$ and $\PZ^0$ bosons will
take place at any order in perturbation theory as long as that
$\PHiggspszero$ Higgs boson is on-shell. Actually, also the
scalar-scalar mixing $\PHiggspszero - G^0$ is generated through a
similar class of one-loop interactions (recall
Fig.~\ref{fig:blobs}b). However, one can see that both mixing
phenomena turn out to be related by the Slavnov-Taylor (ST) identity
%
\begin{eqnarray}
q^2\hat\Sigma_{\AZ}(q^2) + \,M_Z\,\hat\Sigma_{\AG}
(q^2)\Big]_{q^2=M_{\PHiggspszero}^2} = 0\,, \label{eq:slavnov}
\end{eqnarray}
which ultimately stems from the underlying BRS symmetry of the
theory\,\cite{BRS76}. Notice that Eqs.\,(\ref{eq:cond2}) and
(\ref{eq:slavnov}) imply that the real part of the $\Azero-G^0$
mixed self-energy, too, vanishes on-shell. Let us recall that a BRS
transformation is a local Grassmann gauge transformation in which
the infinitesimal gauge parameters read
$\delta\omega^{a}(x)=\eta^{a}(x)\delta\bar{\lambda}$, with
$\eta^{a}$ the FP ghost fields (scalar fields with Fermi statistics)
and where $\delta\bar{\lambda}$ is an infinitesimal Grassmann (i.e.
anticommuting) parameter. In fact, using the notation $\langle
X\rangle\equiv\bra{0} TX \ket{0}$ for a Green's function constructed
out of a product of field operators $X$ on which the time-ordering
operator $T$ acts, the BRS invariance is expressed as
$\delta_{\lambda}\langle X\rangle=0$, where
$\delta_{\lambda}=\delta\bar{\lambda}\,\delta/\delta\bar{\lambda}$.
Recalling the expression for the gauge-fixing Lagrangian
Eq.~(\ref{eq:sv4}), we consider the BRS invariance of the 2-point
Green's function constructed with an anti-FP ghost field
$\bar{\eta}^0$ and a neutral \CP-odd Higgs boson. We have
\begin{multline}
0=\langle\,\delta_{\lambda}\, \left(
\bar{\eta}^0\,\PHiggspszero\right)\rangle
= -\langle\, \mathcal{F}^0\,\PHiggspszero\,\rangle\delta\bar{\lambda} \\
= -\langle\partial^\mu\,\PZ^0_\mu
\PHiggspszero + M_Z\,G^0\PHiggspszero\rangle\delta\bar{\lambda}\, \Rightarrow \\
\langle -\PZ^0_\mu\,\partial^\mu\, \PHiggspszero +
M_Z\,G^0\PHiggspszero \rangle = 0 \label{eq:slavdemo1}.
\end{multline}
Here we have used the standard BRS transformation of the anti-FP
ghost field, which is simply related to the gauge fixing term
$\mathcal{F}^0$ as follows: $\delta_{\lambda} \bar{\eta}^0= -
\mathcal{F}^0\,\delta\bar{\lambda}$. Moreover, the second term of
the BRS variation vanishes because $\delta_{\lambda}A\propto
A\,(\tau_a/2)\eta^{a}(x)\delta\bar{\lambda}$ and we cannot have
Green's functions with external ghost legs.

Fourier transforming the expression (\ref{eq:slavdemo1}), we can
rewrite it in momentum space,
\begin{eqnarray}
-i\,q^\mu\,\Delta_\mu^{\AZ} +
M_Z\,\Delta^{\AG}\Big]_{q^2=M_{\PHiggspszero}} = 0
\label{eq:slavdemo2},
\end{eqnarray}
with $\Delta^{XY}_{\mu}$ denoting the renormalized (non-amputated)
2-point Green's functions introduced in
Eqs.~\eqref{eq:greenAZ}-\eqref{eq:greenAG}. The ST-identity
(\ref{eq:slavnov}) is just a transcription of (\ref{eq:slavdemo2})
in terms of the renormalized mixed self-energies. In words, the
identity states that an on-shell $\PHiggspszero$ does not mix
neither with the neutral gauge boson $\PZ^0$ nor with the neutral
Goldstone boson $G^0$, provided that the conditions in
Eqs.~(\ref{eq:cond1}) and ~(\ref{eq:cond2}) hold. A similar ST
identity can be derived in the charged Higgs sector (see
Ref.\,\cite{Coarasa:1996qa}), and although we do not need it in the
context of our calculation we quote it for completeness. In the
present notation, it reads
\begin{eqnarray}
q^2\hat\Sigma_{H^{\pm}W^{\pm}}(q^2) \pm
i\,M_W\,\hat\Sigma_{\PHiggs^{\pm}G^{\pm}}
(q^2)\Big]_{q^2=M_{H^{\pm}}^2} = 0 \label{eq:slavnov2}.
\end{eqnarray}
For an extended discussion of these identities, see
Ref.~\cite{Collins84}, and in particular for the off-shell regime
and its form in generalized non-linear gauges~\cite{Baro:2008bg}.

A comment on the physical definition of $\tan\beta$ is now in order.
Recall that we have assumed that the relation (\ref{eq:vacuumv1v2})
holds between the counterterms of the VEV's. Such relation says that
the renormalized value of $\tan\beta$ is defined to be the ratio of
the renormalized values of $v_2$ and $v_1$. This is tantamount to
enforce the Higgs tadpoles to
vanish\,\cite{Dabelstein:1994hb,Chankowski:1992er},
\begin{eqnarray}
 T_{\PHiggsheavyzero\,\{\PHiggslightzero\}} + \delta\,t_{\PHiggsheavyzero\,\{\PHiggslightzero\}} =
 0\,.
\label{eq:tad}
\end{eqnarray}
Indeed, the tadpoles are absent at the tree-level and must be absent
too at one-loop if $v_1$ and $v_2$ are to be the renormalized VEV's
characterizing the true ground state of the potential at the given
order of perturbation theory.

A note of caution may be pertinent at this point: renormalized
quantities are not always physical quantities, although sometimes
they are; for example, on-shell renormalized masses are of course
physical masses, and this is indeed what is stipulated by the
fundamental on-shell renormalization conditions (\ref{eq:massc}).
However, the renormalized VEV's $v_i$ are not directly measurable
quantities. So, even though the above definition of $\tan\beta$ may
look quite natural, it is not entirely physical yet, in the sense
that it is not tied to a physical observable at this point. In order
to link $\tan\beta$ to a genuine physical quantity we have to
specify the process from which $\tan\beta$ could be extracted from
the experiment. For example, in \cite{Coarasa:1996qa} it is derived
from the charged Higgs decay $H^{\pm}\to\tau^{\pm}\nu_{\tau}$,
whereas in other cases it is collected from other observables, e.g.
from $A^0\to\tau^+\tau^{-}$\,\cite{Dabelstein:1994hb} -- see also
\cite{Baro:2008bg}. The ultimate connection of the current
definition of $\tan\beta$ with physics demands to take account of
the contribution from the process-dependent quantum effects, as
explained in detail e.g. in \cite{Coarasa:1996qa}. Notice that the
fact that this reference focuses on the MSSM is not relevant for
this feature. All that said, let us stress that these
process-dependent effects are not crucially important for the
present analysis because they do \textit{not} involve the triple
Higgs boson couplings under study. Therefore, since we wish to place
ourselves in the region of parameter space where these couplings are
most favored, the process-dependent corrections cannot noticeably
alter the bulk supply of the significant renormalization effects on
the cross-sections (\ref{2h}) computed under these conditions. Thus,
it is not crucial for this study of the leading triple Higgs boson
self-interactions to deal with the details of the particular process
from which one could eventually extract the value of $\tan\beta$,
and for this reason we will not include these corrections in our
analysis\,\footnote{The situation is, in the quantitative aspect,
quite similar to the analysis of the branching ratio of the charged
Higgs decay of the top quark ($t\to H^{+}b$) both in the
2HDM\,\cite{Coarasa:1997ky} and in the MSSM (``a window to virtual
SUSY''-- see Ref.\,\cite{Coarasa:1996qa}). The potential quantum
effects related to the intrinsic dynamics of this decay can be so
large that the process-dependent corrections associated to the
definition of $\tan\beta$ pale in comparison and are, therefore,
unessential to demonstrate the very origin of the quantum
corrections.}. In short, in all practical respects the definition of
$\tan\beta$ employed here, which is based on the condition
(\ref{eq:vacuumv1v2}), is well-suited to illustrate the leading
renormalization effects produced by the Higgs boson
self-interactions on the 2HDM cross-sections (\ref{2h}).

Combining the renormalization conditions ~(\ref{eq:cond1}) and
~(\ref{eq:cond2}), and using (\ref{eq:litran}), (\ref{eq:self}) and
(\ref{eq:cAZ}), we are lead to
\begin{eqnarray}\label{eq:constantsA}
&&\Re e\self'_{\PHiggspszero\PHiggspszero}(M^2_{\PHiggspszero})
+\sin^2\beta\,\delta Z_{\Phi_1}+\cos^2\beta\,\delta Z_{\Phi_2}=0
\\
&& \Re e\self_{\PHiggspszero\,\PZ^0}(M^2_{\PHiggspszero})
-\frac12\,M_Z\sin 2\beta \left(\delta\,Z_{\Phi_2} -
\delta\,Z_{\Phi_1}\right)=0\nonumber\,.
\end{eqnarray}
The WF renormalization constants for the Higgs doublets can now be
solved explicitly as follows:
\begin{eqnarray}
 \delta\,Z_{\Phi_1} &=& - \Re e\self'_{\PHiggspszero}(M^2_{\PHiggspszero}) -
 \frac{1}{M_Z\,\tan\beta}\,
 \Re e\self_{\PHiggspszero\,\PZ^0}(M^2_{\PHiggspszero})\,, \nonumber \\
 \delta\,Z_{\Phi_2} &=& - \Re e\self'_{\PHiggspszero}(M^2_{\PHiggspszero})
+ \frac{\tan\beta}{M_Z}\,
 \Re e\self_{\PHiggspszero\,\PZ^0}(M^2_{\PHiggspszero})\,.
\label{eq:constants}
\end{eqnarray}
From (\ref{eq:tb}) and (\ref{eq:constants}) the specific form of the
counterterm associated to our definition of $\tan\beta$ finally
ensues:
\begin{equation}\label{deltatanbetaexpl}
\frac{\delta\tan\beta}{\tan\beta}=\frac{1}{M_Z\,\sin 2\beta} \Re
e\self_{\PHiggspszero\,\PZ^0}(M^2_{\PHiggspszero})\,.
\end{equation}

Up to now, we have established a set of conditions that allows us to
renormalize five out of the seven free parameters (\ref{freep}) of
the 2HDM. These are the conditions (\ref{eq:vacuumv1v2}),
(\ref{eq:massc}) and (\ref{eq:cond2}). Notice that the additional
one (\ref{eq:cond1}) is a field renormalization condition and,
therefore, it does not affect the definition of the parameters
themselves -- although it certainly affects the renormalization of
the Green's functions and cross-sections. The two renormalized
parameters that remain to be defined are the\ \CP-even mixing angle
$\alpha$ and the coupling $\lambda_5$. As far as $\alpha$ is
concerned, we can split its bare value in the usual form
$\alpha^{(0)} = \alpha + \delta\alpha$, where we define the
renormalized value $\alpha$ by setting to zero the renormalized
$\PHiggslightzero - \PHiggsheavyzero$ mixing self energy at the
scale of the Higgs mass in the final state. If e.g. the final state
is the lightest $\CP$-even Higgs boson (as in Figs.
\ref{fig:self1}-\ref{fig:counter}), then
\begin{equation}
\Re e\hat{\Sigma}_{\PHiggslightzero\PHiggsheavyzero}(q^2 =
M_{\PHiggslightzero}^2)=0\,, \label{eq:rmassmix}
\end{equation}
whereas if it is the heavy $\CP$-even Higgs boson then
$M^2_{\PHiggslightzero}$ is replaced by $M^2_{\PHiggsheavyzero}$. In
what follows we presume that the final state is the lightest
$\CP$-even state. It is not difficult to see that the
renormalization condition (\ref{eq:rmassmix}) fixes the mixing angle
counterterm $\delta\alpha$ in terms of the mixing mass counterterm
$\delta m_{\hzero\Hzero}^2$ in the expression of the renormalized
mixed self-energy $\hat{\Sigma}_{\hzero\Hzero} $ -- see
Eq.\,(\ref{eq:sigmacp1}). Obviously, (\ref{eq:rmassmix}) implies
\begin{equation}\label{deltamhH}
\delta m_{\hzero\Hzero}^2 =\Re e\Sigma_{\hzero\Hzero}(M^2_{\hzero})
+ \frac12\,\delta\,Z_{\hzero\Hzero}\,(M^2_{\hzero} -
M^2_{\Hzero})\,,
\end{equation}
with
$\delta\,Z_{\hzero\Hzero}=\sin{2\alpha}(\delta\tan\beta/\tan\beta)$.
On the other hand, one may show from the renormalized Higgs
potential that
\begin{equation}\label{deltaalpha}
\delta\alpha=\frac{\delta m_{\hzero\Hzero}^2}{M^2_{\hzero} -
M^2_{\Hzero}}\,.
\end{equation}
Actually, the condition \eqref{eq:rmassmix} is bound to the fact
that the external Higgs bosons ($\hzero,\Hzero$) are to be on-shell
to all orders in perturbation theory. To better assess the
significance of our renormalization conditions in the $\CP$-even
Higgs sector, we note that the quantum corrections in this sector
induce the following non-trivial inverse propagator matrix:
\begin{eqnarray}
\Delta^{-1}_{\hzero\Hzero} &=& \left(\begin{array}{cc} p^2 -
M^2_{\Hzero} + \hat{\Sigma}_{\Hzero\Hzero} &
\hat{\Sigma}_{\hzero\Hzero} \\
\hat{\Sigma}_{\hzero\Hzero} & p^2 - M^2_{\hzero} +
\hat{\Sigma}_{\hzero\hzero}
\end{array}\right) \nonumber\, . \\
\label{eq:propmix}
\end{eqnarray}
 \noindent The physical Higgs boson masses\footnote{Let us recall
that all Higgs boson masses are input parameters within the general
2HDM, unlike the MSSM case.} ensue from the real part of the roots
of $\mbox{det}\,\Delta^{-1}_{\hzero\Hzero} =
0$, i.e. of the algebraic equation
\begin{eqnarray}
\left(p^2 - M^2_{\Hzero} + \hat{\Sigma}_{\Hzero\Hzero}\right)\,
\left(p^2 - M^2_{\hzero} + \hat{\Sigma}_{\hzero\hzero}\right) -
\hat{\Sigma}^2_{\hzero\Hzero} = 0\,.\nonumber\\
\label{eq:mixonshell}
\end{eqnarray}
The above equation makes clear that both conditions
~\eqref{eq:massc} and~\eqref{eq:rmassmix} are needed in order to
settle the correct on-shell properties for the $\hzero$ boson.
Similarly, we have Eq.~\eqref{eq:massc} and $\Re e\,
\hat\Sigma_{\PHiggslightzero\PHiggsheavyzero}(q^2 =
M_{\PHiggsheavyzero}^2)=0$ in the case of $\Hzero$.

The set of renormalization conditions that we have introduced so far
enable us to anchor the full collection of counterterms within the
Higgs sector, in particular the mass counterterms
$\delta\,m^2_{G^0}, \delta m^2_{\Azero G^0}$ that remain unsettled
in Eq.~\eqref{eq:self}. In terms of the tadpole counterterms and
$\delta\tan\beta$ (all of them already fixed in this renormalization
scheme), we may write their explicit expressions as follows:
\begin{eqnarray}
\delta\,m^2_{G^0} &=& \frac{-e}{2\,\sw M_W}\,\left[
\sin(\beta-\alpha)\,\delta t_{\hzero}
+  \cos(\beta-\alpha)\,\delta t_{\Hzero}\right] \nonumber \\
\label{eq:ctg0g0} \\
\delta\,m^2_{\Azero G^0} &=& \frac{e}{2\,\sw
M_W}\,\left[\sin(\beta-\alpha)\,\delta t_{\Hzero}
- \cos(\beta-\alpha)\, \delta t_{\hzero} \right] \nonumber \\
&& - \sin\beta\cos\beta\,M^2_{\Azero}\,\delta\tan\beta\,.
\label{eq:cta0g0}
\end{eqnarray}
The only parameter left is $\lambda_5$. However, this parameter is
not involved at the tree-level in any of the cross-sections under
study, which are entirely dependent on just the electroweak gauge
boson couplings at the lowest order. Therefore, there cannot be any
one-loop UV divergent quantity that needs to be absorbed into the
renormalization of the parameter $\lambda_5$. In other words, this
parameter is not renormalized at one-loop for the processes under
consideration. It goes without saying that the situation would be
different at higher orders, e.g. at 2-loop order and beyond. But we
need not go that far to illustrate the main message of this paper
and its potential implications. In Section~VII, we shall briefly
return to this point.

\section{Neutral Higgs-boson pair production at 1 loop in the linear collider:
theoretical setup} \label{sec:general}

Due to its vantage point as a high precision experimental machine,
and hence owing to its complementarity with the LHC, it is of
cardinal importance to understand in detail the phenomenology of the
Higgs sector in the context of a linear collider (both within the SM
and its renormalizable extensions). Quite an effort has already been
devoted to this goal in the literature, specially within the MSSM,
see section \ref{sec:HiggsLinear} and references therein. Analyses
of Higgs boson production in the LHC within the 2HDM have been
considered e.g. in\ \cite{Moretti:2004wa,Kanemura:2009mk}, and the
results are as encouraging as the corresponding calculation for the
linear
colliders\,\cite{Ferrera:2007sp,Hodgkinson:2009uj,Osland:1998hv}
except that in the latter case the experimental environment is much
cleaner and the real possibilities to discriminate the nature of the
Higgs bosons are correspondingly higher.

Exclusive processes in the linear collider can be specially
significant. The simplest neutral Higgs boson processes of this kind
in a linac involve the production of two Higgs bosons in the final
state, Eq.\,(\ref{2h}). At the tree-level, they are of order
$\mathcal{O}(\alpha^2_{ew})$ and are mediated by $\PZ^0$ exchange.
The next-to-simplest ones are the triple Higgs boson channels
(\ref{3H}). These have been studied in \cite{Ferrera:2007sp}, and in
section \ref{sec:conclusions} we will compare their cross-sections
with the 2H ones under the same set of conditions.
%
\begin{figure*}[htb]
\centerline{ \resizebox{!}{6.5cm}{\includegraphics{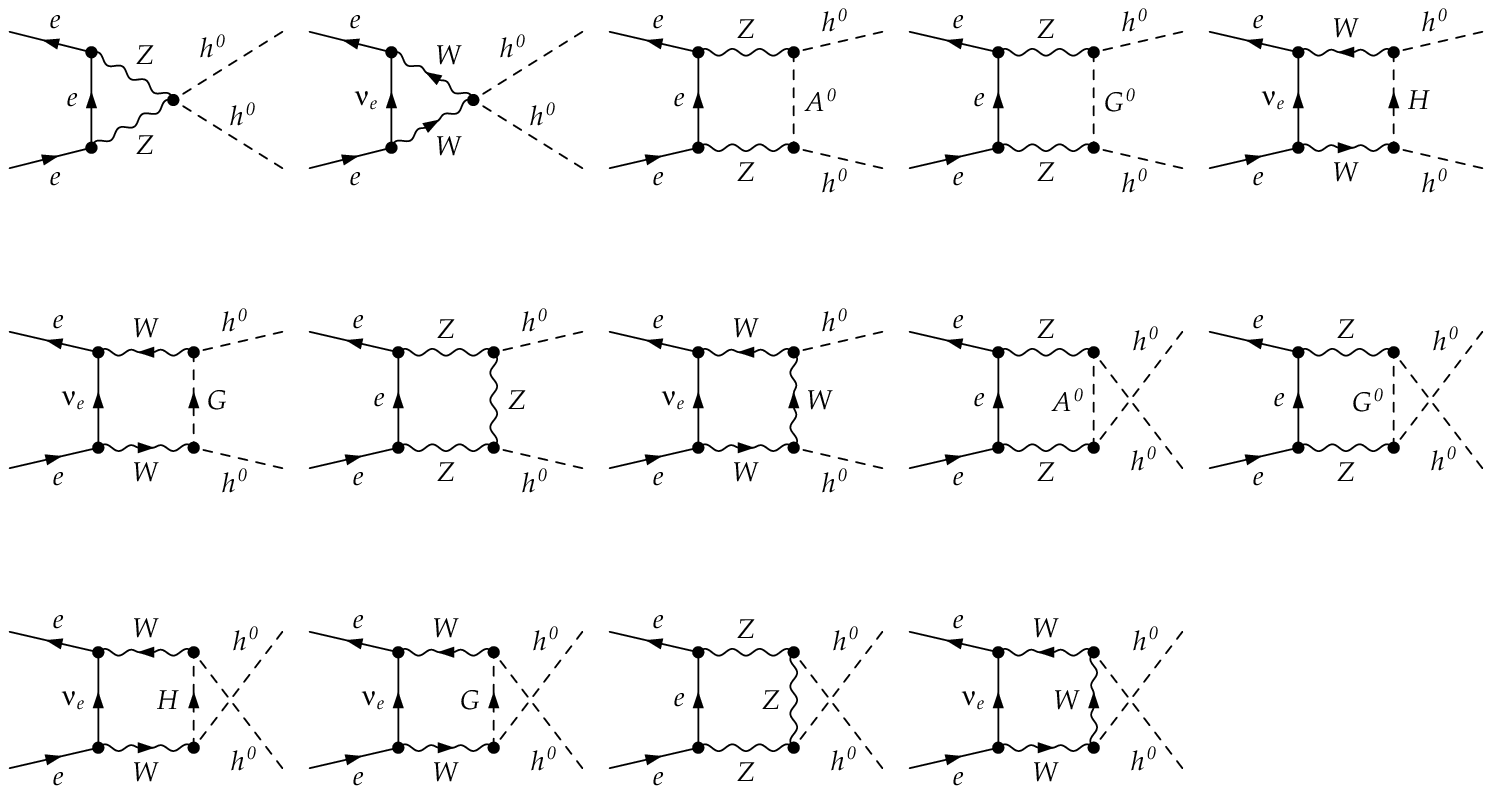}}}
\caption{\CP-conserving box diagrams corresponding to the process
$\HepProcess{\APelectron\Pelectron \HepTo
\PHiggslightzero\PHiggslightzero}$ in the 2HDM. A similar collection
of diagrams describes the analogous process in the MSSM.}
\label{fig:box_hh}
\end{figure*}
%

\vspace{2cm}
\begin{figure*}[htb]
\centerline{ \resizebox{!}{14.5cm}{\includegraphics{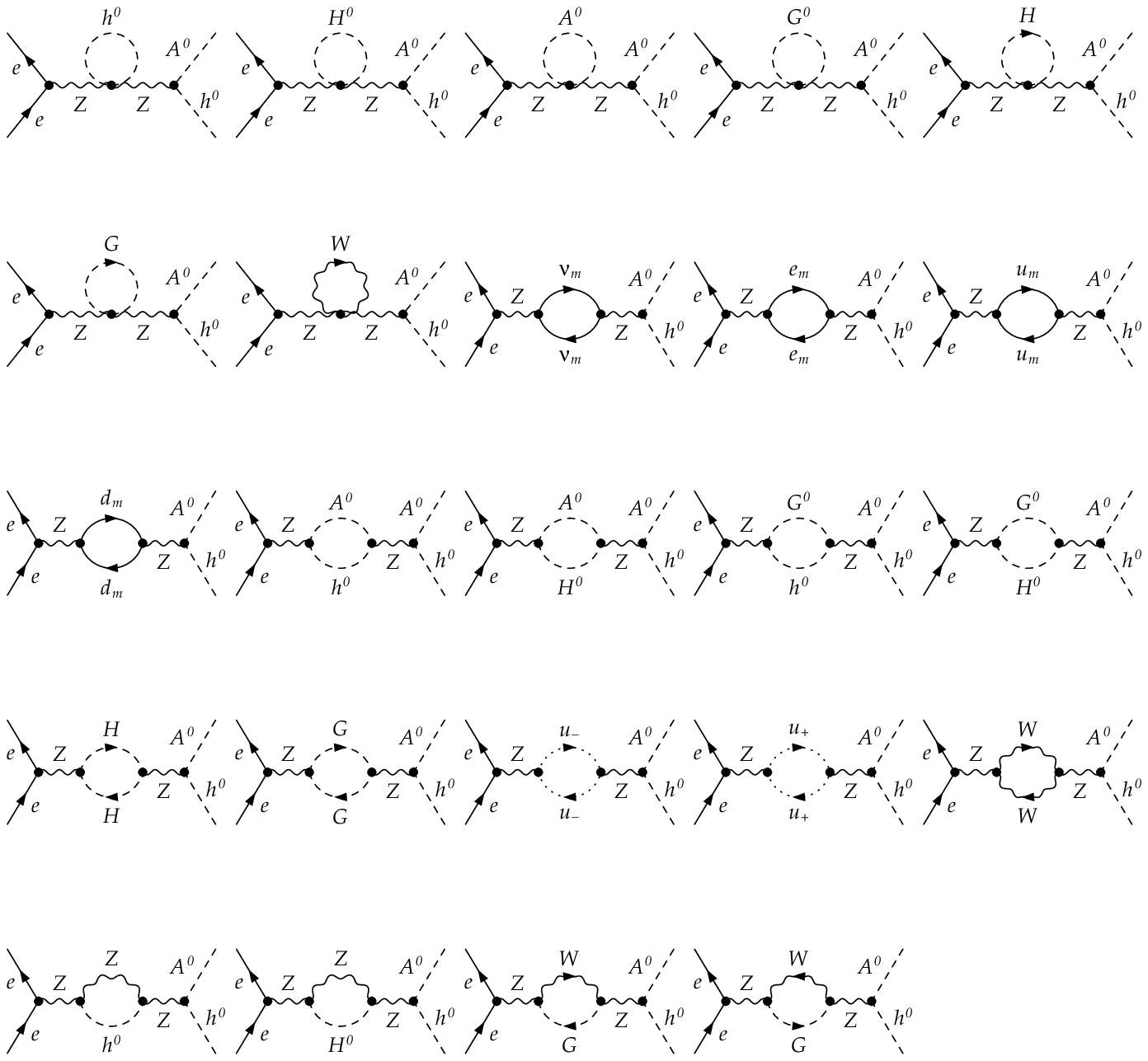}}}
\caption{Set of Feynman diagrams contributing to the $\eeAl$ process
at the one-loop level within the 2HDM. These diagrams account for
the vacuum polarization effects on the $\PZ^0$ boson propagator.}
\label{fig:self1}
\end{figure*}

Notice that due to\,\ \CP-conservation --  or, what is more, due to
Bose-Einstein symmetry -- processes with only two identical neutral
Higgs bosons in the final state cannot proceed at the tree-level
(neither in the SM nor in any of its extensions). Specifically, the
combinations $2H = \PHiggslightzero \PHiggslightzero,
\PHiggsheavyzero\PHiggsheavyzero, \PHiggslightzero\PHiggsheavyzero$
of \CP-even Higgs bosons can only be generated through
\CP-conserving box-diagrams at $\mathcal{O}(\alpha^4_{ew})$, see
Fig.~\ref{fig:box_hh}. Such diagrams render up to $\sigma_{2H}\sim
5\times 10^{-6}$ pb at $\sqrt{s}=1.5 \,\TeV$
\,\cite{Ferrera:2007sp}, which would translate into a rate of
scarcely a few events per $500 \,\invfb$ of integrated luminosity,
thus a too handicapped rate to provide feasible detection signals.
It is worth pointing out that, since all couplings appearing in the
box-type diagrams of Fig.~\ref{fig:box_hh} are fixed by the gauge
symmetry, the obtained results do not depend on whether we consider
the SM, the MSSM or the general (unconstrained) version of the 2HDM.

Being these couplings generated from the electroweak kinetic terms
(\ref{eq:kindoublets}), they are fully determined by the gauge
symmetry, and as a consequence they are formally the same both in
the MSSM and in the 2HDM. It means that, if we aim at distinguishing
between the two basic types (supersymmetric and non-supersymmetric)
of Higgs boson scenarios, the study of the quantum corrections is
mandatory here.  Even at one-loop order, this implies to cope with a
formidable calculation.


\begin{figure*}[htb]
\centerline{ \resizebox{!}{8cm}{\includegraphics{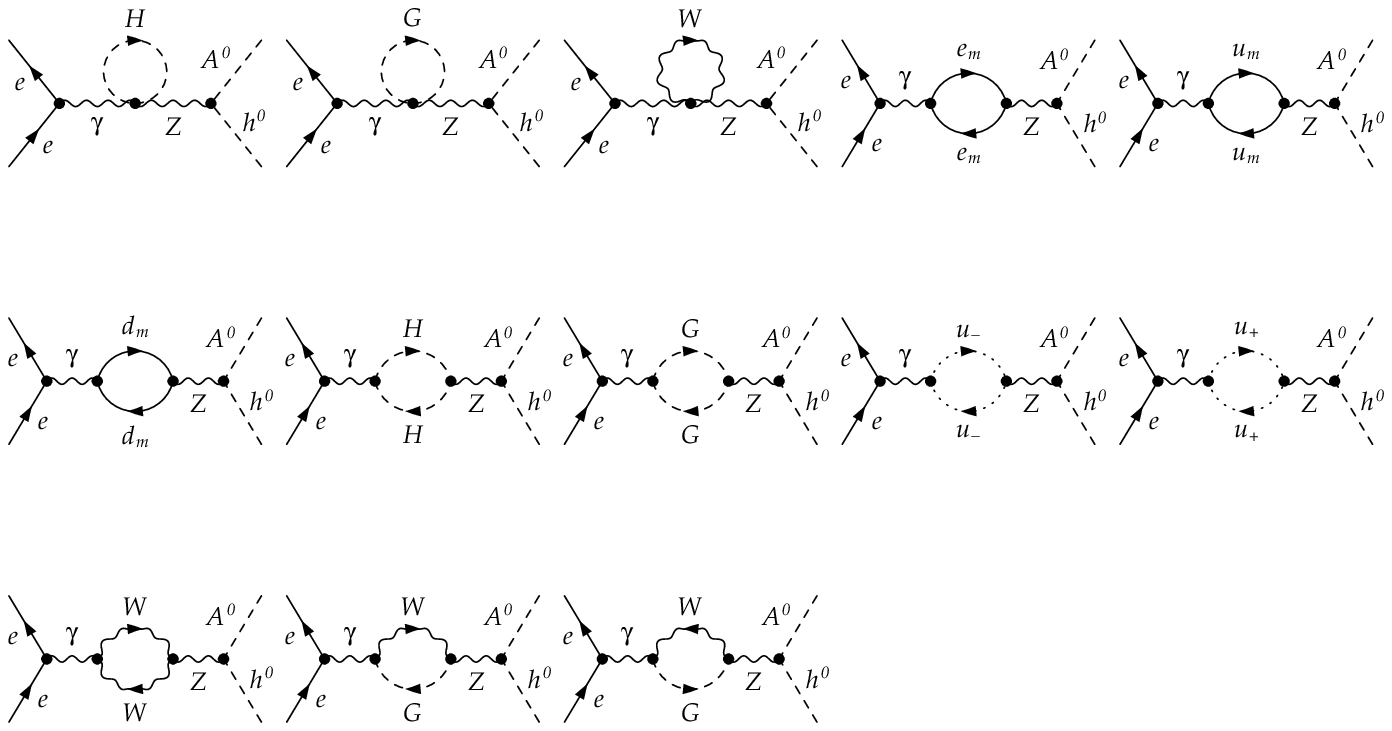}}}
\caption{Set of Feynman diagrams contributing to the $\eeAl$ process
at the one-loop level within the 2HDM. They account for the vacuum
polarization effects on the $\PZ^0-\Pphoton$ mixing propagator.}
\label{fig:self2}
\end{figure*}

Even though the detection of the 2H processes (\ref{2h}) would
signify an unmistakable sign of physics beyond the SM, none of them
is sensible to the trilinear Higgs boson couplings at the leading
order. It means that the order of magnitude of the cross-sections
($1-30$ fb) is not significantly enhanced with respect to, say, the
MSSM ones. What can be remarkably affected in the general 2HDM,
however, is the size of the radiative corrections, i.e. the virtual
quantum effects on these processes. For the triple Higgs boson
channels (\ref{3H}) and the gauge boson fusion ones (\ref{2HX}),
instead, already the tree-level cross-sections can be greatly
modified by the triple Higgs boson self-interactions. Indeed, one
can achieve sizable production rates up to roughly $1\picobarn$,
whereas the maximum MSSM payback lies around
$10^{-3}-10^{-2}\,\picobarn$ (see e.g. Tables 5 and 2 of
Refs.\,\cite{Ferrera:2007sp} and \cite{Hodgkinson:2009uj},
respectively). These results can be traced back to the intrinsically
different nature of the 3H self-coupling in the supersymmetric model
(a pure gauge coupling) as compared to the non-supersymmetric ones.

The use of the two-body processes (\ref{2h}) as a tool to probe the
nature of Higgs bosons in the final states is nevertheless possible
through the analysis of the radiative corrections. Closely related
to them are the two-body final states $\APelectron\Pelectron \to
\PZ^0\PHiggslight$ (with $\PHiggslight = \hzero\, ,\Hzero$), which
have long been known to be complementary to the former in the
MSSM\,\cite{Djouadi:1992pu}. We will not consider the latter in the
present analysis\,\cite{NDJ10}.

The theoretical setup for the one-loop calculation starts from the
basic bare interaction vertices extracted from
Eq.~(\ref{eq:kindoublets}):
\begin{eqnarray}
\lag_{\vzlA} = \frac{e\,\cos(\beta-\alpha)}{2\,\sw\,\cw}
\,Z_\mu^0\,\PHiggspszero\,\overleftrightarrow{\partial}^\mu\,\PHiggslightzero\,,
\label{eq:lag11}
\end{eqnarray}
\begin{eqnarray}
\lag_{\vzHA} = -\frac{e\,\sin(\beta-\alpha)}{2\,\sw\,\cw}
\,Z_\mu^0\,\PHiggspszero\,\overleftrightarrow{\partial}^\mu\,\PHiggsheavyzero\,.
 \label{eq:lag12}
\end{eqnarray}
In Fourier space, they read:
\begin{eqnarray}
\Gamma_{\PHiggspszero\PHiggslightzero\PZ^0} &:&
\frac{e\,\cos(\beta-\alpha)} {2\,\sw\,\cw} \, (p-p')^\mu\,,
\label{eq:coup11}
\end{eqnarray}
\begin{eqnarray}
\Gamma_{\PHiggspszero\PHiggsheavyzero\PZ^0} &:&
-\frac{e\,\sin(\beta-\alpha)} {2\,\sw\,\cw} \, (p-p')^\mu\,,
\label{eq:coup12}
\end{eqnarray}
$p$ and $p'$ being the 4-momenta of the \CP-odd and even Higgs boson
pointing outwards, respectively.

\begin{figure*}[htb]
\centerline{\resizebox{!}{12.5cm}{\includegraphics{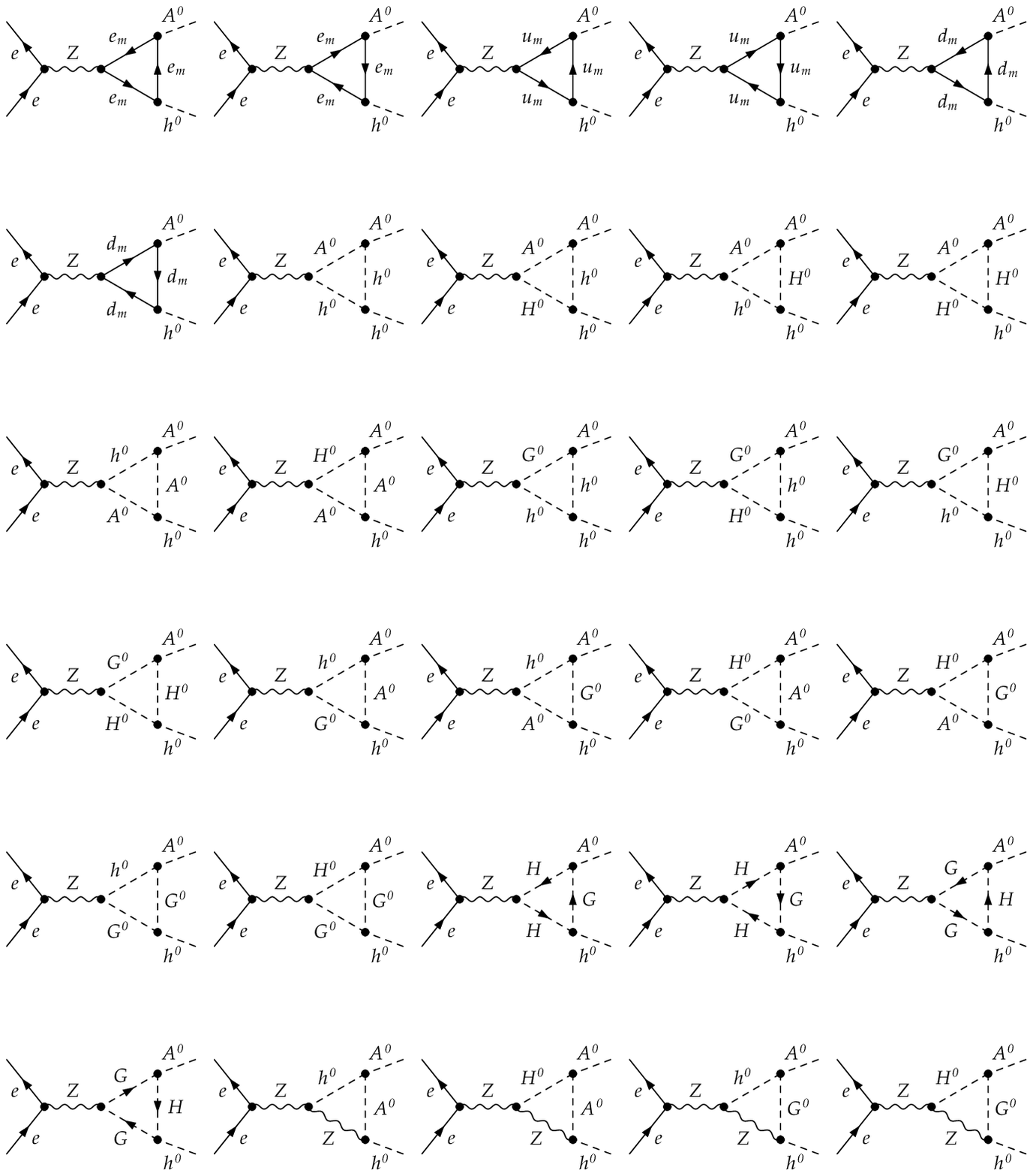}}}
\caption{Set of Feynman diagrams contributing to the $\eeAl$ process
at the one-loop level within the 2HDM. These diagrams describe part
of the quantum corrections to the $\vzlA$ interaction vertex.}
\label{fig:vert11}
\end{figure*}

\begin{figure*}[h]
\centerline{\resizebox{!}{15cm}{\includegraphics{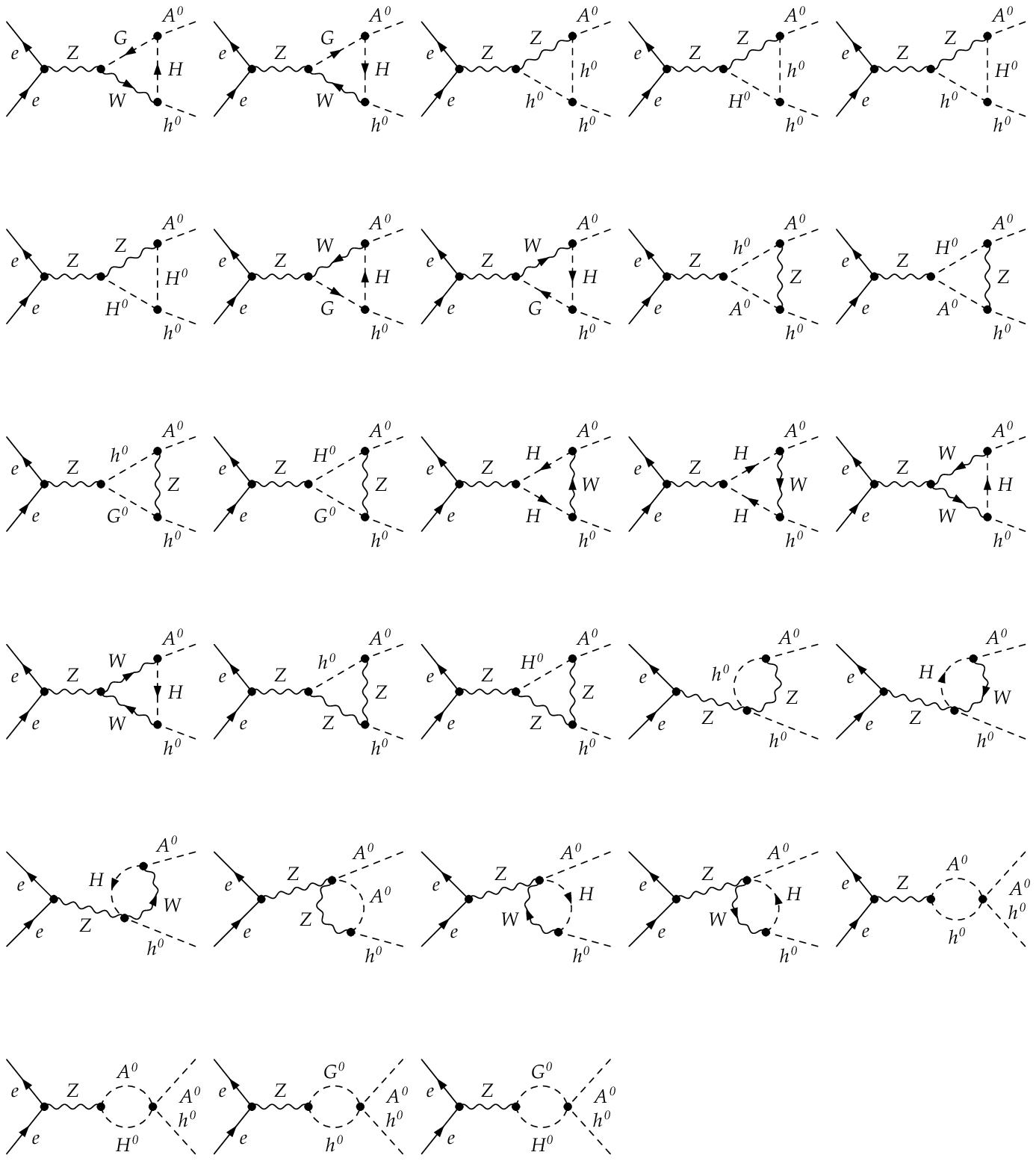}}}
\caption{Set of Feynman diagrams contributing to the $\eeAl$ process
at the one-loop level within the 2HDM. These diagrams describe the
second part of the quantum corrections to the $\vzlA$ interaction
vertex.} \label{fig:vert12}
\end{figure*}

\begin{figure*}[p]
\centerline{ \resizebox{!}{14cm}{\includegraphics{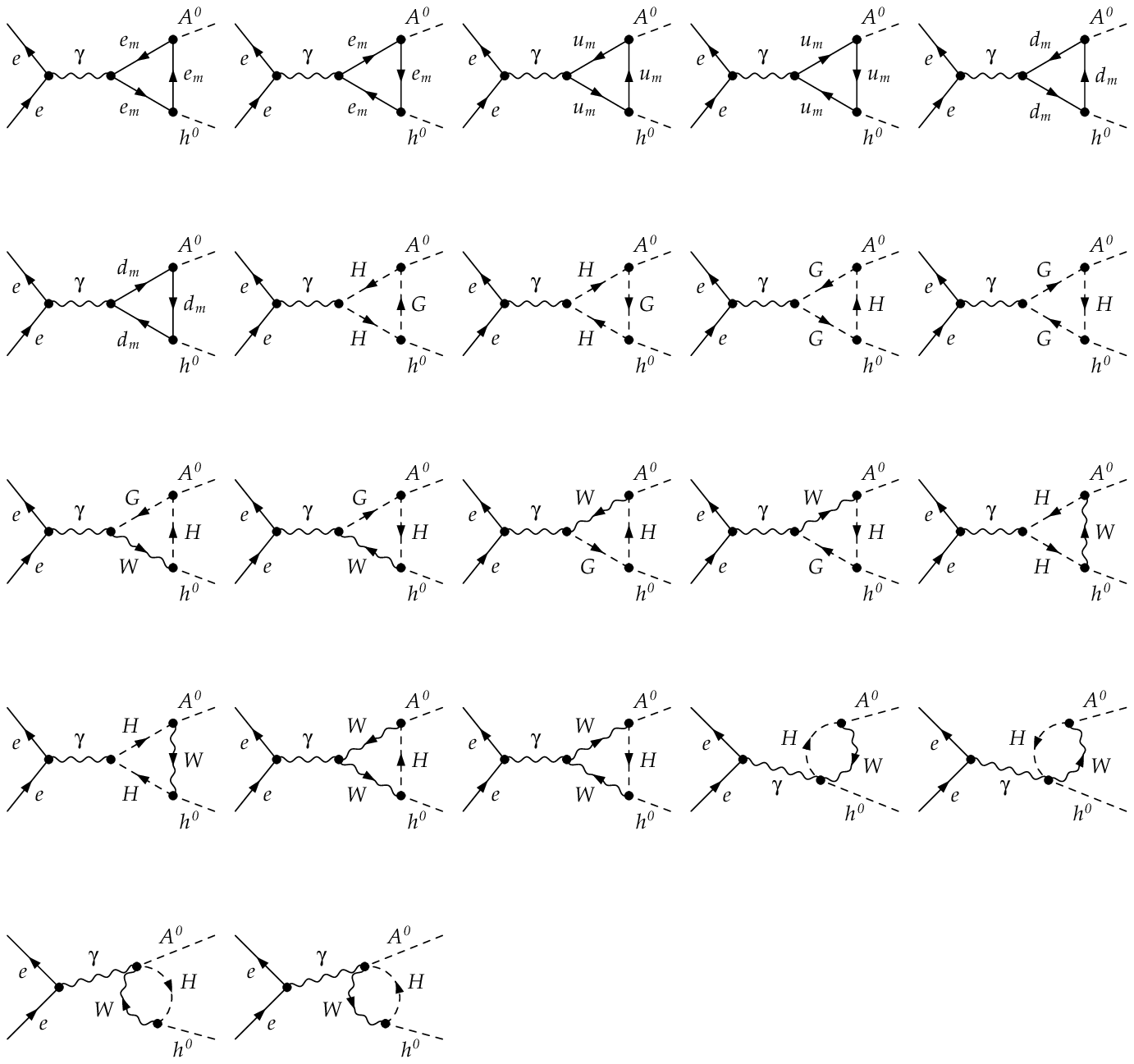}}}
\caption{Set of Feynman diagrams contributing to the $\eeAl$ process
at the one-loop level within the 2HDM. These diagrams describe the
loop-induced $\Pphoton\hzero\Azero$ interaction vertex.}
\label{fig:vert2}
\end{figure*}

\begin{figure*}[p]
\centerline{\resizebox{!}{2.cm}{\includegraphics{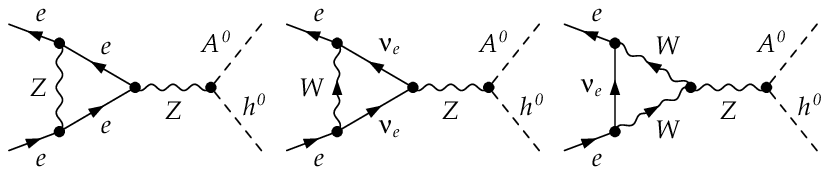}}}
\caption{Set of Feynman diagrams contributing to the $\eeAl$ process
at the one-loop level within the 2HDM. These diagrams describe the
quantum corrections to the $\APelectron\Pelectron\PZ^0$ interaction
vertex.} \label{fig:vert3}
\end{figure*}

\begin{figure*}[thb]
\centerline{ \resizebox{!}{2.25cm}{\includegraphics{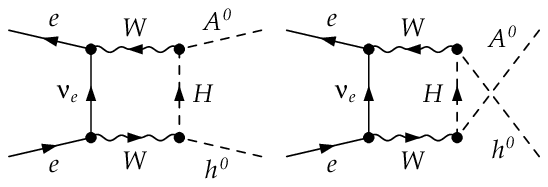}}}
\caption{Set of Feynman diagrams contributing to the $\eeAl$ process
at the one-loop level within the 2HDM. These diagrams describe the
box-type quantum corrections. } \label{fig:box}
\end{figure*}

\begin{figure*}[htb]
\begin{center}
\resizebox{!}{1.5cm}{\includegraphics{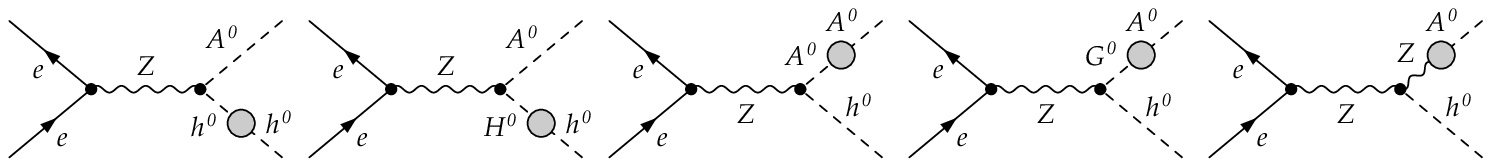}} \caption{Set of
Feynman diagrams contributing to the $\eeAl$ process at the one-loop
level within the 2HDM. These diagrams describe the finite WF
corrections to the Higgs-boson external legs. \label{fig:extern}}
\end{center}
\end{figure*}

\begin{figure*}[hbt]
\centerline{ \resizebox{!}{1.5cm}{\includegraphics{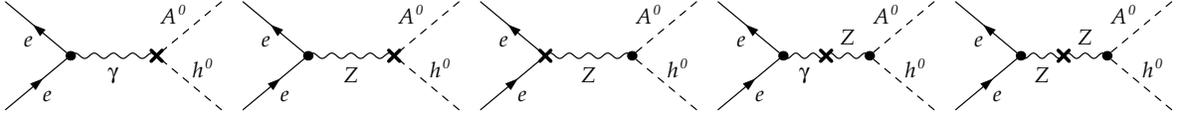}}}
\caption{Counterterm diagrams for the renormalization of the $\eeAl$
process. \label{fig:counter}}
\end{figure*}

Indeed, one-loop quantum corrections to the $\eeAl$ amplitude are
driven by a large set of Feynman diagrams involving the following
subsets of contributions: i) the self-energy corrections to the
$\PZ^0$ propagator and the $\Pphoton - \PZ^0$ mixing propagator (cf.
Figs.~\ref{fig:self1}-\ref{fig:self2}); ii) the vertex corrections
to the $\vzlA$ interaction  (cf.
Figs.~\ref{fig:vert11}-\ref{fig:vert12}); iii) the loop-induced
$\Pphoton\,\PHiggslightzero\,\PHiggspszero$ interaction (cf.
Fig.~\ref{fig:vert2}); iv) the vertex corrections to the
$\APelectron\Pelectron\PZ^0$ interaction  (cf.
Fig.~\ref{fig:vert3}); v) the box-type diagrams  (cf.
Fig.~\ref{fig:box}); vi) the finite wave-function renormalization of
the external Higgs-boson legs  (cf. Fig.~\ref{fig:extern}); and,
finally, the counterterm diagrams (cf. Fig.~\ref{fig:counter}). Of
course an equivalent set of diagrams would be needed to describe the
complementary process $\eeAH$, but we refrain from displaying them
explicitly.

One might expect that the cross-sections under study could be
sensitive to the model differences between type-I and type-II 2HDM's
through the Yukawa coupling corrections (cf. e.g. the first two rows
of diagrams in Fig.~\ref{fig:vert11}). However, the leading effects
are concentrated on the triple Higgs boson self-interactions, and
being the latter identical for type-I and type-II,  the model
differences are tiny in the relevant regions of the parameter space.

Let us also mention that for our current purposes we can safely
discard the pure QED corrections to the $\APelectron\Pelectron\PZ^0$
vertex. These QED corrections and the pure weak ones factorize into
two subsets which are separately UV finite and gauge-invariant.
Moreover, these photonic contributions are fully insensitive, at the
order under consideration, to the relevant 3H self-couplings on
which we are focusing in this work. For the processes under
consideration, where the Higgs bosons in the final state are
electrically neutral, the one-loop QED effects are confined to the
initial $e^+e^-$ vertex. In practice, the net outcome of the
accompanying initial-state real photon bremsstrahlung is to lower
the effective center-of-mass energy, and hence to modify the shape
of the cross sections as a function of $\sqrt{s}$. These QED effects
are the only source of infrared (IR) divergences in our calculation,
and since we do not consider them the obtained scattering amplitude
is IR finite. Finally, let us also emphasize that its inclusion
would not change significantly the overall size of the radiative
corrections, as in the regions of the parameter space where the 3H
self-couplings are dominant the QED effects are comparatively
negligible (as we have checked explicitly). In short, they are
unessential at this stage to test the presence of the new dynamical
features triggered by the 2HDM in the processes under consideration.

The UV divergent contributions arising from the set of Feynman
diagrams that we have just described can be absorbed by means of the
associated counterterm diagrams of Fig.~\ref{fig:counter}. In
particular, we need to include corresponding counterterms to
renormalize the $\vzlA$ and $\Azero\Hzero A^0$ basic interaction
vertices at the one-loop level. The analytical expressions ensue
from Eqs.~(\ref{eq:lag11}-\ref{eq:lag12}) after splitting parameters
and fields in the usual way into renormalized ones plus
counterterms. The final result reads as follows:
\begin{widetext}
\begin{eqnarray}
    \delta\,\lag_{\vzlA} = \frac{e\,\cos\,(\beta-\alpha)}{2\,\sw\,\cw} \PZ^0_\mu \PHiggspszero\,
    \overleftrightarrow{\partial^\mu}\,\PHiggslightzero &\Big[& \frac{\delta\,e}{e}
    +\frac{\sw^2 - \cw^2}{\cw^2}\,\frac{\delta \sw}{\sw} -
    \sin\beta\cos\beta\,\tan{(\beta-\alpha)}\,
    \frac{\delta\,\tan\beta}{\tan\beta} + \nonumber \\ &&
    + \frac{1}{2}\,\delta\,Z_{\PHiggslightzero} +\frac{1}{2}\,\delta\,
    Z_{\PHiggspszero} + \frac{1}{2}\,\delta\,Z_{\PZ^0} -
    \frac{1}{2}\,\tan{(\beta-\alpha)}\,\delta Z_{\PHiggsheavyzero\PHiggslightzero} +
   \nonumber \\
    &&
    +\frac{1}{2}\,\tan{(\beta-\alpha)}\,\delta Z_{\PHiggspszero\,G^0}
    +\frac{1}{2}\,\tan{(\beta-\alpha)}\,\delta Z_{\PHiggspszero\PZ^0}
    \,\Big]\,.
    \nonumber \\
    \label{eq:counter11}
     \nonumber \\
     \delta\,\lag_{\vzHA} = -\frac{e\,\sin\,(\beta-\alpha)}{2\,\sw\,\cw} \PZ^0_\mu \PHiggspszero\,
    \overleftrightarrow{\partial^\mu}\,\PHiggsheavyzero &\Big[& \frac{\delta\,e}{e}
   +\frac{\sw^2 - \cw^2}{\cw^2}\,\frac{\delta \sw}{\sw} +
    \sin\beta\cos\beta\,\cot{(\beta-\alpha)}\,
    \frac{\delta\,\tan\beta}{\tan\beta} + \nonumber \\ &&
   + \frac{1}{2}\,\delta\,Z_{\Hzero} +\frac{1}{2}\,\delta\,
    Z_{\PHiggspszero} + \frac{1}{2}\,\delta\,Z_{\PZ^0} -
    \frac{1}{2}\,\cot{(\beta-\alpha)}\,\delta Z_{\PHiggsheavyzero\PHiggslightzero} -
   \nonumber \\
    &&
    -\frac{1}{2}\,\cot{(\beta-\alpha)}\,\delta Z_{\PHiggspszero\,G^0}
    -\frac{1}{2}\,\cot{(\beta-\alpha)}\,\delta Z_{\PHiggspszero\PZ^0}
    \,\Big]\,.
    \nonumber \\
    \label{eq:counter12}
\end{eqnarray}
\end{widetext}
%
%

When writing down the above counterterms, one has to keep track
explicitly of the UV divergences stemming from the Higgs-Higgs
($\PHiggslightzero - \PHiggsheavyzero$), Higgs-Goldstone
($\PHiggspszero - G^0$) and Higgs-vector ($\PHiggspszero - \PZ^0$)
mixing effects which arise at the quantum level. By the same token,
and despite there are no $\Pphoton\PHiggslightzero\PHiggspszero$/
$\Pphoton\PHiggsheavyzero\PHiggspszero$ tree-level couplings, we
must also introduce appropriate counterterms in order to dispose of
the associated UV-divergences that appear on account of the
$\Pphoton-\PZ^0$ mixing at the quantum level. The relevant terms are
\begin{eqnarray}
\delta\,\lag_{\Pphoton\PHiggslightzero\PHiggspszero} &=&
\frac{\cos(\beta-\alpha)}{2\,\sw\,\cw}\,\frac{\delta\,Z_{\PZ^0\Pphoton}}{2}
\,A_\mu\,\PHiggspszero\,\overleftrightarrow{\partial^\mu}\PHiggslightzero \nonumber \\
\end{eqnarray}
and
\begin{eqnarray}
\delta\,\lag_{\Pphoton\PHiggsheavyzero\PHiggspszero} &=&
-\frac{\sin(\beta-\alpha)}{2\,\sw\,\cw\,}\,\frac{\delta\,Z_{\PZ^0\Pphoton}}{2}
\,A_\mu\,\PHiggspszero\,\overleftrightarrow{\partial^\mu}\PHiggsheavyzero\,.
\label{eq:counter22}
\end{eqnarray}
Likewise, we must introduce the corresponding counterterms for the
$\APelectron\Pelectron\PZ^0$ vertex, as well as for the $\PZ^0 -
\PZ^0$ and $\PZ^0 - \Pphoton$ self energies, for which we import the
same expressions that hold in the standard on-shell scheme for the
SM, see e.g.~\cite{Bohm:1986rj,Hollik95}, but of course including in
their calculation the loop contributions from the 2HDM Higgs bosons
(cf. Figs. \ref{fig:self1} and \ref{fig:self2}).
Let us denote by ${\cal M}^{(1)}$ the overall set of one-loop
contributions to $\plA$, and by $\delta\,{\cal M}^{(1)}$ the
associated counterterm amplitude. As we have discussed previously in
terms of Feynman diagrams, we can split the one-loop amplitudes into
the following (UV-finite) subsets;
\begin{align}
{\cal M}^{one-loop}_{\eeAl} = & \quad {\cal M}^{1,\PZ^0 - \PZ^0} & \mathcal{O}(\alpha^2_{ew})\nonumber \\
& +  {\cal M}^{\Pphoton - \PZ^0} &  \mathcal{O}(\alpha_{ew}\,\alpha_{em})\nonumber  \\
& +  {\cal M}^{\APelectron\Pelectron\PZ^0} & \mathcal{O}(\alpha^2_{ew})\nonumber  \\
& +  {\cal M}^{\vzlA} & \mathcal{O}(\alpha_{ew}\lambda^2_{3H})\nonumber \\
& +  {\cal M}^{\Pphoton\PHiggslightzero\PHiggspszero} & \mathcal{O}(\alpha_{em}\,
\lambda^2_{3H}) \nonumber \\
& +  {\cal M}^{\mbox{box}} & \mathcal{O}(\alpha^2_{ew})\nonumber \\
& +  {\cal M}^{\mbox{WF,\hzero}} & \mathcal{O}(\alpha_{ew}\,\lambda^2_{3H})\nonumber \\
& + {\cal M}^{\hzero\Hzero} & \mathcal{O}(\alpha_{ew}\,\lambda^2_{3H})\nonumber \\
& +  {\cal M}^{\Azero\PZ^0,\Azero G^0} &
\mathcal{O}(\alpha_{ew}\,\lambda^2_{3H}) \label{eq:loopsplit}
\end{align}
and similarly for the case of the heavy Higgs boson, $\eeAH$. In the
above equations, we understand that each amplitude is supplemented
with the corresponding counterterm. The complete expression
$\mathcal{M}^{one-loop} = {\cal M}^{(1)} + \delta\,{\cal
M}^{(1)}$ is free of UV divergences, as we
have explicitly checked. Moreover, at this point of the discussion
we are keeping track explicitly of the $\lambda_{3H}$ factors, in
order to highlight those contributions which can be enhanced by the
3H self-couplings. For example, among the interference terms between
the tree-level amplitude and the one-loop amplitude
(\ref{eq:loopsplit}) we have the class of terms
$\mathcal{O}(\alpha_{ew}^3)$ and also the class of terms
$\mathcal{O}(\alpha_{ew}^2\,\lambda^2_{3H})$, both classes of the
same order in perturbation theory, although the latter are expected
to be comparatively enhanced. We will check it numerically in the
next section.

Taking into account the above expressions, the overall (one-loop
corrected) scattering amplitude for the process $\eeAl$ reads as
follows:
\begin{eqnarray}
{\cal M}_{\eeAl} &=& \sqrt{\hat{Z}_{\hzero}}\, \,{\cal
M}_{\eeAl}^{(0)}
 + {\cal M}^{(1)}_{\eeAl} \nonumber \\ && + \delta\,{\cal
 M}^{(1)}_{\eeAl}\,.
\label{eq:amp11}
\end{eqnarray}
In the first term on the \textit{r.h.s.} of this equation, we have
appended  a \textit{finite} WF constant $\hat{Z}_{\hzero}$ for the
Higgs boson external leg. Actually, only the $\hzero$ field receives
a non-trivial contribution of this sort after we have chosen
$\hat{Z}_{\Azero}=1$. Indeed, while we are entitled to do so for the
\CP-odd field on account of the field renormalization condition
\eqref{eq:cond2}, this is not possible for $\hzero$. The reason is
that we use a WF renormalization constant for each Higgs doublet,
not for each field, see Eq.\,(\ref{eq:doublets}), and thus the
relations (\ref{eq:litran}) give no further room to normalize to one
the residue of the propagator for the other fields. The non-trivial
finite renormalization contribution $\hat{Z}_{\hzero}\neq 1 $ for
the other external Higgs boson in this process, $\hzero$, must be
computed explicitly, the result being
\begin{eqnarray}
     \hat{Z}_{\hzero} &=& \frac{1}{1 + \Re e\,\rs'_{\hzero}(q^2) -
       \left(\frac{\Re e\,\rs^2_{\hzero\Hzero}(q^2)}{q^2 - M_H^2 + \Re e\,\rs_{\Hzero}(q^2)}\right)'}\Bigg]_{q^2 =
       M^2_{\hzero}}
\label{eq:wf2}
\end{eqnarray}
Retaining from this expression only the
$\mathcal{O}(\alpha_{ew}\,\lambda^2_{3H})$ contributions to the
scattering amplitude, we are left with
\begin{eqnarray}
&&\sqrt{\hat{Z}_{\hzero}}\, \,{\cal M}_{\eeAl}^{(0)} =
\nonumber \\
&=&
\left[1 - \frac{\Re e\,\rs'_{\hzero}(M^2_{\PHiggslightzero})}{2} \right]\,{\cal M}_{\eeAl}^{(0)} + \mathcal{O}(\alpha^3_{ew})\,, \nonumber \\
\label{eq:wf3}
\end{eqnarray}

At the same time, our definition of the renormalized \CP-even mixing
angle implies that the $\hzero-\Hzero$ contribution to the one-loop
amplitude (\ref{eq:loopsplit}) vanishes identically:
\begin{eqnarray}
{\cal M}^{\hzero\Hzero}_{\eeAl} &=& \hat{Z}_{\hzero\Hzero}\,{\cal M}_{\eeAl}^{(0)} \nonumber \\
&=& -\,\frac{\Re e
\,\rs_{\hzero\Hzero}(M_{\hzero}^2)\,{\cal M}_{\eeAl}^{(0)}}{M_{\hzero}^2 - M_{\Hzero}^2 + \Re e\,\rs_{\Hzero}(M_{\hzero}^2)}\, = 0\,. \nonumber \\
\label{eq:wf31}
\end{eqnarray}
In turn, the finite correction associated to the on-shell $\Azero$
leg can be expressed in terms of the renormalized mixing
self-energies in the guise

\begin{eqnarray}
i{\cal M}^{\Azero\PZ^0+\Azero G^0}_\mu &=& -
\frac{e\,\sin(\beta-\alpha)}{\sw\cw}\,\frac{\hat{\Sigma}_{\Azero
G^0}(M_{\Azero}^2)}{M_{\Azero}^2 - M_Z^2}\,p_\mu
\nonumber \\
&& - \frac{\,M_Z\,\sin(\beta-\alpha)}{\sw\cw}\,
\frac{\hat{\Sigma}_{\Azero \PZ^0}(M_{\Azero}^2)}{M_{\Azero}^2 - M_Z^2}\,p_\mu  \nonumber \\
&=& -\frac{e\,\sin(\beta-\alpha)}{\sw\cw\,(M_{\Azero}^2 - M_Z^2)}\,p_\mu\, \nonumber \\
\qquad \qquad \qquad
&& \left[\hat{\Sigma}_{\Azero G^0}(M_{\Azero}^2) + M_Z\,\hat{\Sigma}_{\Azero \PZ^0}(M_{\Azero}^2)\right] \nonumber \\
&=&
-\frac{e\,\sin(\beta-\alpha)}{\sw\cw}\,\frac{\hat{\Sigma}_{\Azero
G^0}(M_{\Azero}^2)} {M_{\Azero}^2}\,p_\mu\,, \nonumber \\
\label{eq:wf4}
\end{eqnarray}
where in the last step we have made use of the Slavnov-Taylor
identity \eqref{eq:slavnov}. The 4-momentum $p_\mu$ stands for an
outgoing $\Azero$ boson. Finally, the expression for the
renormalized mixing self-energy $\Azero-G^0$ appearing in
Eq.\,(\ref{eq:wf4}) is obtained from (\ref{eq:self}) after
substituting in it the explicit form of the counterterm
(\ref{eq:cta0g0}).

%

Notice that since Eq.~\eqref{eq:cond2} decrees that the real part of
the function $\hat{\Sigma}_{\Azero \PZ^0}(M_{\Azero}^2)$ must
vanish, in the above amplitude~\eqref{eq:wf4} the only non-vanishing
contribution shall stem from the imaginary part, $\Im
m\hat{\Sigma}_{\Azero G^0}(M_{\Azero}^2)$. A similar expression may
be derived for the $\Hzero\Azero$ channel:
\begin{eqnarray}
i{\cal M}^{\,\Azero\PZ^0,\Azero G^0}_\mu =
-\frac{e\,\cos(\beta-\alpha)}{\sw\,\cw}\, \frac{\hat{\Sigma}_{\Azero
G^0}(M_{\Azero}^2)} {M_{\Azero}^2}\,p_\mu\,. \label{eq:wf5}
\end{eqnarray}
The complete $\eeAl$ amplitude at order $\mathcal{O}(\alpha^3_{ew})$
can finally be rearranged in the following way:
\begin{eqnarray}
&& {\cal M}_{\plA} =
{\cal M}^{(0)}_{\plA}\,  \times \nonumber\\
&& \times \left[ 1 -
\frac{\Re\,e\,\rs'_{\hzero}\left(M^2_{\PHiggslightzero}\right)}{2}
-\tan{(\beta-\alpha)}\frac{\hat{\Sigma}_{\Azero G^0}(M_{\Azero}^2)}
{M_{\Azero}^2}\right] \nonumber \\ && + {\cal M}^{(1)}_{\plA} +
\delta\,{\cal M}^{(1)}_{\plA} \label{eq:amp12}.
\end{eqnarray}
\noindent where the tree-level amplitude takes the explicit form:
\begin{eqnarray}
&& {i\cal M}^{(0)}_{\plA} =
-\frac{2\,\alpha_{em}\,\pi\,\cos(\beta-\alpha)}{\sw\cw\,(s -
M_Z^2)}\,\times
\nonumber \\
&& \times \bar{v}(p_1,\eta_1)\,(\slashed{k}_2 -
\slashed{k}_1)\,(g_L\,P_L + g_R\,P_R)\,u(p_2,\eta_2)\,
\label{eq:amp2}.
\end{eqnarray}
In the above expression, the notation reads as follows:
$\APelectron$, with 4-momentum $p_1$ and helicity $\eta_1$;
$\Pelectron$, with 4-momentum $p_2$ and helicity $\eta_2$;
$\PHiggspszero$, with 4-momentum $k_1$; and $\PHiggslightzero$, with
4-momentum $k_2$. We have also introduced the left and right-handed
weak couplings of the $\PZ^0$ boson to the electron, ${g_L} = (-1/2
+ \swd)/\cw\sw$, ${g_R} = \sw/\cw$ and the left and right-handed
projectors $P_{L,R}=(1/2)(1\mp\gamma_5)$. Let us also mention that
we have neglected the contribution of the $\PZ^0$-width, as it is
completely irrelevant at the working energies of the linear
colliders ($s\gg M_{\PZ^0}^2$). Finally, the total cross section is
obtained after squaring the matrix element, performing an averaged
sum over the polarizations of the colliding $\APelectron\Pelectron$
beams and integrating over the scattering angle:
\begin{eqnarray}
\sigma(\plA) &=&
\int\,d(\cos\theta)\,\frac{\lambda^{\frac{1}{2}}(s,M^2_{\PHiggslightzero},
M^2_{\PHiggspszero})}{32\,\pi\,s^2}\,\times \nonumber \\ &&
\bar{\Sigma}_{\eta_1,\eta_2}\,|M_{\plA}|^2 \label{eq:cs},
\end{eqnarray}
where we have introduced the standard definition of the K\"ahlen
function $\lambda(x,y,x) = x^2 + y^2 + z^2 - 2xy -2xz -2yz$.

Before closing this section, let us stress once more that a similar
set of formulae hold for the companion process $\eeAH$, which can be
related to the equations above through the correspondence
$\cos(\beta-\alpha) \to -\sin(\beta-\alpha)$ and $M_{\hzero} \to
M_{\Hzero}$.


\section{Neutral Higgs-pair production at 1 loop: numerical analysis}
\label{sec:results}
\subsection{Computational setup}

In this section, we describe in detail the results that are obtained
from the numerical analysis of the processes
$\APelectron\Pelectron\to \hzero\Azero/\Hzero\Azero$ at the
one-loop level. The basic quantities of interest are two: i) the
predicted cross section \eqref{eq:cs} at the Born-level
$\sigma^{(0)}$ and at one-loop $\sigma^{(0+1)}$; and ii) the
relative size of the one-loop radiative corrections, which we track
through the parameter
\begin{equation}\label{deltar}
\delta_r = \frac{\sigma^{(0+1)}-\sigma^{(0)}}{\sigma^{(0)}}\,.
\end{equation}
In practice, the computation has been performed with the help of the
standard algebraic and numerical packages  \emph{FeynArts},
\emph{FormCalc} and \emph{Looptools} \cite{feynarts} for the
generation of the Feynman diagrams, the analytical calculation and
simplification of the scattering amplitudes as well as the numerical
evaluation of the cross section.
\begin{table}[htb]
\begin{center}
\begin{tabular}{|c||c|c|}
 & $\cos(\beta-\alpha)$ & $\sin(\beta-\alpha)$  \\ \hline
$\alpha = \beta$ & 1 & 0 \\
$\alpha = \beta - \pi/2$ & 0 & 1 \\
$\alpha = \beta - \pi/3$ & $\frac{1}{2}$ & $\frac{\sqrt{3}}{2}$ \\
$\alpha =  \beta - \pi/6$ & $\frac{\sqrt{3}}{2}$ &  $\frac{1}{2}$ \\
$\alpha = \frac{\pi}{2}$  & $\sin\beta$ & $-\cos\beta$ \\
$\alpha = 0$  & $\cos\beta$ & $\sin\beta$ \\
\end{tabular}
\caption{Range of values for the $\vzlA$ and $\vzHA$ tree-level
couplings -- see equations (\ref{eq:coup11}) and (\ref{eq:coup12})
--- that are used throughout the numerical analysis
\label{tab:coup}.}
\end{center}
\end{table}
Aiming at a wider survey of the regions of phenomenological interest
in the 2HDM parameter space, we will explore the relevant range of
values for both $\cos(\beta-\alpha)$ and $\sin(\beta-\alpha)$ by
picking up the representative configurations quoted in
Table~\ref{tab:coup}. In particular, in the case $\alpha = \beta -
\pi/2$  the $\hzero\hzero\hzero$ coupling takes on the SM form, as
warned before. Furthermore, the case $\alpha = \pi/2$ corresponds to
one of the so-called \emph{fermiophobic} scenarios; namely, for type
I 2HDM, the lightest \CP-even Higgs couples to all fermions as
$\cos\alpha/\sin\beta$ times the SM coupling (cf. Table
\ref{typeIandII}); therefore, if $\alpha=\pm\pi/2$ the lightest
Higgs decouples from all the fermions\,\footnote{In the two
additional realizations of the 2HDM\,\cite{Barger:1989fj} (i.e.
those beyond the canonical ones indicated in Table \ref{typeIandII})
-- usually denoted as type-I' and type-II' models--, the leptons
couple differently from the quarks of the same third weak-isospin
component, and as a result not even the type I' models can exhibit
full fermiophobia. }. Full fermiophobia is of course not possible
for type-II models (although it is partly possible). In particular,
for the MSSM this phenomenon is impossible and its detection would
be a direct signal of non-SUSY physics. Let us clarify, however,
that \textit{per se} the fermiophobic scenarios are not particularly
relevant for the current analysis because the Yukawa couplings do
not play a central role here (as we will discuss below). However,
our cross-sections are indeed sensitive to the particular value of
$\alpha$ that we select, and in this sense the two fiducial choices
$\alpha = \beta$ and $\alpha = \pi/2$ represent two distinctive
regimes for our numerical analysis, as we will see.

For practical reasons, throughout our calculation we will be
interested in projecting out the large radiative corrections from
regions of the parameter space for which the predicted production
rates at the tree-level are already sizable. This means that we
shall mainly dwell on regimes nearby the optimal value of the
tree-level coupling for the $\l \A$ channel ($\beta \simeq \alpha$)
and for the $\H \A$ one ($\beta \simeq \pi/2 + \alpha$).

\jump In Table~\ref{tab:masses} we quote the different Higgs-boson
mass spectra that we shall consider in our numerical analysis. These
mass sets have been designed to cover a wide span of
phenomenologically motivated regimes. Because of their mass
splittings, Sets I-III can only be realized within a general 2HDM,
whereas Sets IV-VI reflect characteristic benchmark scenarios of
Higgs boson masses within the MSSM; they have been generated through
the MSSM parameter inputs specified in Table~\ref{tab:parmssm}. Note
that Sets I to III (similarly, Sets IV to VI) are organized through
increasing values of the Higgs boson masses. Furthermore, let us
also highlight that, due to the presence of a light charged Higgs
boson, Sets I-II and IV-V are adequate only for type-I 2HDM, whereas
Sets III and VI are valid either for type-I or type-II 2HDM.

\begin{table}[htb]
\begin{center}
\begin{tabular}{|c||c|c|c|c|}
 & $M_{\PHiggslightzero}\,[\GeV]$ & $M_{\PHiggsheavyzero}\,[\GeV]$ & $M_{\PHiggspszero}\,[\GeV]$ & $M_{\PHiggs^\pm}\,[\GeV]$  \\ \hline
Set I & 100 & 150 & 140 & 120  \\
Set II & 130 & 150 & 200 & 160  \\
Set III & 150 & 200 & 260 & 300  \\
Set IV & 95 & 205 & 200 & 215  \\
Set V & 115 & 220 & 220 & 235 \\
Set VI & 130 & 285 & 285 & 300 \\
\end{tabular}
\caption{Choices of Higgs masses that used throughout our
computation. Due to the value of $M_{\PHiggs^\pm}$, Sets I-II and
IV-V are only suitable for type-I 2HDM, whilst Sets III and VI can
describe either type-I or type-II 2HDM. Moreover, Sets IV-VI have
been specially devised to reproduce the characteristic splitting of
the Higgs masses within the MSSM at one-loop. \label{tab:masses}}
\end{center}

\end{table}

\begin{table}[htb]
\begin{center}
\begin{tabular}{|c|c|c|c|}
 & Set IV & Set V & Set VI \\ \hline
$\tan\beta$ & 3.7 & 20.0 & 20.0 \\
$M_{\Azero}$ & 200 & 220 & 285 \\
$M_{SUSY}$ & 300 & 650 & 800 \\
$\mu$ & 300 & 300 & 300 \\
$M_2$ & 300 & 200 & 300 \\
$X_t \equiv A_t - \mu/\tan\beta$ & -300  & -300 & -1100
\end{tabular}
\end{center}
\caption{Choices of MSSM parameters which give rise to the Higgs
mass sets IV-VI in Table~\ref{tab:masses}, thus mimicking the
characteristic one-loop mass splittings within the Higgs sector of
the MSSM. The numerical value for these masses has been obtained
with the aid of the program \emph{FeynHiggs} by taking the full set
of EW corrections at one-loop\,\cite{feynHiggs}. Universal trilinear
couplings ($A_t = A_b = A_\tau$) and GUT relations for the gaugino
soft-SUSY breaking mass terms are assumed throughout.
\label{tab:parmssm}}
\end{table}

\begin{figure*}[htb]
\begin{center}
\begin{tabular}{ccc}
 \hspace{-0.6cm}\resizebox{!}{7cm}{\includegraphics{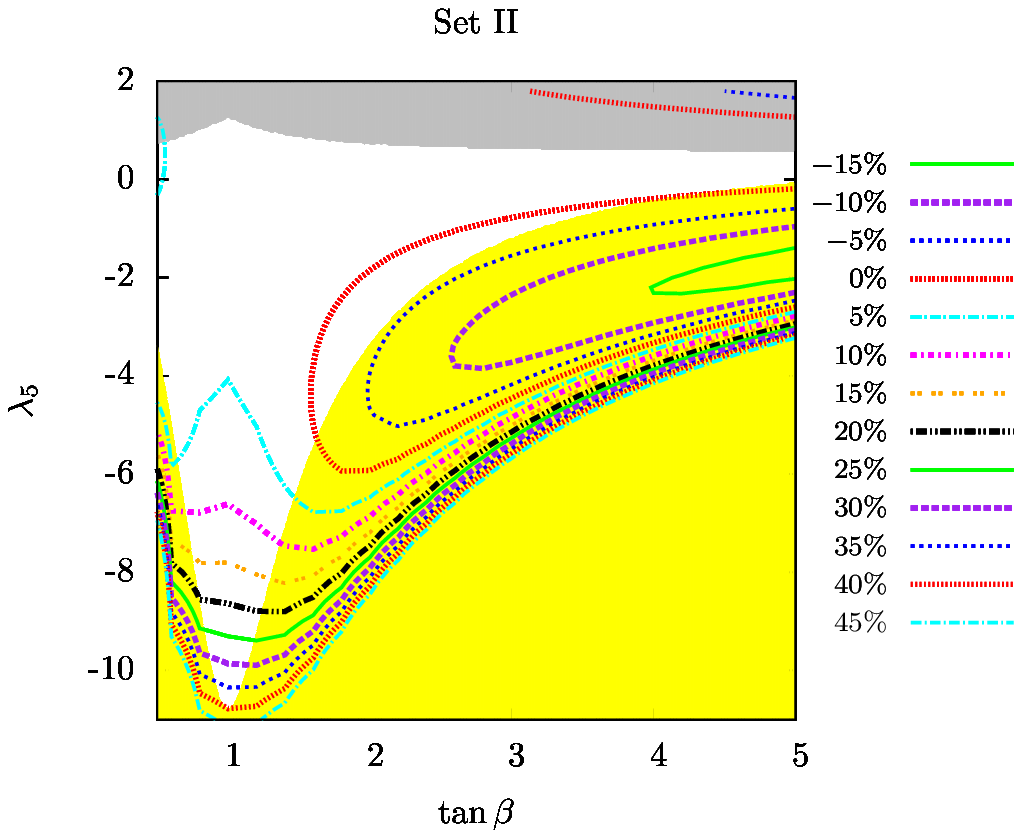}} & \hspace{0.5cm} &
 \hspace{ -0.6cm}\resizebox{!}{6.5cm}{\includegraphics{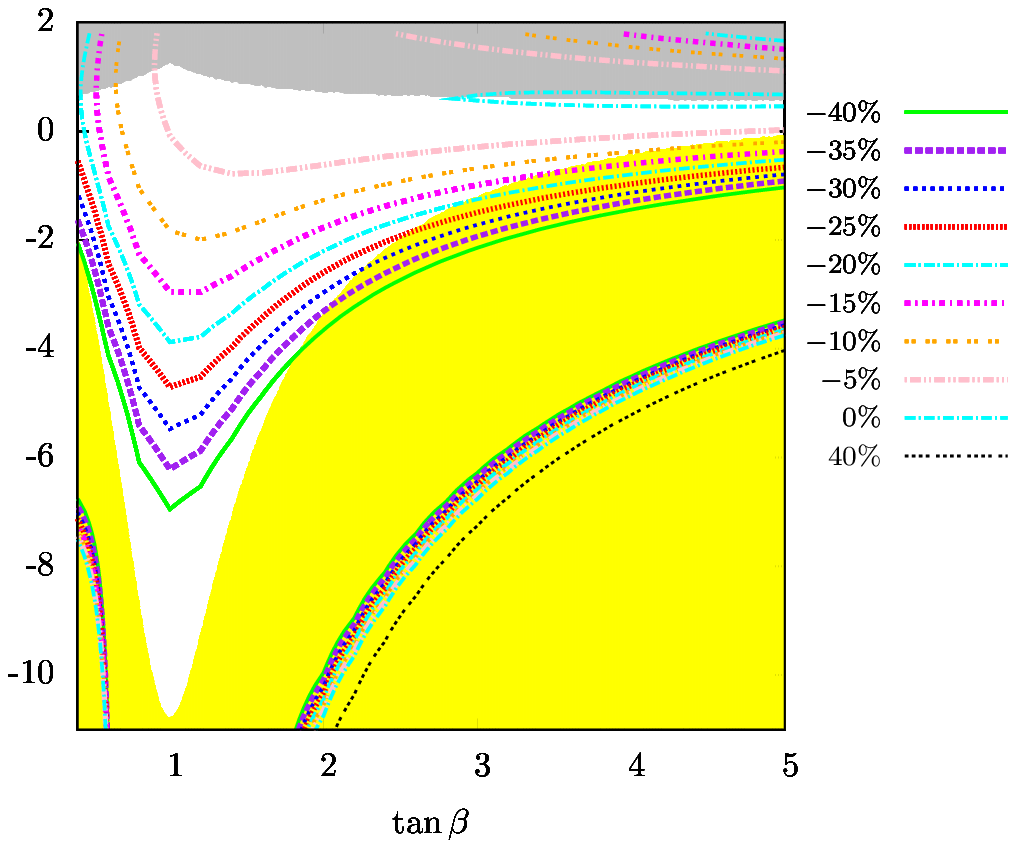}} \\
\end{tabular}
\caption{Contour plots of the quantum corrections $\delta_r$ (in \%)
-- defined in Eq. (\ref{deltar}) -- to the $\eeAl$ cross-section as
a function of $\tan\beta$ and $\lambda_5$, for Set II of Higgs boson
masses (cf. Table~\ref{tab:masses}), $\alpha = \beta$ and $\sqrt{s}
= 500\,\GeV$ (left panel), $\sqrt{s} = 1000\,\GeV$ (right panel).
The shaded area at the top stands for the region excluded by the
vacuum stability bounds, whereas the disjoint shaded area occupying
most of the central and lower region signals the domain excluded by
the tree-level unitarity bounds. \label{fig:scan-s2}}
\end{center}
\end{figure*}
\begin{figure*}[htb]
\begin{center}
\begin{tabular}{ccc}
 \hspace{-0.6cm}\resizebox{!}{7cm}{\includegraphics{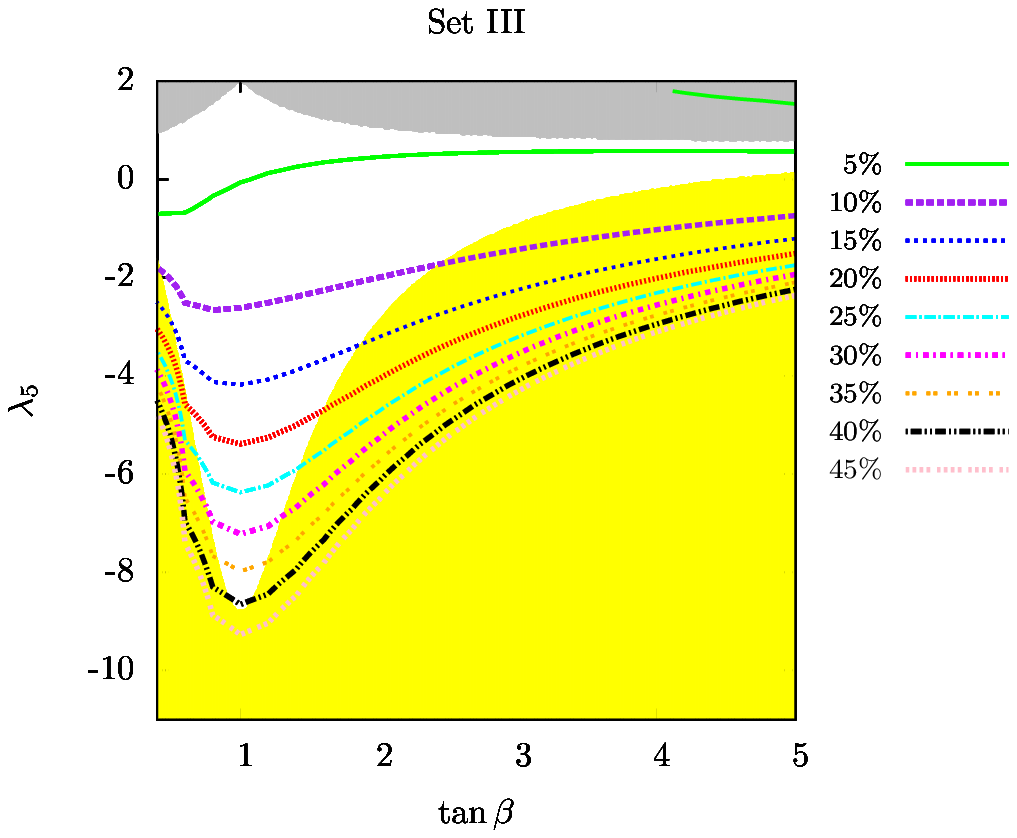}} & \hspace{0.5cm} &
 \hspace{ -0.6cm}\resizebox{!}{6.5cm}{\includegraphics{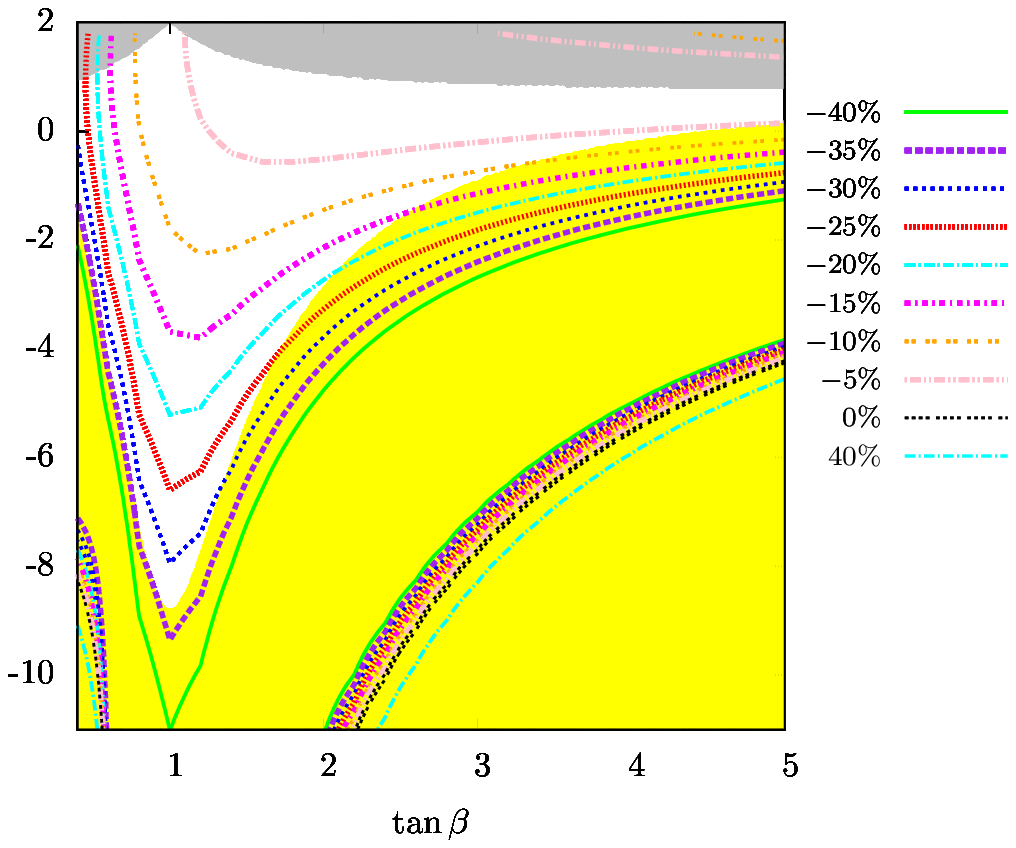}} \\
\end{tabular}
\caption{As in Fig.\,\ref{fig:scan-s2}, but for the Set III of Higgs
boson masses (cf. Table~\ref{tab:masses}), $\alpha = \beta$ and
$\sqrt{s} = 500\,\GeV$ (left panel), $\sqrt{s} = 1000\,\GeV$ (right
panel). \label{fig:scan-s3}}
\end{center}
\end{figure*}
\begin{figure*}[htb]
\begin{center}
\begin{tabular}{ccc}
 \hspace{-0.6cm}\resizebox{!}{7cm}{\includegraphics{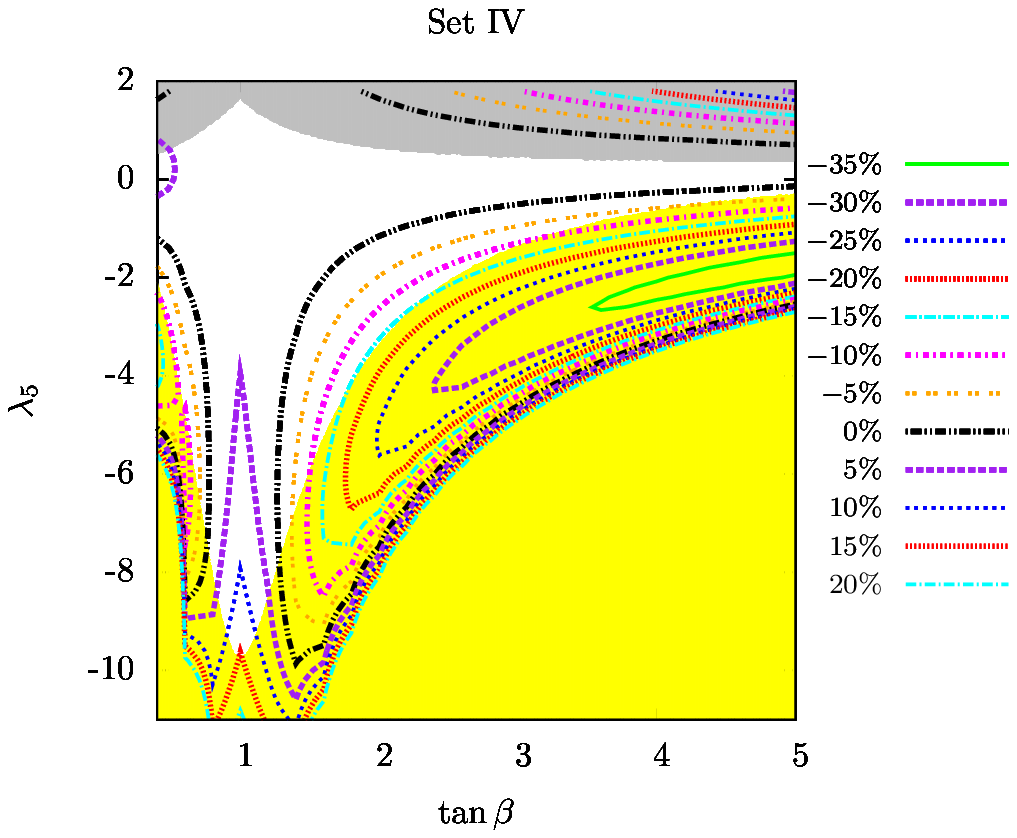}} & \hspace{0.5cm} &
 \hspace{ -0.6cm}\resizebox{!}{6.5cm}{\includegraphics{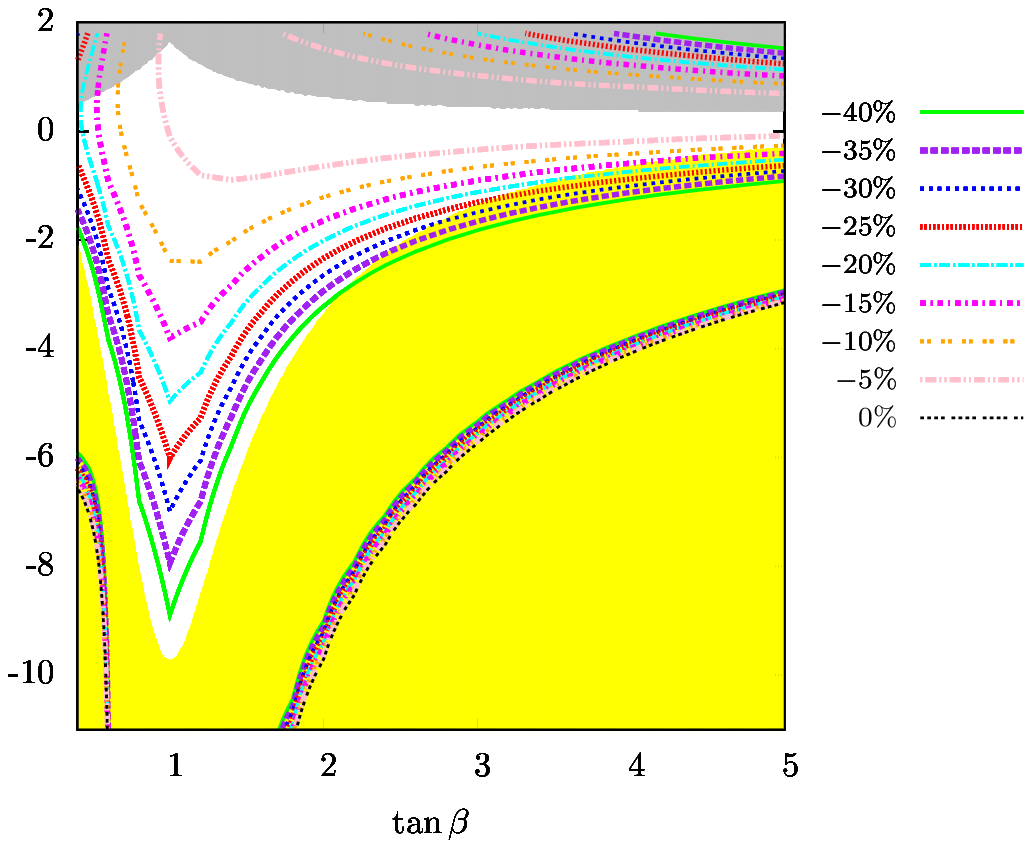}}  \\
\end{tabular}
\caption{As in Fig.\,\ref{fig:scan-s2}, but for the Set IV of Higgs
boson masses (cf. Table~\ref{tab:masses}), $\alpha = \beta$ and
$\sqrt{s} = 500\,\GeV$ (left panel), $\sqrt{s} = 1000\,\GeV$ (right
panel). \label{fig:scan-s4}}
\end{center}
\end{figure*}
In turn, the value of $\lambda_5$ is severely restrained by the
theoretical constraints stemming from the tree-level unitarity and
vacuum stability, i.e. from equations (\ref{eq:vacuum_conditions})
and (\ref{eq_uncond})-(\ref{eq:arbounds}). The bottom line is that
$\lambda_5> 0$ is strongly disfavored, and the permissible values
lie mostly in the range
\begin{equation}\label{eq:lambda5bound}
-11 \lesssim \lambda_5 \lesssim 0\,.
\end{equation}
The value of $\tan\beta$ that is preferred so as to optimize the
cross-sections in the above range is  $\tan\beta \simeq 1$. We will
carefully derive these constraints in what follows and shall examine
their implications on the numerical analysis of the cross-sections
under study.

\subsection{$\plA$}

\begin{table*}[htb]
\begin{center}
\begin{tabular}{|c|c||c|c|c|c||c|c|c|c|} \hline
\multicolumn{2}{|c||}{\,} & \multicolumn{4}{|c||}{$\sqrt{s} = 500\,\GeV$}
& \multicolumn{4}{|c|}{$\sqrt{s} = 1\,\TeV$} \\ \hline
\multicolumn{2}{|c||}{\,} &$\alpha = \beta$ & $\alpha = \beta - \pi/3$ & $\alpha = \beta - \pi/6$
& $\alpha = \pi/2$
 &$\alpha = \beta$ & $\alpha = \beta - \pi/3$ & $\alpha = \beta - \pi/6$
& $\alpha = \pi/2$
\\ \hline \hline \multirow{2}{1cm}{Set I} &
$\sigma_{max}\,[\femtobarn]$ & 34.13 & 13.12 & 24.96 & 18.93  & 2.89 & 0.99 & 1.64 & 0.99 \\
\cline{2-10}
 & $\delta_r\,[\%]$ & 3.27 & 58.83 & 0.70 & 14.57 & -72.99 & -62.69 & -79.58 & -81.47\\ \hline \hline
\multirow{2}{1cm}{Set II} & $\sigma_{max}\,[\femtobarn]$ & 26.71 &
7.34 & 20.05 & 13.10 & 4.08 &
0.85 & 2.70 & 1.56  \\ \cline{2-10}
 & $\delta_r\,[\%]$ & 31.32 & 44.43 & 31.42 & 28.81  & -58.42 & -65.14 & -63.28 & -68.11\\ \hline \hline
\multirow{2}{1cm}{Set III} & $\sigma_{max}\,[\femtobarn]$ & 11.63 &
3.60 & 9.08 & 6.36 & 6.11 &
1.39 & 4.22 & 2.86   \\ \cline{2-10}
 & $\delta_r\,[\%]$ & 35.17 & 67.59 & 40.68 & 47.86  & -30.16 & -36.36 & -35.62 & -34.58 \\ \hline \hline
\multirow{2}{1cm}{Set IV} & $\sigma_{max}\,[\femtobarn]$ & 27.44 &
12.12 & 18.37 & 15.41 & 5.99 &
1.19 & 2.65 & 1.67   \\ \cline{2-10}
 & $\delta_r\,[\%]$ & 12.86 & 99.42 & 0.76 & 26.81 & -40.53 & -52.78 & -64.96 & -66.81\\ \hline \hline
\multirow{2}{1cm}{Set V} & $\sigma_{max}\,[\femtobarn]$ & 23.03 &
9.17 & 16.90 & 13.21 & 6.35 &
1.35 & 3.48 & 2.26  \\ \cline{2-10}
 & $\delta_r\,[\%]$ & 21.87 & 94.22 & 19.25 & 39.80 & -34.22 & -44.05 & -51.98 & -53.21\\ \hline \hline
\multirow{2}{1cm}{Set VI} & $\sigma_{max}\,[\femtobarn]$ & 8.78 &
2.33 & 6.86 & 4.63 & 7.33 &
1.63 & 4.78 & 3.34  \\ \cline{2-10}
 & $\delta_r\,[\%]$ & 19.17 & 26.21 & 24.13 & 25.58  & -14.08 & -23.71 & -25.28 & -21.70 \\ \hline \hline
\end{tabular}
\caption{Maximum total cross section $\sigma^{(0+1)}(\eeAl)$ at
$\sqrt{s} = 500\,\GeV$ and $\sqrt{s} = 1\,\TeV$, together with the relative size of the
radiative corrections, defined by $\delta_r$ in Eq. (\ref{deltar}),
for the different sets of Higgs bosons masses quoted in
Table~\ref{tab:masses}. The results are obtained at fixed $\tan\beta
= 1$ and different values of $\alpha$, with $\lambda_5$ at its
largest attainable value, cf. Table~\ref{tab:maxl5} .
\label{tab:xslight}}
\end{center}
\end{table*}

\begin{table}[htb]
\begin{center}
\begin{tabular}{|c|c|}
 &\ $\ \lambda_5\ $ \\ \hline
\ Set I \ \  &\  -11\ \\
\ Set II\ \ & \ -10 \ \\
\ Set III\ \ &\ -8\ \\
\ Set IV\ \ & \ -9\ \\
\ Set V\ \ &\ -9\ \\
\ Set VI\ \ &\ -6\ \\
\end{tabular}
\end{center}
\caption{Approximate maximum (negative) value of the parameter
$\lambda_5$ which can be attained for each of the mass sets in
Table~\ref{tab:masses}, in accordance with the stringent constraints
arising from unitarity and vacuum stability bounds. As both sets of
constraints depend explicitly on the input Higgs masses, we find a
different maximum $|\lambda_5|$ for each of the mass sets.
\label{tab:maxl5}}
\end{table}
A fairly generous overview of the numerical results for the
Higgs-pair production through $\plA$ at one-loop is provided
systematically in Figs~\ref{fig:scan-s2} - \ref{fig:scan-s4} and in
Table~\ref{tab:xslight}. In the latter, we
display the predicted values for the total cross-section
$\sigma(\plA)$, together with the relative size of the associated
quantum corrections (\ref{deltar}) for each of the Higgs-boson mass
sets in Table~\ref{tab:masses}. The results are obtained for fixed
$\tan\beta = 1$ and different values of $\alpha$ indicated in the
table. The fiducial ILC center-of-mass energy is taken to be either
$\sqrt{s}= 500\,\GeV$ or $\sqrt{s} =
1\,\TeV$. For each mass set,
$\lambda_5$ is allowed to take its largest (negative) attainable
value indicated in Table~\ref{tab:maxl5}. As we will confirm
throughout the numerical analysis, this prescription insures
maximally enhanced quantum corrections.
%

\begin{figure*}[htb]
\begin{center}
\includegraphics[scale=0.5]{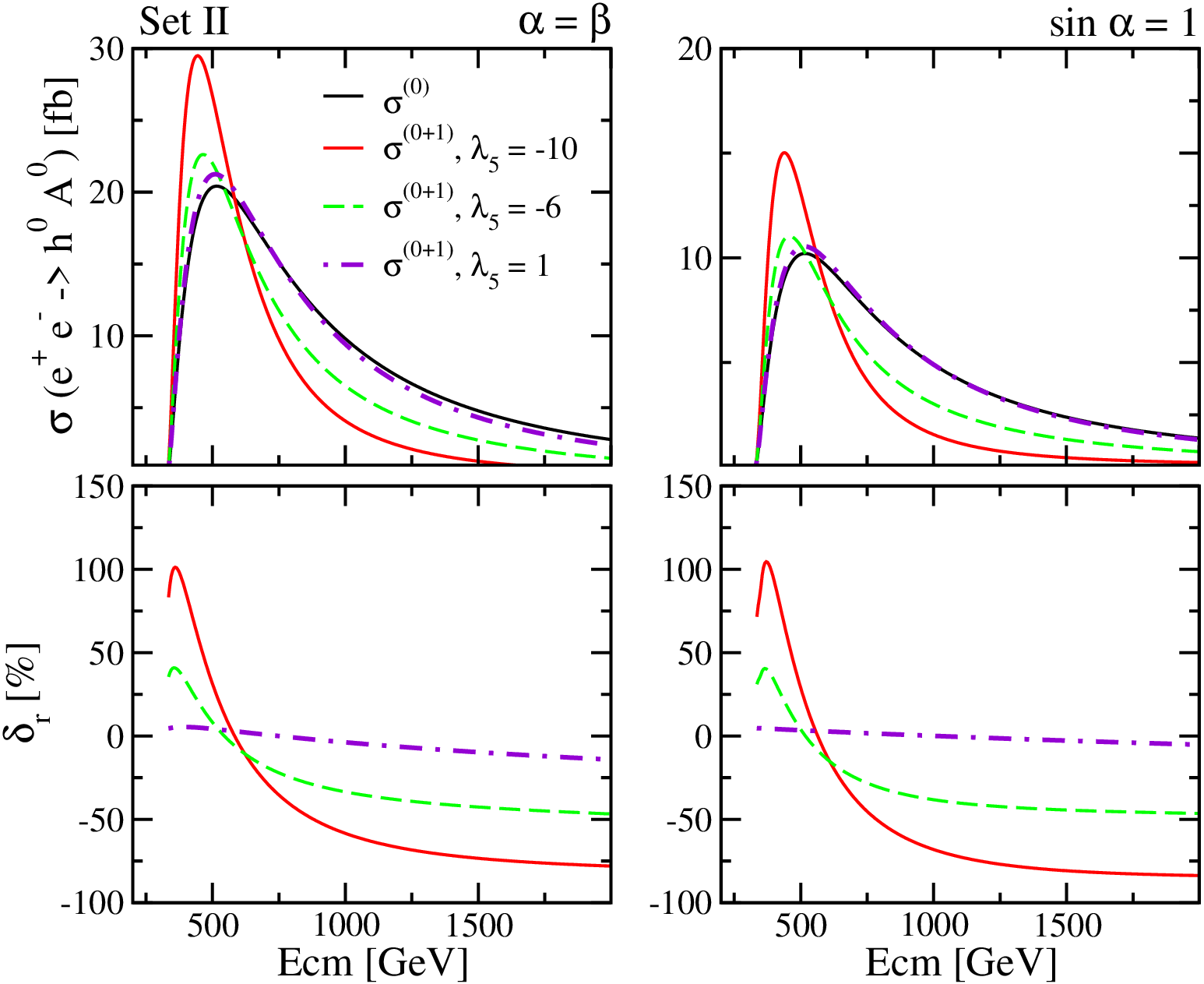}
\caption{Total cross section $\sigma(\eeAl)$ (in \femtobarn) at the
tree-level and at one-loop (upper panels) and relative one-loop
correction $\delta_r$ (in \%) -- see Eq.\,(\ref{deltar})-- (lower
panels) as a function of $\sqrt{s}$ for Set II of Higgs boson
masses, cf. Table~\ref{tab:masses}. Shown are the results obtained
within three different values of $\lambda_5$, at $\tan\beta=1$ and
for $\alpha=\beta$ (left) and $\alpha=\pi/2$ (right) -- the latter
defining the so-called \emph{fermiophobic} limit for the $\hzero$
boson (for type I 2HDM). \label{fig:overs_s2}}
\end{center}
\end{figure*}
\begin{figure*}[htb]
\begin{center}
\includegraphics[scale=0.5]{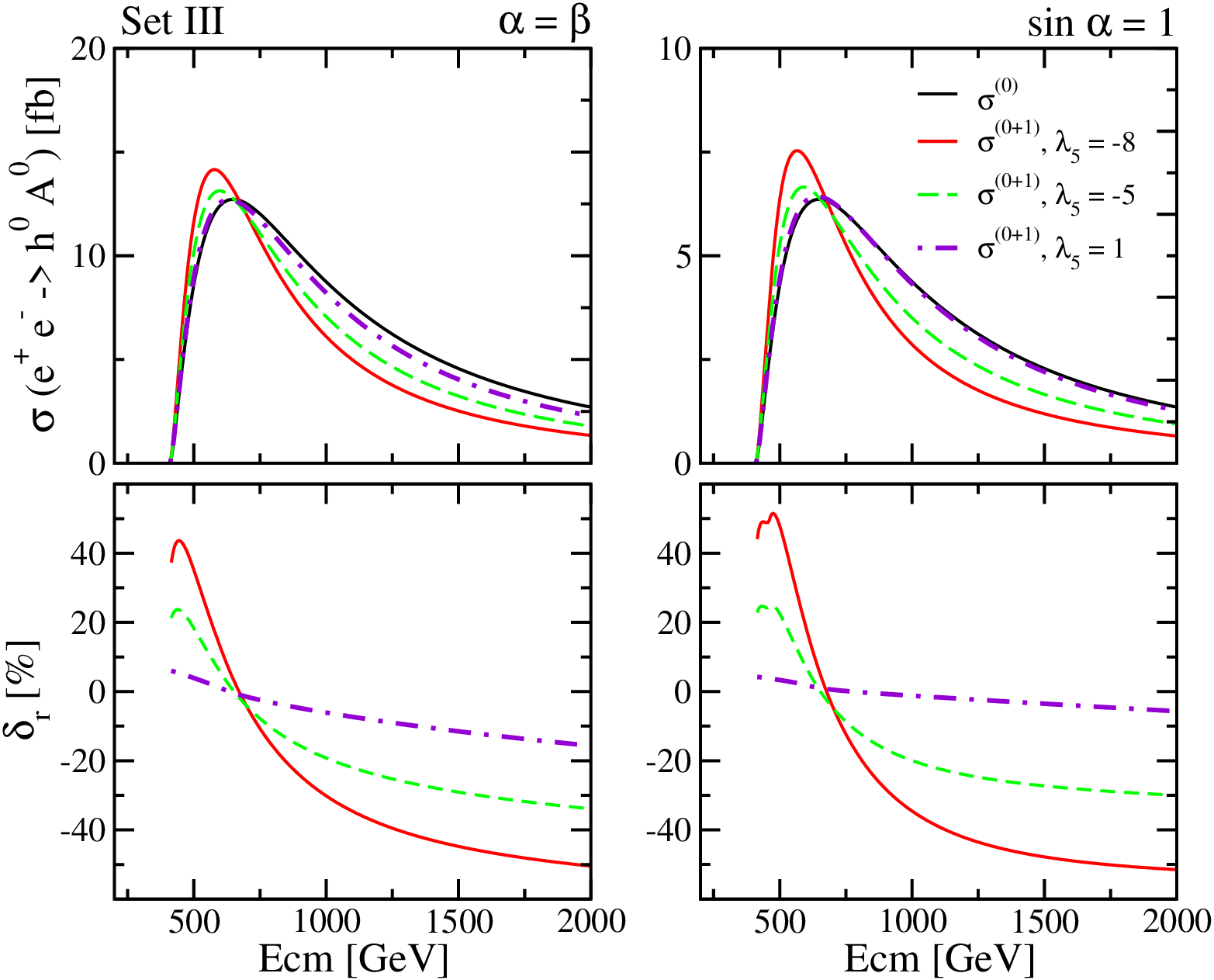}
\caption{Total cross section $\sigma(\eeAl)$ (upper panels) and
relative one-loop correction $\delta_r$ (lower panels) as a function
of $\sqrt{s}$ for Set III of Higgs boson masses, cf.
Table~\ref{tab:masses}. Shown are the results obtained within three
different values of $\lambda_5$, at $\tan\beta=1$ and for
$\alpha=\beta$ (left) and $\alpha=\pi/2$ (right).
\label{fig:overs_s3}}
\end{center}
\end{figure*}
\begin{figure*}[htb]
\begin{center}
\includegraphics[scale=0.5]{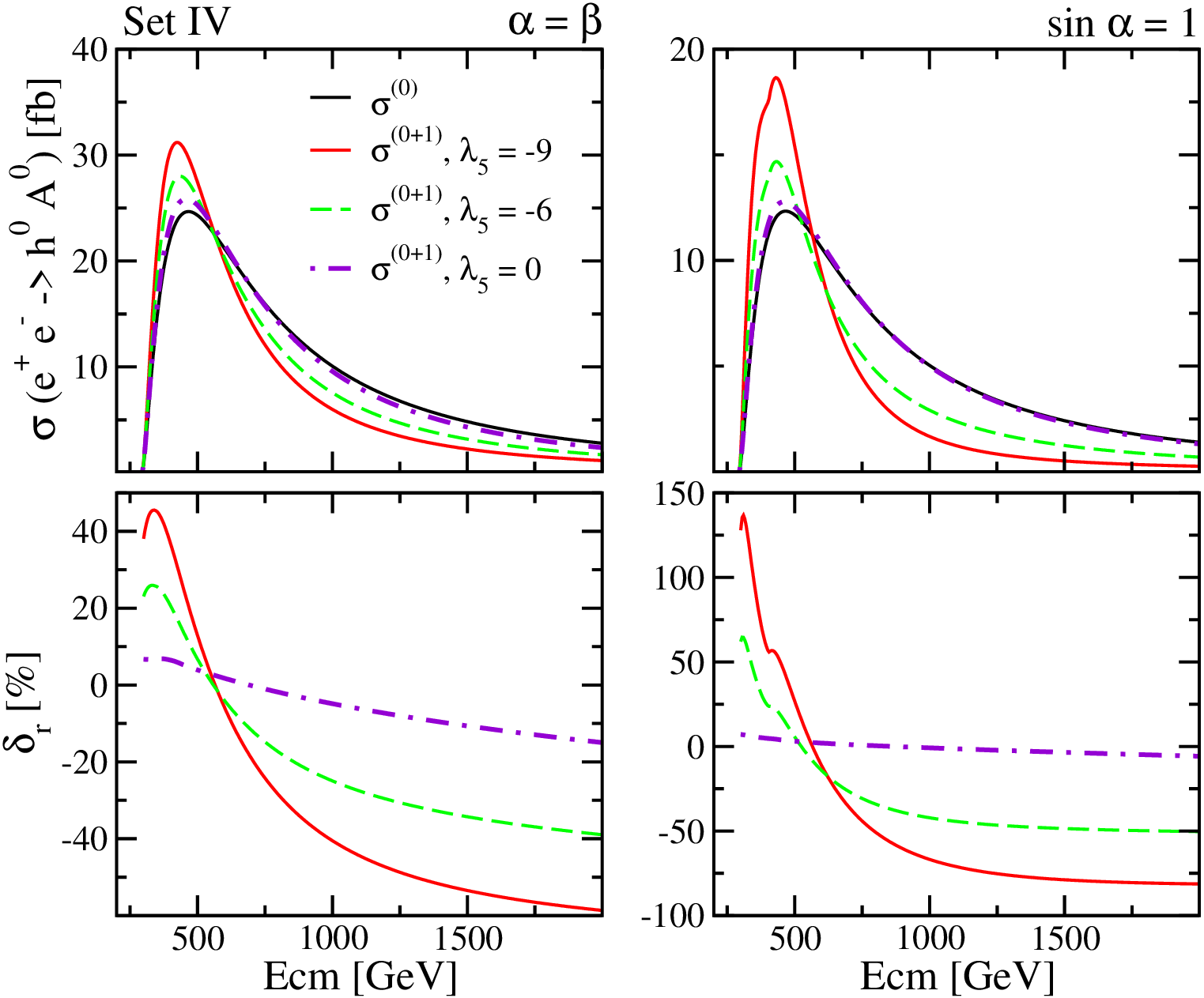}
\caption{Total cross section $\sigma(\eeAl)$ (upper panels) and
relative one-loop correction $\delta_r$ (lower panels) as a function
of $\sqrt{s}$ for Set IV of Higgs boson masses, cf.
Table~\ref{tab:masses}. Shown are the results obtained within three
different values of $\lambda_5$, at $\tan\beta=1$ and for
$\alpha=\beta$ (left) and $\alpha=\pi/2$ (right).
\label{fig:overs_s4}}
\end{center}
\end{figure*}

The corresponding loop-corrected cross sections
$\sigma^{(0+1)}(\plA)$ stay in the approximate range $2-30$ fb for
$\sqrt{s} = 500\,\GeV$ -- this would entail up to barely $10^3-10^4$
events in the standard segment $500\,\invfb$ of integrated
luminosity. At $\sqrt{s} = 1\,\TeV$, however, the predicted yields
deplete down to $\mathcal{O}(1)\,\femtobarn$ in most cases, although this
would still give a turnover of a few hundred events at the end of
the luminosity shift.

As for the radiative corrections themselves, they are also
explicitly quoted in
Table~\ref{tab:xslight}. One can see that
they can be remarkably large, being either positive (at ``low''
energies, viz. for $\sqrt{s} \sim 500\,\GeV$) or negative (for
$\sqrt{s}\sim 1\,\TeV$ and above). In the most favorable instances,
such corrections can boost the cross-section value up to $\delta_r
\sim 100\,\%$ at $\sqrt{s}=500\,\GeV$, whereas the corresponding
suppression in the high energy range ($\sqrt{s}=1\,\TeV$) can attain
$\delta_r \sim - 80\%$ owing to a severe destructive interference
between the tree-level and the one-loop amplitudes.

In Figs~\ref{fig:scan-s2} - \ref{fig:scan-s4} we explore in more
detail the behavior of the radiative corrections and their interplay
with the theoretical constraints associated to the perturbative
unitarity and vacuum stability conditions. Although the range
$\tan\beta\geqslant 1$ is usually the preferred one from the
theoretical point of view, we entertain the possibility that
$\tan\beta$ can be slightly below $1$ in order to better assess the
behavior around this value. In these plots, we depict $\delta_r$ (in
$\%$) for two choices of energies, $\sqrt{s}= 500\,\GeV$ (left
panels) and $\sqrt{s}= 1\,\TeV$ (right panels). For each of these
figures we show independent contour plots corresponding respectively
to the Sets II-IV of Higgs boson masses (cf. Table~\ref{tab:masses})
in the $\tan\beta - \lambda_5$ plane.
The first of these sets involves an assorted spectrum of Higgs boson
masses, which is impossible to reproduce in the MSSM, whereas the
second set closely mimics a typical MSSM-like Higgs mass spectrum,
in which we recognize the characteristic (approximate) degeneracy of
the heavy Higgs boson masses: $M_{\Hzero} \simeq M_{\Azero} \simeq
M_{\PHiggs^{\pm}}$.
We find that these scenarios
are well representative of the phenomenological trends shown by all
Higgs mass spectra under analysis. By setting $\alpha = \beta$ we
optimize the tree-level $\vzlA$ coupling and in this way we insure a
sizable value of the lowest order cross-section, which is of course
a good starting point for having a chance to eventually measure
quantum corrections on it.

From these plots, it is eye-catching that the unitarity constraints
are the most restrictive ones, and tend to disfavor large values of
$\tan\beta>1$ or values $\tan\beta\ll 1$. This is natural because
the triple and quartic Higgs self-couplings rocket fast with large
and small $\tan\beta$, see Eq.\,(\ref{eq:param2hdm}) and Table
\ref{tab:trilinear}. At the same time, these unitarity constraints
place a lower bound on another highly sensitive parameter of this
study, viz. $\lambda_5$, whereby  $\lambda_5$ cannot be smaller than
$-11$ or thereabouts (in other words, $|\lambda_5|\lesssim 11$).
Besides, this parameter is sharply stopped from above near
$\lambda_5\gtrsim 0$ as a direct consequence of the vacuum stability
condition (cf. the shaded band in the upper part of the plots).
Thus, the combined set of bounds build up a characteristic physical
domain in the parameter space of Figs~\ref{fig:scan-s2} -
\ref{fig:scan-s4} which pretty much looks like a deep valley flanked
with sharp cliffs on each side, and centered at $\tan\beta=1$. As a
result, large departures from this central value are incompatible
with large values of $|\lambda_5|$. In particular, the range from
$\tan\beta = 3$ onwards is circumscribed to a narrow-edged neck
around $\lambda_5 \simeq 0$. In addition, the heavier the Higgs mass
spectrum, the more severe the unitarity bounds are. Let us also
remark that all these constraints over $|\lambda_5|$ exhibit a mild
dependence on the actual value of the\ \CP-even mixing angle
$\alpha$ in the regimes considered here, whereas our cross-sections
do vary significantly with it.

In short, from the analysis of
Figs.~\ref{fig:scan-s2}-\ref{fig:scan-s4} two basic regimes of
phenomenological interest can be sorted out: 1) On the one hand, we
find scenarios in which $\lambda_5<0$ and where this coupling can
stay moderately large in absolute value (viz. $|\lambda_5|\sim
5-10$) while $\tan\beta \simeq 1$; this would correspond to the
relatively narrow allowed stretch on the left-hand-side of these
plots. In such configurations, a subset of 3H self-couplings are
remarkably singled out -- the more negative is $\lambda_5$, the
greater is the coupling strength. Such enhancement, which is
transferred to the $\eeAl$ amplitude through a set of Higgs
boson-mediated one-loop vertex diagrams (see Fig.~\ref{fig:vert11}),
translates into sizable quantum corrections which become more
vigorous with growing $|\lambda_5|$. For instance, from
Fig.~\ref{fig:scan-s2} we see that around $\tan\beta\simeq 1$,
$\delta_r$ reaches $\sim +10\,\%$ at $\lambda_5 = -6$, whilst it
becomes $+30\,\%$ for $\lambda_5 = -10$. Furthermore, we encounter
that positive radiative corrections as big as $\delta_r \sim +40\%$
at $\sqrt{s}= 500\,\GeV$ may switch drastically into large negative
effects of the same order $\sim - 40\%$ (hence suppressing
significantly the cross-section) when the center-of-mass energy is
increased up to $\sqrt{s}= 1\TeV$ --  see e.g. the right panel of
Fig.~\ref{fig:scan-s2}. Obviously, positive radiative corrections
are preferred in practice because they invigorate the physical
signal. Therefore, the startup energy $\sqrt{s}= 500\,\GeV$ of the
ILC seems to be the ideal regime to probe these particular quantum
effects rather than moving to higher energies;
%
%
\begin{table*}[t]
\begin{center}
\begin{tabular}{|c||c|c|c|c|c|c|} \hline
 & $\PZ^0/\Pphoton - \PZ^0$ $[\%]$& box $[\%]$ & $\vzlA$ $[\%]$&
$\Pphoton\PHiggslightzero\PHiggspszero$ $[\%]$ &
$\APelectron\Pelectron\PZ^0$ $[\%]$ & WF $[\%]$ \\ \hline \hline Set
I & 0.82 & -10.09 & 16.56 & 0.11 & 7.60 & -7.00  \\ \hline Set II &
0.77 & -9.84 & 44.95 & 0.04 & 7.64 & -7.69 \\ \hline Set III & 0.71
& -6.14 & 35.15 & -0.23 & 7.40 & -0.57 \\ \hline Set IV & 0.70 &
-7.37 & 20.21 & -0.30 & 7.59 & -5.79  \\ \hline Set V & 0.68 & -7.12
& 26.32 & -0.30 & 7.62 & -3.49 \\ \hline Set VI & 0.66 & -6.26 &
15.06 & -0.20 & 7.69 & 2.64 \\ \hline
\end{tabular}
\caption{Relative contribution to the total cross section
$\delta_{r,i} = (\sigma^{(0+i)}-\sigma^{(0)})/\sigma^{(0)}$ from
each topology of one-loop diagrams normalized to the tree-level
rate. The results are derived at fixed $\sqrt{s}=500\,\GeV$ for
$\tan\beta =1$, $\alpha = \beta$ and choosing the value of
$\lambda_5$ that optimizes the quantum corrections for each mass
regime, see Table~\ref{tab:maxl5}. \label{tab:top}}
\end{center}
\end{table*}

2) On the other hand, a very different behavior occurs in the
neck-shaped region of the contour plots at moderate $\tan\beta>1$,
as it is apparent in Figs.~\ref{fig:scan-s2}-\ref{fig:scan-s4}. Here
the enhancement capabilities of the leading 3H self-couplings are
rather meager with respect to the aforementioned large-$|\lambda_5|$
scenario. The reason is that for $\tan\beta \gtrsim 3$ the coupling
$|\lambda_5|$ is nailed down to take very low values.

In the absence of a significant yield from the Higgs boson-mediated
loop corrections, we might expect alternative sources of quantum
effects to pop up. The first ones coming to mind are of course those
originating in the Higgs-fermion Yukawa sector. However, it turns
out that the Yukawa interactions are not particularly efficient in
the regions explored in Figs.~\ref{fig:scan-s2}-\ref{fig:scan-s4}.
Indeed, from Table \ref{typeIandII} it is apparent that none of the
Higgs boson couplings to fermions is spurred on for
$\tan\beta\gtrsim 1$ (in any of the two canonical 2HDM's). Only the
regions with $\tan\beta < 1$ -- which are disfavored by the general
theoretical expectations -- and $\cos\alpha \sim 1$ could contribute
here mainly via  the Higgs-top interaction, whose strength is
$\hzero\APtop\Ptop \sim m_t\,\cos\alpha/\sin\beta$ regardless of
considering type-I or type-II models. Additionally, the Higgs-bottom
Yukawa coupling ($\sim m_b\,\tan\beta$ within type-II 2HDM), could
furnish a competitive source of enhancement, as it does e.g. in the
MSSM. In the present case, however, no trace is left of such effect
owing to our main focus on the range $\tan\beta\gtrsim 1$, although
this situation could change in the region $\tan\beta<1$ (see below).
\begin{figure*}[htb]
\begin{center}
\includegraphics[scale=0.5]{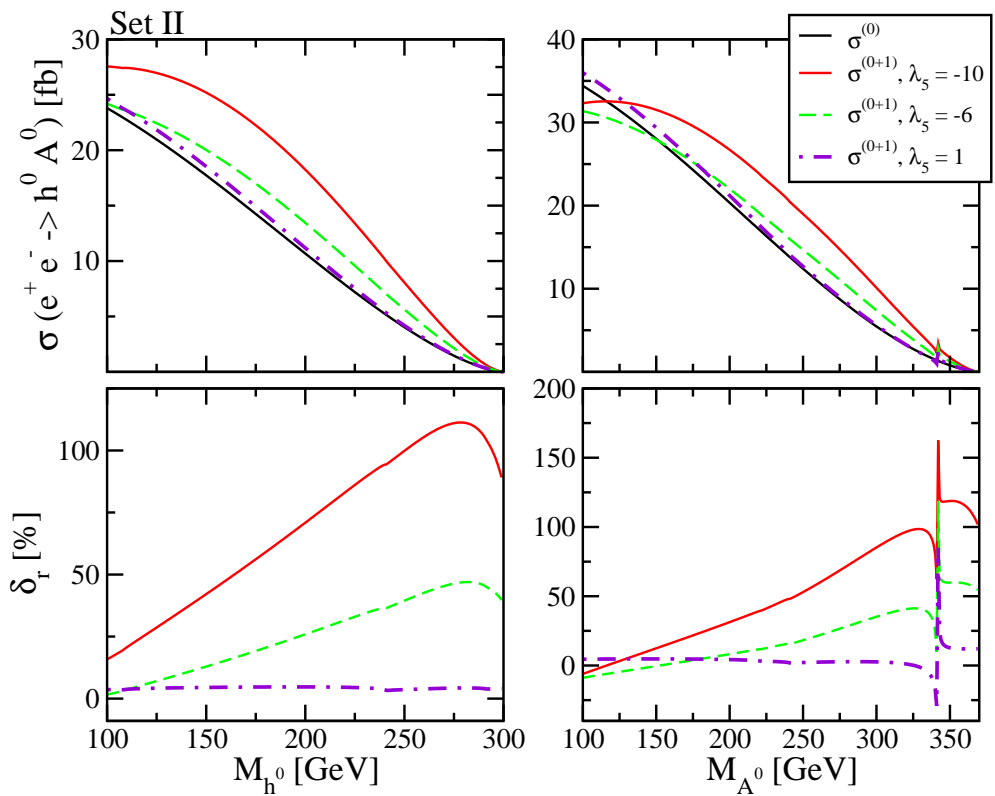}
\caption{Total cross section $\sigma(\eeAl)$ (upper panels) and
relative one-loop correction $\delta_r$ (lower panels) at fixed
$\sqrt{s}=500\,\GeV$, given as a function of $M_{\PHiggslightzero}$
(left) and $M_{\PHiggspszero}$ (right) for Set II of Higgs boson
masses, cf. Table~\ref{tab:masses}; $\tan\beta = 1$, $\alpha=\beta$
and three different values of $\lambda_5$. \label{fig:overmh0_s2}}
\end{center}
\end{figure*}
\begin{figure*}[htb]
\begin{center}
\includegraphics[scale=0.5]{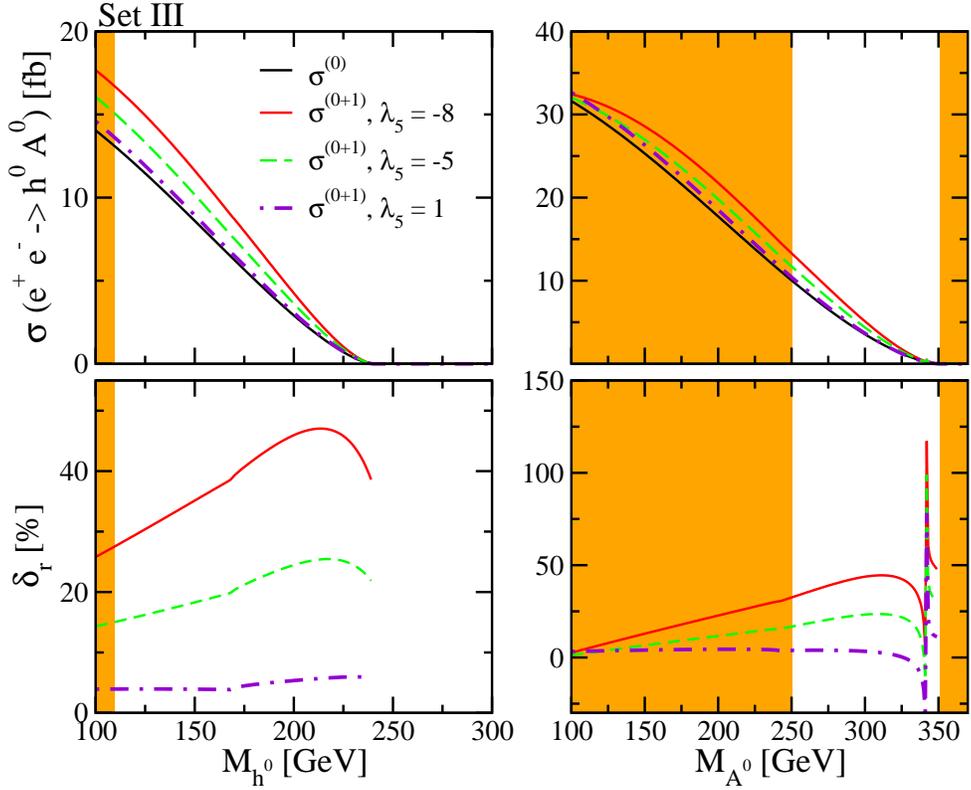}
\caption{Total cross section $\sigma(\eeAl)$ (upper panels) and
relative one-loop correction $\delta_r$ (lower panels) at fixed
$\sqrt{s}=500\,\GeV$, given as a function of $M_{\PHiggslightzero}$
(left) and $M_{\PHiggspszero}$ (right) for Set III of Higgs boson
masses, cf. Table~\ref{tab:masses}; $\tan\beta = 1$, $\alpha=\beta$
and three different values of $\lambda_5$. The dashed regions are
excluded by the constraints over the $\delta\rho$ parameter.
\label{fig:overmh0_s3}}
\end{center}
\end{figure*}
\begin{figure*}[htb]
\begin{center}
\includegraphics[scale=0.5]{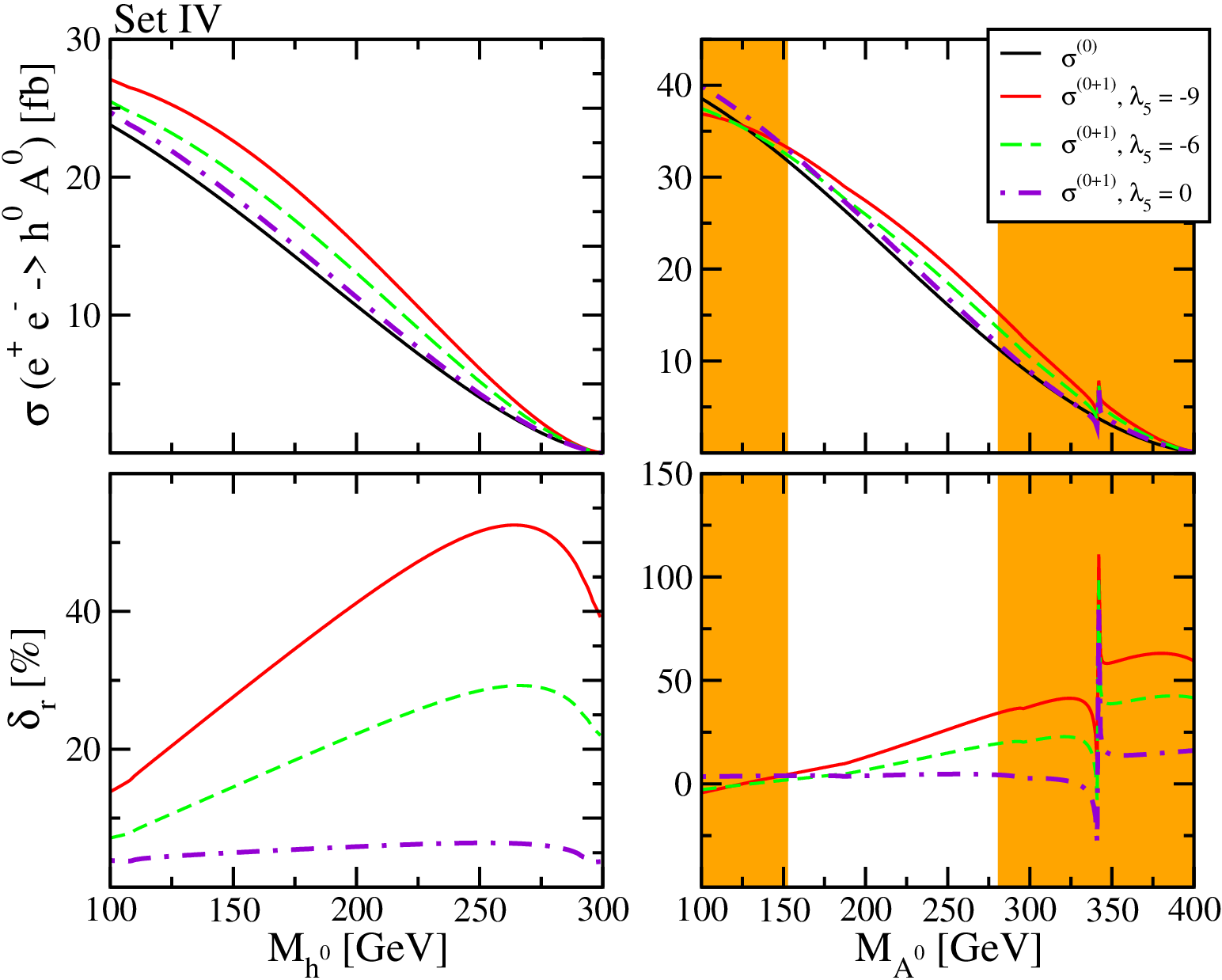}
\caption{As in Fig. \ref{fig:overmh0_s3}, but for Set IV of Higgs
boson masses, cf. Table~\ref{tab:masses}. \label{fig:overmh0_s4}}
\end{center}
\end{figure*}

Closely related with the previous comment, we should also stress the
fact that there are no outstanding differences between the obtained
results from type-I and type-II 2HDM's in the context of this
analysis. Indeed, to start with the Higgs-quark Yukawa interactions
do not enter at the tree-level in the processes (\ref{2h}) under
consideration; second, the trilinear couplings are common to both
types of models; and third, we have already emphasized that the
one-loop differences that could manifest through the distinct form
of these Yukawa couplings in type-I and type-II models are virtually
obliterated in the physical region permitted by the constraints. All
in all, perhaps the most distinctive feature between both models
boils down, in this context, to just the requirement that
$M_{\Hpm}\gtrsim 300\,\GeV$ for type-II 2HDM's as a result of the
low-energy $B$-meson physics constraints, which tend to
substantially raise the average mass of the Higgs spectrum for this
class of models -- cf. Sets III and VI of in Table~\ref{tab:masses}.
The basic differences in the physics of $\eeAl$ in both models
thereby narrow down to mass differences between their characteristic
Higgs-boson mass spectra \footnote{Recently, a combined
analysis of different $B$-meson physics constraints over the 2HDM
parameter space suggests that values of $\tan\beta \sim 1$ could be
disfavored for charged Higgs masses $M_{\Hpm}$ too near to $300$ GeV
(and certainly below) \,\cite{mahmoudi}. However, the level of
significance is not high and, moreover, our leading quantum
corrections are basically insensitive to the charged Higgs mass.
Therefore, a shift of $M_{\Hpm}$ slightly upwards restores the
possibility of $\tan\beta \sim 1$ at, say, $2\sigma$ without
altering significantly our results (as we have checked.)}.

A very similar pattern is met in the other explored scenarios, the
only differences being ascribable to the distinct Higgs boson mass
spectrum that each of these scenarios accommodate. In particular,
heavier Higgs boson masses may have a twofold impact: i) dynamical,
in that they soften the influence of Higgs-mediated one-loop
diagrams -- and hence the strength of the 3H self-couplings --
inasmuch as these contributions become suppressed by inverse powers
of the Higgs masses; and ii) kinematical, in that they shift the
peak of the $\sigma(\sqrt{s})$ curve, since the latter is correlated
with the production threshold of the Higgs pair.

All the above mentioned phenomenological features are transparent in
Figs.~\ref{fig:overs_s2}-\ref{fig:overs_s4}, in which we explore the
evolution of the cross section as a function of the center-of-mass
energy. We include, in each plot, the tree-level contribution
$\sigma^{(0)}$ and the loop-corrected one $\sigma^{(0+1)}$ for
different values of $\lambda_5$. Worth noticing is that by testing
the influence of $\lambda_5$ on the cross-sections, we are testing
the bulk capability of the 3H self-interactions, and hence their
potential fingerprint on the cross-sections at the quantum level. In
the lower panels, we track the related behavior of the quantum
correction $\delta_r$ with $\sqrt{s}$ for the same set of
$\lambda_5$ values. The plots are generated at fixed $\tan\beta=1$
and for two different values of the tree-level $\vzlA$ coupling:
namely, $\alpha=\beta$ (where the tree-level coupling is maximum)
and $\alpha = {\pi}/{2}$ (corresponding to the aforementioned
\emph{fermiophobic} limit, in which $\hzero$ fully decouples from
the fermionic sector for type I models). In
Figs.\,\ref{fig:overs_s2},\ref{fig:overs_s3} and \ref{fig:overs_s4},
we display the numerical results obtained for Sets II, III and IV of
Higgs boson masses, respectively. A similar numerical output is
obtained for the remaining mass sets in Table~\ref{tab:masses},
showing no relevant departure from the phenomenological trend that
we have recorded as yet.

Let us focus e.g. on Fig.~\ref{fig:overs_s2} for a while. The
tree-level curve exhibits the expected behavior with $\sqrt{s}$, as
it scales with the s-channel $\PZ^0$-boson propagator $\sim
1/(s-M_Z^2)$. The maximum cross section is achieved at $\sim
500\,\GeV$. The same pattern is followed by the full loop-corrected
cross-section, although the dependence with $\sqrt{s}$ is sharper.
In fact, the range where $\delta_r>0$ is much briefer than the one
where $\delta_r<0$, although the former is characterized by very
significant quantum effects; so much so, that they may skyrocket up
to $\lesssim 100\,\%$ (at the largest allowed values of
$|\lambda_5|$) around the critical energy domain where
$\sqrt{s}=500\,\GeV$. Such effects could hardly be missed in the
first runs of the ILC. In contrast, when we move from $\sqrt{s} =
500\,\GeV$ to $\sqrt{s} = 1000\,\GeV$ the resulting cross section
$\sigma^{(0+1)}$ drops by at least $50\%$. The impact of the
enhanced 3H self-couplings can be easily read off from the dramatic
differences in the one-loop production rates when the value of
$\lambda_5$ is varied in the theoretically allowed range
(\ref{eq:lambda5bound}). Thus, when $|\lambda_5|$ is pulled down at
fixed $\tan\beta$, the quantum effects are rapidly tamed, in
correspondence with the fact that the 3H self-couplings are no
longer stirred up.

The drastic decline of the triple Higgs boson self-interactions in
this domain cannot be counterbalanced by the contribution of the
Yukawa couplings, as they are not very significant in the region of
the parameter space where $\tan\beta$ is of order one. At this
point, it comes to mind our former observation about the potential
role that the region $\tan\beta<1$ could play. Even though it is not
favored on theoretical grounds, we can not exclude the possibility
that ${\cal O}(0.1)<\tan\beta<1$, and in fact the unitarity and
vacuum stability bounds cannot dismiss in block this range when
$\lambda_5$ is sufficiently small. Thus, for example, in the
conditions of Fig.\,\ref{fig:scan-s2}, and under the assumption that
$\lambda_5 = 0$, we have checked that $\tan\beta \sim 0.2$ is
allowed by the aforesaid constraints, and in this region we
encounter quantum effects as large as $+30\%$. They are the result
of the combined effort from the diagrams involving the top quark
Yukawa coupling and those involving the trilinear Higgs boson
couplings -- which, in this case, become enhanced by a pure
$\tan\beta<1$ effect, rather than by an oversized $\lambda_5$.

It is also interesting to track carefully the fractional payoff $\,
\delta_r = (\sigma^{(0+i)}-\sigma^{(0)})/\sigma^{(0)}$ from the
various one-loop topologies ($i=1,2,3,...$) of Feynman diagrams
contributing to the process $\plA$. The corresponding results are
displayed in Table~\ref{tab:top} under the given conditions.
Included there are the gauge invariant subsets of diagrams of
various types, namely $\PZ^0 - \PZ^0/\Pphoton$ vacuum polarization
effects, box diagrams, the renormalized $\vzlA$ and
$\APelectron\Pelectron\PZ^0$  vertex corrections, the loop-induced
form factor $\Pphoton\hzero\Azero$, and finally the finite WF
corrections to the external $\hzero$ leg. We can see that the gauge
boson self-energies provide tiny contributions of order $1\%$ at
most. The box diagrams, in their stead, are non-negligible (of order
$10\%$), albeit negative. At the same time we detect mutually
destructive effects between e.g. the finite WF correction and the
$\APelectron\Pelectron\PZ^0$ vertex.

Most conspicuously, and above all, we revalidate the dominant role
of the Higgs boson mediated loop-diagrams (triggered by 3H
self-interactions) as the truly leading source of quantum
corrections to the cross-section. These dominant effects are
concentrated on the fundamental $\vzlA$ vertex. This vertex, which
is purely gauge-like at the lowest order of perturbation theory,
becomes drastically promoted at the one-loop level as a result of
the dominant quantum effects sourced by the 3H self-interactions. We
may cast the overall renormalization of such vertex as
follows\,\footnote{A similar discussion is of course also applicable
to the complementary $\vzHA$ coupling and the corresponding
production process $\APelectron\Pelectron\to\Hzero\Azero$, see the
next subsection.}:

\begin{align}
\Gamma^0_{\vzlA} &  \rightarrow Z^{1/2}_{\l}\,\left(\Gamma^0_{\vzlA}
+ \Gamma^1_{\vzlA}\right) \nonumber \\ & \simeq
Z^{1/2}_{\l}\,\left(\Gamma^0_{\vzlA} +
\Gamma^{1,3H}_{\vzlA}\right)\,,
\label{eq:strengtcorrectedlight}
\end{align}
%
%
where $\Gamma^{1,3H}$ denotes the leading subset of (Higgs-mediated)
one-loop corrections. The latter can be encapsulated in a UV-finite
subset of one-loop Feynman diagrams and subsequently reabsorbed into
a modified or effective $\vzlA$ coupling. Simple power counting and
educated guess provides the following estimate:
\begin{eqnarray}
\Gamma^{eff}_{\vzlA} \sim
\Gamma^{0}_{\vzlA}\frac{\lambda^2_{3H}}{16\pi^2\,s}\,f(M^2_{\hzero}/s,
M^2_{\Azero}/s) \label{eq:eff}
\end{eqnarray}
where $f(M^2_{\hzero}/s, M^2_{\Azero}/s)$ is a dimensionless form
factor that accounts for the complete one-loop amplitude. For all
its cumbersome structure, we do know at least that $f$ dies out with
the center-of-mass energy of the process and encodes the sign flip
of the radiative corrections when we increase the center-of-mass
energy from $\sqrt{s} = 0.5\,\TeV$ to $\sqrt{s} = 1\,\TeV$.  The
prefactor $1/16\pi^2$ stands for the trademark numerical suppression
attached to one-loop form factors, whilst $\lambda_{3H}$ denotes a
generic 3H self-coupling. Noteworthy is the fact that this coupling
can reach values as high as $\lambda_{3H}/M_W\simeq
|\lambda_5|/e\simeq 30$ and still under the stringent constraints
imposed by unitarity.

It is enlightening to try to estimate the expected size of the
maximum corrections. From Eq.~\eqref{eq:eff} we can roughly gauge
the maximum relative contribution, e.g. at the startup energy
$\sqrt{s}= 500\,\GeV$:

\begin{eqnarray}
\delta_r &=& \frac{\sigma^{(0+1)}-\sigma^{(0)}}{\sigma^{(0)}} =
\frac{\left<2\Re e{\cal M}^{(0)}{\cal M}^{(1)}+|{\cal M}^{(1)}|^2\right>}{\left<|{\cal M}^{(0)}|^2\right>}\nonumber \\
&\sim& 2\,\frac{|\lambda_{3H}|^2}{16\,\pi^2\,s}\,f(M^2_{\hzero}/s,
M^2_{\Azero}/s)\Big]_{\sqrt{s}=500} \simeq 50\%\,,
\label{eq:estimate}
\end{eqnarray}
where $<...>$ stands for the various operations of averaging and
integration of the squared amplitudes. In the above equation, we
have neglected the square of the one-loop amplitude (see, however,
just below) and set $f \sim 1$, which is a reasonable assumption
since the current value of the center-of-mass energy, $\sqrt{s}=
500\,\GeV$, is far from the range in which one-loop corrections are
proved to be negligible. The latter result of
Eq.~\eqref{eq:estimate} falls in the right ballpark of the optimal
values for $\delta_r$ that we have identified from our exact
numerical analysis, for Higgs boson masses of the order of a few
hundred GeV and maximum allowed $|\lambda_5|$ values (so as to push
$\lambda_{3H}$ close to the unitarity border). By means of the same
formula we can also estimate the contribution to $\delta_r$ arising
from the squared of the one-loop amplitude, which we have neglected
before. We find
\begin{eqnarray}
\delta_r^{(2)} &=& \frac{\left\langle|{\cal
M}^1|^2\right\rangle}{\left\langle|{\cal M}^0|^2\right\rangle} \sim
\left(\frac{\lambda^2_{3H}}{16\pi^2\,s}\right)^2\big{\vert}_{\sqrt{s}=500}
\simeq 5\%\,, \label{eq:estimate2}
\end{eqnarray}
thus roughly $10\%$ of the one-loop leading effect. It turns out
that this is again in good agreement with the numerical calculation
\footnote{We will further dwell on the potential relevance of such
$\mathcal{O}(\alpha^4_{ew})$ effects in Section
~\ref{sec:conclusions}.}. All in all we conclude that, as long as 3H
couplings are sizable, quantum effects will be manifest -- and
potentially very large.

\begin{table*}[htb]
\begin{center}
\begin{tabular}{|c|c||c|c|c|c||c|c|c|c|} \hline
\multicolumn{2}{|c||}{\,} & \multicolumn{4}{|c||}{$\sqrt{s} = 500\,\GeV$}
& \multicolumn{4}{|c|}{$\sqrt{s} = 1\,\TeV$} \\ \hline
\multicolumn{2}{|c||}{\,} &$\alpha = \beta - \pi/2$ & $\alpha = \beta - \pi/3$ & $\alpha = \beta - \pi/6$
& $\alpha = 0$  &$\alpha = \beta - \pi/2$ & $\alpha = \beta - \pi/3$ & $\alpha = \beta - \pi/6$
& $\alpha = 0$
\\ \hline \hline \multirow{2}{1cm}{Set I} &
$\sigma_{max}\,[\femtobarn]$ & 40.54 & 20.33 & 2.29 & 8.82 & 2.51 & 1.30 & 0.16 & 0.59 \\
\cline{2-10}
 & $\delta_r\,[\%]$ & 51.83 & 1.50 & -65.72 & -33.96 & -75.59 & -83.10 & -93.62 & -88.50 \\ \hline \hline
\multirow{2}{1cm}{Set II} & $\sigma_{max}\,[\femtobarn]$ & 26.32 &
17.53 & 4.39 & 10.29 & 3.59 &
2.52 & 0.69 & 1.55  \\ \cline{2-10}
 & $\delta_r\,[\%]$ & 48.45 & 31.82 & -1.03 & 16.10 & -62.62 & -65.09 & -71.39 & -67.75\\ \hline \hline
\multirow{2}{1cm}{Set III} & $\sigma_{max}\,[\femtobarn]$ & 5.00 &
3.28 & 0.87 & 1.95 & 5.17 &
3.98 & 1.41 & 2.75  \\ \cline{2-10}
 & $\delta_r\,[\%]$ & 71.81 & 50.22 & 19.75 & 34.15 & -36.80 & -35.14 & -30.77 & -32.80\\ \hline \hline
\multirow{2}{1cm}{Set IV} & $\sigma_{max}\,[\femtobarn]$ & 20.20 &
10.78 & 1.98 & 3.77 & 3.23 &
4.32 & 1.14 & 2.16   \\ \cline{2-10}
 & $\delta_r\,[\%]$ & 102.73 & 44.21 & -20.53 & -24.43 & -64.02 & -35.78 & -49.13 & -51.97\\ \hline \hline
\multirow{2}{1cm}{Set V} & $\sigma_{max}\,[\femtobarn]$ & 11.42 &
3.75 & 0.98 & 2.13 & 4.03 &
1.69 & 1.04 & 1.82  \\ \cline{2-10}
 & $\delta_r\,[\%]$ & 115.61 & -5.53 & -26.13 & -19.69 & -52.56 & -73.46 & -51.19 & -57.22\\ \hline \hline
\multirow{2}{1cm}{Set VI} & $\sigma_{max}\,[\femtobarn]$& \multicolumn{4}{|c||}{below} & 4.76 &
4.12 & 1.44 & 2.61  \\ \cline{2-2} \cline{7-10}
& $\delta_r\,[\%]$ &\multicolumn{4}{|c||}{threshold} & -26.91 & -15.71 & -11.69 & -19.95\\ \hline \hline
\end{tabular}
\caption{Maximum total cross section $\sigma^{(0+1)}(\eeAH)$
at $\sqrt{s} = 500\,\GeV$ and $\sqrt{s} = 1\,\TeV$, together with the relative size
$\delta_r$ of the radiative corrections, for the different sets of
Higgs bosons masses quoted in Table~\ref{tab:masses}. The results
are obtained at fixed $\tan\beta = 1$ and different values of
$\alpha$, with $|\lambda_5|$ at its largest attainable value, cf.
Table~\ref{tab:maxl5}. \label{tab:xsheavy}}
\end{center}
\end{table*}

Some more comments are in order concerning important features
contained in Figs.~\ref{fig:overs_s2}--\ref{fig:overs_s4}. As we
move from Fig.~\ref{fig:overs_s2} to \ref{fig:overs_s3}, hence as we
transit from a relatively light to a heavier Higgs boson spectrum
(cf. Table~\ref{tab:masses}), we encounter that the resulting cross
sections, both at the leading and at one-loop level, are pulled down
by roughly a factor of 2. Unsurprisingly, this is signaling the
exhaustion of the available phase space. By the same token, the
peak-shaped maximum of the $\sigma(\sqrt{s})$ curve gets shifted
forward by about $100\,\GeV$, in correspondence to the production
threshold of the Higgs pair. Moreover, the heavier the Higgs bosons,
the more suppressed the Higgs-mediated loops become, and hence the
lower is the influence of 3H self-couplings. This is precisely what
we confirm numerically in Fig.~\ref{fig:overs_s3} (lower panels), in
which we can appreciate how $\delta_r$ reaches $\sim 40\%$ at most
-- in contrast to $\delta_r \sim 100\%$ that was spotted for Set II
(cf. Fig.~\ref{fig:overs_s2}). In turn, Set VI (in
Fig.~\ref{fig:overs_s4}) embodies again a rather light Higgs boson
spectrum which resembles that of Set II, even though the present
choice of masses is obtained upon artificially engineering a SUSY
setup. Here, owing to the fact that $M_{\PHiggs^\pm}$ is heavier
than in Set II, unitarity conditions settle a more stringent upper
bound on $|\lambda_5|$, and hence on the maximum enhancement of the
triple neutral Higgs couplings. As a result, radiative corrections
in this case are still sizable, but retreat to the $\sim 40\%$
level.

\begin{figure*}[htb]
\begin{center}
\includegraphics[scale=0.5]{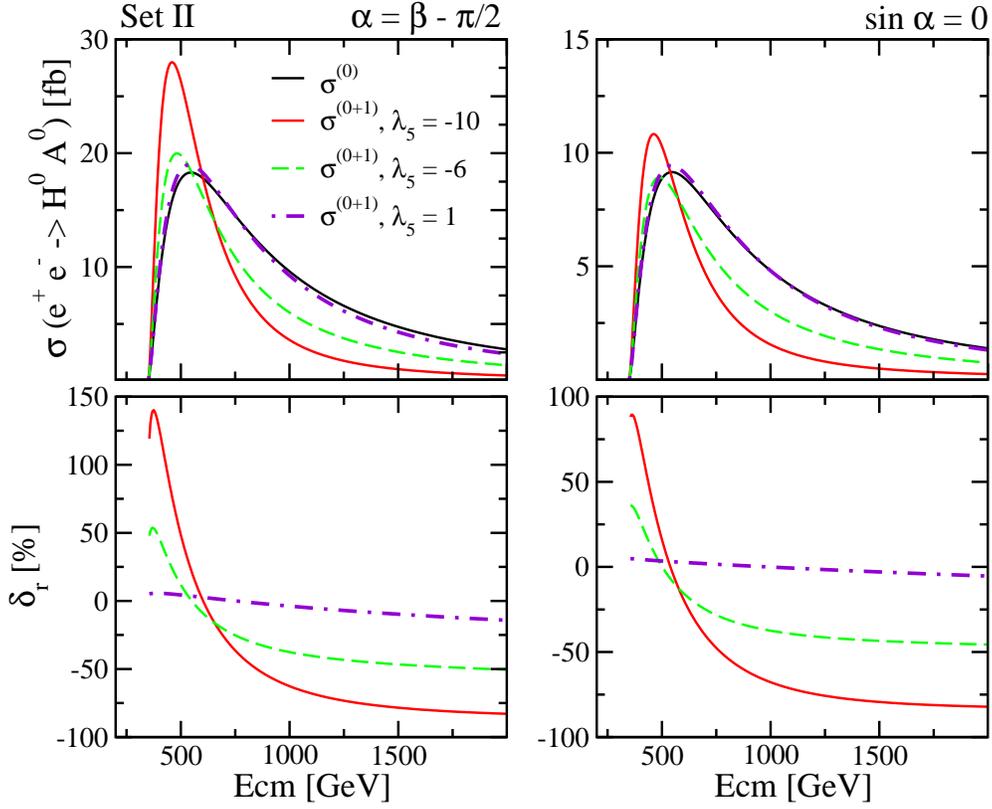}
\caption{Total cross section $\sigma(\eeAH)$ (upper panels) and
relative one-loop correction $\delta_r$ (lower panels) as a function
of $\sqrt{s}$ for Set II of Higgs boson masses, cf.
Table~\ref{tab:masses}. Shown are the results obtained within three
different values of $\lambda_5$, at $\tan\beta=1$ and for
$\alpha=\beta-\pi/2$ (left) and $\alpha=0$ (right) -- the latter
defining the so-called \emph{fermiophobic} limit for the $\Hzero$
boson (for type-I 2HDM). \label{fig:overs_s2_heavy}}
\end{center}
\end{figure*}
\begin{figure*}[htb]
\begin{center}
\includegraphics[scale=0.5]{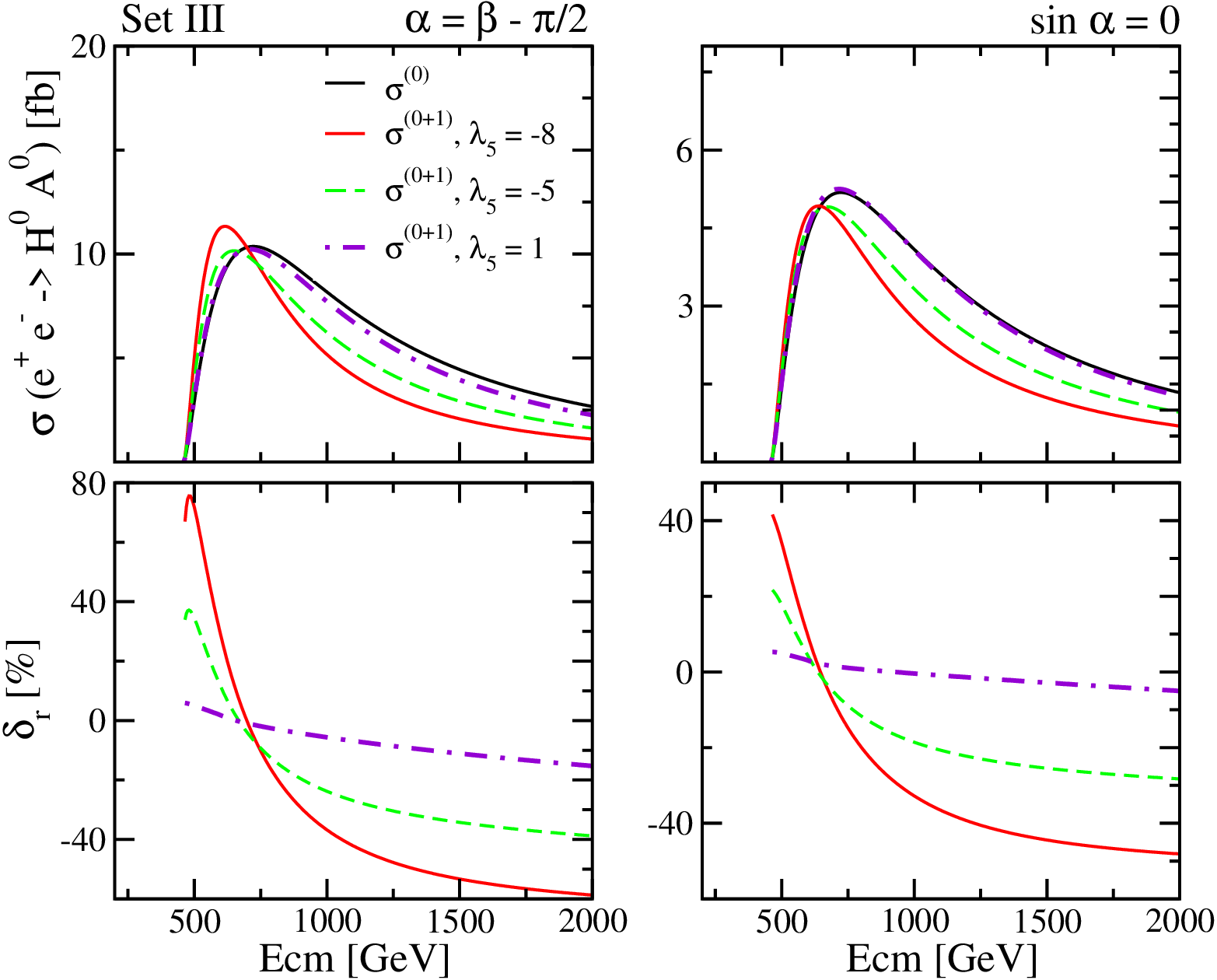}
\caption{As in Fig. \ref{fig:overs_s2_heavy}, but for Set III of
Higgs boson masses, cf. Table~\ref{tab:masses}.
\label{fig:overs_s3_heavy}}
\end{center}
\end{figure*}

\begin{figure*}[htb]
\begin{center}
\includegraphics[scale=0.5]{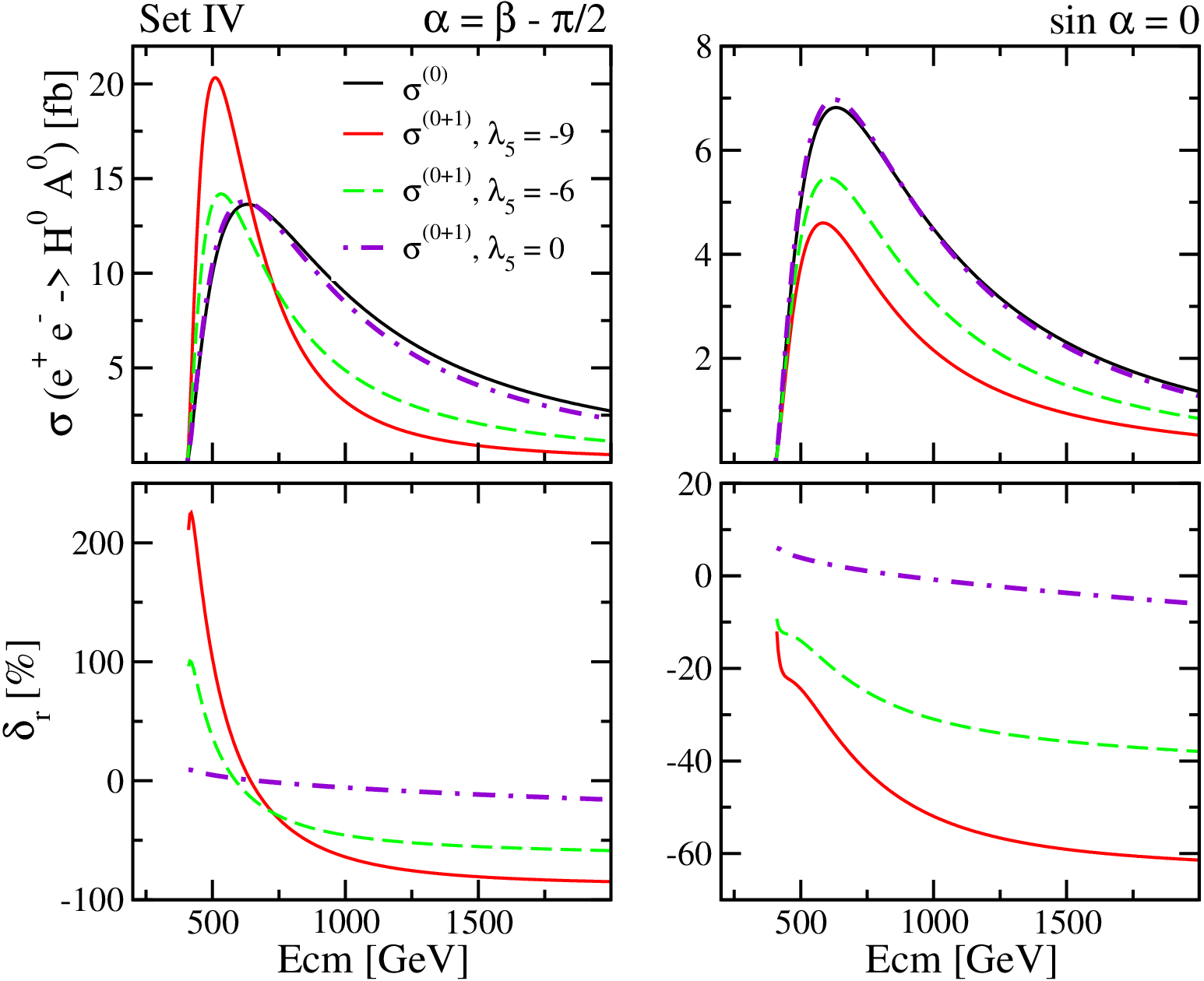}
\caption{As in Fig. \ref{fig:overs_s2_heavy}, but for Set IV of Higgs
boson masses, cf. Table~\ref{tab:masses}.
\label{fig:overs_s4_heavy}}
\end{center}
\end{figure*}

Of great relevance is the study of the behavior of $\sigma(\plA)$ as
a function of the Higgs boson masses. We present these results
conveniently in Figs.~\ref{fig:overmh0_s2}--\ref{fig:overmh0_s4},
where again we superimpose both the tree-level and the
loop-corrected cross-sections for the process $\plA$.  The
dependence with $M_{\hzero}$ is indicated on the left panels and
that with $M_{\Azero}$ on the right panels. The numerical results
are shown for the Sets II,III and IV and have been obtained, once
more, by setting $\tan\beta = 1$, $\alpha = \beta$, and making
allowance for $|\lambda_5|$ to take on the highest permissible
value in each set (see Table~\ref{tab:maxl5}). The center-of-mass
energy is settled at the fiducial value $\sqrt{s} = 500\,\GeV$.

The plots under consideration evince that the raise of the Higgs
boson masses exerts a twofold effect on $\sigma$, one kinematical
and the other dynamical. The former simply means that owing to the
reduction of the available phase space the cross-section obviously
falls down. We can check it in the figures, where we see that both
the tree-level $\sigma^{(0)}$ and the loop-corrected
$\sigma^{(0+1)}$ indeed decrease monotonically with growing
$M_{\hzero}$ and $M_{\Azero}$. As far as the quantum corrections
themselves are concerned, the situation is a bit more subtle.
Although heavier Higgs bosons imply larger propagator suppressions
of the Higgs-mediated one-loop diagrams, as we have just described
in the previous paragraph, the 3H self-couplings can partially
offset this situation since they get invigorated in such
circumstances. This counterbalance feature is manifest in the
explicit expressions for the 3H couplings in
Table~\ref{tab:trilinear} -- the uttermost limitation to their
growing being the unitarity constraints. In particular, 3H couplings
involving neutral fields (which can be shown to carry the dominant
effect on $\plA$) turn out to grow either with $M_{\hzero}$ or
$M_{\Azero}$. Although both effects partially balance each other, we
are ultimately left with the expected decoupling behavior in the
limit of large Higgs boson masses, with the proviso that such
decoupling takes place in a softer way than that of the Born-level
cross section. In fact, in
Figs~\ref{fig:overmh0_s2}--\ref{fig:overmh0_s4} we can verify that
the one-loop corrected cross-section remains most of the time on top
of the tree-level result when we increase the Higgs boson masses,
and moreover the slope of the former is clearly milder, particularly
for $M_{\hzero} \lesssim\, 200\,\GeV$. This translates into a growing
trend of $\delta_r$ up to values of $M_{\hzero} \simeq \,275\,\GeV$. Let
us also mention in passing that the sudden spike in the
$\delta_r(M_{\Azero})$ curve at $M_{\Azero} \simeq \,340 \,\GeV$ (which
is also barely visible for $\sigma$) corresponds to the
$\APtop\Ptop$ pseudo-threshold in the one-loop vertex amplitude (cf.
the first row of diagrams in Fig. \ref{fig:vert11}).

Analogous trends are encountered for sets III and IV of Higgs boson
masses, the only reportable difference being that the decoupling
behavior of $\sigma^{(0+1)}$ with $M_{\hzero},M_{\Azero}$ is more
pronounced, signaling that the increase of these masses does not
boost the 3H self-couplings as efficiently as in the case of Set II.
The shaded regions in Fig.~\ref{fig:overmh0_s3}-\ref{fig:overmh0_s4}
correspond to mass ranges already excluded by experimental
measurements of the $\delta\rho$ parameter (cf. section
\ref{sec:2HDM}).

%

\subsection{$\eeAH$}

On general grounds, the basic phenomenological features are common
to the $\hzero\Azero$ and $\Hzero\Azero$ channels, with the only
proviso that, in the latter case, the resulting production rates are
slightly smaller simply because  $M_{\Hzero} > M_{\hzero}$. The final
cross-sections are nevertheless comparable and reach up to a few dozen fb. In
Table~\ref{tab:xsheavy} we collect, in
a nutshell, the essential results from a dedicated numerical
analysis of $\sigma(\pHA)$ for all sets of Higgs boson masses (cf.
Table~\ref{tab:masses}) and different values of the tree-level
$\vzHA$ coupling, again at fixed $\tan\beta = 1$ and using the
largest permissible value of $|\lambda_5|$ in each set (see
Table~\ref{tab:maxl5}). The relative size of the quantum corrections
$\delta_r$ is also included in Table~\ref{tab:xsheavy}.
As in the previous channel, we have used
fiducial center-of-mass energies $\sqrt{s}=500,\,1000\,\GeV$
respectively. The maximum production rates at $\sqrt{s} = 500\,\GeV$
are achieved for relatively light Higgs boson masses (Set I) and for
$\alpha = \beta-\pi/2$, in which case the tree-level $\vzHA$
interaction is not suppressed, see Eq. (\ref{eq:lag12}). Within this
setup, one is left with $\sigma^{(0+1)} \sim 40 \,\femtobarn$.
Indeed the production rates lie at the $\mathcal{O}(10\,\femtobarn)$
level in a broad range around $\alpha = \beta-\pi/2$ for all the
mass sets under analysis. If we nevertheless consider $\alpha =
\beta-\pi/6$, when the tree-level coupling gets suppressed by
$\sin(\beta-\alpha)=1/2$, the resulting cross sections are still at
the level of a few fb -- each femtobarn rending 500 events per 500
$\invfb$ of integrated luminosity.

Radiative corrections thus may leave a solid imprint into the final
predicted $\sigma^{(0+1)}$ of the current process. In particular,
for $\alpha = \beta - \pi/2$ quantum effects rubber-stamp a
characteristic $50-100\%$ boost upon the tree-level cross-section,
which could hardly be missed. Other choices of $\alpha$ amount to
negative one-loop contributions in which $\vzHA$ is dramatically
suppressed. For instance, in the case of Set I for $\alpha = \beta -
\pi/6$, it renders $\delta_r \sim -65\%$. For center-of-mass
energies from around $\sqrt{s} = 1\TeV$ onwards, a powerful
destructive interference is seen to be under way between the Born
and the one-loop amplitudes. This translates into large (negative)
quantum corrections which may well reach $\delta_r \sim -50-100\%$,
thus literally stamping out the signal and pulling it down at
the $\sim\femtobarn$ level or even below. To better illustrate this feature, in
Figs.~\ref{fig:overs_s2_heavy}--\ref{fig:overs_s4_heavy} we plot the
behavior of $\sigma(\APelectron\Pelectron \to \Azero\Hzero)$ (top
panels) and $\delta_r$ (bottom panels) as a function of $\sqrt{s}$
for the representative Sets II-IV of Higgs boson masses of
Table~\ref{tab:masses} and for different values of the tree-level
coupling; viz. $\alpha = \beta-\pi/2$ (left panels), which
corresponds to the maximum of the tree-level $\vzHA$ coupling, and
$\sin\alpha =0$ (right panels), corresponding to the
\emph{fermiophobic} limit for $\Hzero$ boson within type-I 2HDM (cf.
Table \ref{typeIandII}). Again we fix $\tan\beta = 1$ and we vary
$\lambda_5$ within the range allowed by unitarity and vacuum
stability restrictions. As it was already observed in the analysis
of the $\hzero\Azero$ channel, heavier Higgs-boson spectra furnish
smaller production rates. We refrain from extending the discussion,
which is entirely similar to that of $\plA$.

To our knowledge, no pre-existing study of the one-loop effects on
the neutral Higgs boson channels (\ref{2h}) can be found in the
literature. Let us, however,  briefly comment on the prospects for
the charged Higgs-pair channel, $\APelectron\Pelectron \to \PHiggs^+
\PHiggs^-$. This was considered some time ago in
references~\cite{Arhrib:1998gr,kraft99,Guasch:2001hk}. These old
studies were devoted to the computation of
$\sigma(\APelectron\Pelectron \to \PHiggs^+ \PHiggs^-)$ at one-loop
in both the MSSM and the 2HDM. In the latter case, they also uncover
very large quantum effects (of order $100\%$ or even higher
\footnote{We should nevertheless put a note of caution here in that
the unitarity constraints that were employed in
Ref.~\cite{Arhrib:1998gr,kraft99,Guasch:2001hk} are substantially
less restrictive than those considered throughout our work. In this
regard, an upgraded calculation of $\sigma(\APelectron\Pelectron \to
\PHiggs^+ \PHiggs^-)$ is mandatory before we can judiciously compare
with the results that we have presented here for the neutral Higgs
boson channels. However, due to the increase of the charged Higgs
boson mass bounds in the last few years, the maximum cross-section
rates should correspondingly fall down, thereby making this process
less competitive.}) which are attributed to enhanced 3H
self-interactions involving charged Higgs bosons. Such quantum
effects may lie well above the typical MSSM counterparts, which are
found to render up to $|\delta_r| \sim 20\%$, and mostly negative.
Let us, however, remark that the $\APelectron\Pelectron \to
\PHiggs^+ \PHiggs^-$ channel could operate at sizable cross-sections
levels only for type-I models since the type-II models require a
rather heavy mass spectrum.


\section{Discussion and Conclusions}
\label{sec:conclusions}

In this work,  we have undertaken a thorough investigation of the
pairwise production of neutral Higgs-bosons within the context of
the general two-Higgs-doublet model (2HDM) at the ILC/CLIC linear
colliders (linac) through the basic processes $\eeAlH$. To the best
of our knowledge, this study is the first one existing in the
literature which addresses these important processes at the one-loop
level. To sum up the main objectives of our work, we can assert that
we have concentrated on the following three tasks: i) we have, on
the one hand, singled out the most favorable regimes for the
Higgs-pair production processes once the radiative corrections are
taken into account, specifically within the framework of the
on-shell renormalization scheme with an appropriate definition of
$\tan\beta$ and of the \CP-even mixing angle $\alpha$ at the quantum
level; ii) next, we have carefully quantified the importance of the
leading quantum corrections, and not only so, we have pinpointed
also the origin and relative weigh of the various sources; and,
finally, iii) we have correlated the most powerful source of quantum
effects with the strength of the Higgs boson self-interactions
(essentially the triple Higgs boson self-couplings $\lambda_{3H}$),
to which the processes $\eeAlH$ are highly sensitive through
Higgs-mediated one-loop diagrams. With these aims in mind, we have
systematically searched over the 2HDM parameter space by considering
different Higgs boson mass scenarios and different values of the
tree-level $\vzlA$ and $\vzHA$ couplings. At the same time, we have
carefully explored the very frontiers of the generic 2HDM parameter
space, specially the limits that cannot be trespassed by the
strengths of the various 3H self-couplings.

Among the various subtle triggers that control the size of these
important couplings, one of them stands out with overwhelming
supremacy over the others, and this is the $\lambda_5$ coupling.
This is the only one parameter in the list of seven free parameters
-- cf. Eq.\, (\ref{freep}) -- conforming the structure of the
general (\CP-invariant) Higgs potential of the 2HDM that cannot be
ultimately traded for a physical mass or a mixing angle. This
parameter is, in addition, the only Higgs self-coupling that need
not be renormalized at one loop in the current framework. Despite
its special status of being apparently unfettered and without
obvious physical connotations, one quickly discovers that it becomes
severely restrained by subtle bounds connected with the quantum
field theoretical consistency of the model, namely the stringent
constraints imposed by tree-level unitarity and vacuum stability
conditions. These restrictions, together with the limitations
dictated by custodial symmetry, and of course also those emerging in
the light of the direct collider bounds on the masses of generic
Higgs boson models, not to mention the indirect phenomenological
restraints (in particular, the upper bound on the charged Higgs mass
in type-II models ensuing from radiative $B$-meson decays), do
indeed eventually determine a fairly compact region of parameter
space. In particular, $\lambda_5$ is compelled to be mostly negative
and in the range $|\lambda_5|\lesssim {\cal O}(10)$. At the same
time, $\tan\beta$ cannot be large when $|\lambda_5|$ is also large
without violating the constraints. The bottom line is that the value
that maximizes the processes under consideration is a rather
equitable one: $\tan\beta\simeq 1$.

Notwithstanding the sharp limits placed upon the physical region of
the parameter space, our explicit calculation of radiative
corrections has shown -- and we feel it is truly remarkable -- that
the general 2HDM models are still able to unleash a good deal of
their ``repressed'' power. If only we could catch a smoking gun of
this stupendous potential through the accurate measurements that a
linear collider should be able to render on fundamental processes
like $\APelectron\Pelectron \to\hzero\Azero/ \Hzero\Azero$, we might
be on the verge of experiencing an intense episode of Higgs boson
physics beyond the SM.  In the following, we describe the basic
results and strategies that we could follow to this effect.

We have identified three main scenarios of potential
phenomenological interest in the corresponding parameter space. In
all of them we observe no remarkable dependence either on the
details of the Higgs mass spectrum nor on the pattern of
Higgs-fermion Yukawa couplings (either for type-I or type-II 2HDM),
not even on the particular channel $\hzero\Azero/\Hzero\Azero$ under
study. In what follows we describe each one of these scenarios and
spell out clearly their differences, in particular we provide the
characteristic potential size of the relative quantum correction
$\delta_r$ -- cf. Eq.\,(\ref{deltar}) -- typically associated to
them (cf. Figs.\,\ref{fig:scan-s2}-\ref{fig:scan-s4}):

\begin{itemize}

\item \textbf{Scenario 1}. To start with, regions of $\tan\beta\gtrsim 1$ and large
$|\lambda_5|$ (specifically, $\lambda_5 < 0, |\lambda_5|\simeq
5-10$) support the bulk of the radiative corrections. They may well
reach $\delta_r=\pm (50 - 100)\%$ at the largest attainable values
of $|\lambda_5|$. Furthermore, these effects critically depend on
the actual size of $|\lambda_5|$ and on the center-of-mass energy
$\sqrt{s}$ (cf. Figs.\,\ref{fig:overs_s2}-\ref{fig:overs_s4}), to
wit: $\delta_r$ is positive at ``low'' center-of-mass energies
around $\sqrt{s}=500\pm 100\,\,\GeV$, while it is negative for
$\sqrt{s}> 600\,\GeV$ and certainly also in the uppermost
foreseeable energy segments $\sqrt{s}=1-3\,\TeV$, usually reserved
for an upgraded ILC and specially for CLIC. Thus, interestingly
enough, quantum corrections turn out to drastically invigorate the
cross section of the basic processes $\APelectron\Pelectron
\to\hzero\Azero/\Hzero\Azero$ near the standard booting energy of
the ILC, while they tend to suppress the two-body Higgs boson signal
for a center-of-mass energy some $20-30\%$ beyond this initial
regime -- and of course for the TeV range and above -- as a result
of constructive/destructive interference effects, respectively,
between the tree-level and the one-loop amplitudes. Incidentally, we
clarify that this optimal scenario cannot be iterated for large and
positive values of $\lambda_5$ because the combined set of
constraints (mainly the vacuum stability bounds, in this case)
preclude most of the region $\lambda_5>0$;

\item \textbf{Scenario 2}. On the other hand, we find a tail of subleading effects within a
band of the $(\tan\beta, \lambda_5)$ subspace at moderate
$\tan\beta=1.5-5$ and $|\lambda_5|\lesssim 2 $ wherein the maximum
radiative corrections turn out to stagnate and remain barely at the
level of $\sim 10\%$ or less. Should the parameters of the model
inhabit this region, the practical possibilities to detect these
effects would obviously become thinner than in the previous case.
Still, $5-10\%$ effects are not that alien to the high precision
standards planned for a linac collider, and hence it should not
deter us from the searching task. At this point, the detection
strategies associated to the Higgs boson decay patterns could be
essential (see below);

\item \textbf{Scenario 3}. Finally, there is the region of parameter space where
the coupling $\lambda_5$ is very small or it just vanishes. Here the
quantum corrections can still have a chance provided $\tan\beta<1$,
and indeed we have found domains of this kind preserving all the
basic constraints. While this region is not the most favored one, neither
from the theoretical point of view nor from the experimental data, specially if $\tan\beta\ll 1$,
we should keep it in mind as a possibility. Even for moderately
small $\tan\beta$, say for $\tan\beta\gtrsim 0.2$, it can be the
source of sizeable quantum effects of order $10-30\%$ (and
positive), which are larger than those in Scenario 2 and, in some
cases, even comparable to those in Scenario 1. It is worth noticing
that, in this region, there is an interesting collaborative
interplay between the Higgs boson trilinear self-couplings and the
top quark Yukawa coupling (which is the same in type I or type II
2HDM's). Only in this scenario a Yukawa coupling could play a
significant quantitative role on equal footing with the trilinear
Higgs boson self-couplings.

\end{itemize}

Obviously, the success of the whole Higgs boson search endeavor will
also depend on the pattern of distinctive signatures available to
these particles in the final state. For instance, for
$M_h<2M_V\lesssim 180\,\GeV$, we should basically expect
back-to-back pairs of highly energetic (roughly
$E\sim\sqrt{s}/4\gtrsim 100\,\GeV$) and collimated b-quark and/or
$\tau$-lepton jets from $\Azero,\hzero\to b\bar{b}/\tau^{+}\tau^{-}$
which, in contrast to the LHC, \phantom{}{should not be overshadowed
by the large QCD background which is inherent in hadronic machines.
And of course similarly with the $\Azero\Hzero$ final state. This
much clearer environment notwithstanding, we should be aware of the
fact that other characteristic processes of linac physics, such as
the four-jet cross-section, might be large enough to partially
obscure the Higgs-pair production signal. However, the precise
analysis of the corresponding signal distributions, which would be
required in order to completely assess the real experimental
possibilities, is beyond the scope of the present study.}.
And of course the same reasoning applies to the $\Azero,\Hzero$ final
state.
For
$M_{h}>2M_V$, on the other hand, signatures from $\hzero,\Hzero\to
VV$ and $\Azero\to b\bar{b}/\tau^{+}\tau^{-}$ could also play a role
(depending on the particular choice of $\alpha$ and $\beta$, cf. Table
\ref{tab:coup}). They could ultimately lead to signatures with two
(up to four) charged leptons against a $b\bar{b}$ or
$\tau^{+}\tau^{-}$ pair in the final state (from $W^{\pm}\to
\ell^{\pm}+\text{missing energy}$ and, specially, from $Z\to
\ell^+\ell^-$). Even if the branching ratios of gauge
bosons into leptons are very small ($\mathcal{B}(V\to l l)\sim
0.03$),the predicted number of Higgs-pair events is
large enough so that, in the most favorable scenarios, one could
collect $\mathcal{O}(10^2)$ events for these
complementary signatures, which could be instrumental for a final
tagging and identification of the Higgs bosons. Notice that we use
the fact that the \CP-odd $\Azero$ ``always'' decays into $b\bar{b}$
or, to a lesser extent, to $\tau^{+}\tau^{-}$ (even for
$\tan\beta=1$) as it cannot couple to gauge bosons. So the patterns
of signatures are rather characteristic and could not be missed in
normal circumstances. But there is more than meets the eye in the
whole strategy game, as we shall see later on.

In the most favorable regimes (Scenario 1), the experimental chances
could be spectacular, especially in the case of relatively light
Higgs boson mass spectrum and a large (negative) value of
$\lambda_5$, for which the predicted one-loop cross sections
$\sigma^{(0+1)}(\plA / \Hzero\Azero)$ at $\sqrt{s}= 500\,\GeV$ may
border the $\mathcal{O}(100\,\femtobarn)$ level. This would
translate into barely $5\times 10^4$ events per $500\,\invfb$ of
integrated luminosity at the linac. While the former situation
describes perhaps the most optimistic possibility, cross sections of
$\mathcal{O}(10\,\femtobarn)$ at $\sqrt{s}= 500\,\GeV$ should be
quite common in a wide patch of the 2HDM parameter space, and
correspondingly of $\mathcal{O}(1\,\femtobarn)$ at $\sqrt{s} =
1\,\TeV$ (see Table \ref{tab:masses}), thus delivering rates that
range from a few hundred to a few thousand events respectively. With
this level of statistics (plus a comparable number of events from
the $\Hzero\Azero$ channel) it should be perfectly possible to
insure a comfortable tagging of the Higgs bosons in the clean
environment of a linac machine.

Most important, we have been able to demonstrate that the dominant
quantum effects are, as we expected, primarily nurtured by the
Higgs-mediated one-loop corrections to the $\vzlA/\vzHA$ vertices,
and can be ultimately traced back to the potentially enhanced 3H
self-interactions $\lambda_{3H}$ (cf. Table II). This is indeed a
trademark feature of the 2HDM, with no counterpart in the MSSM -- in
which SUSY invariance highly curtails the structure of the Higgs
boson self-interactions and compels them to be purely gauge-like.
Obviously, probing the structure of Higgs boson self-interactions --
in the present case through the analysis of radiative corrections on
direct production processes -- is a most useful strategy to
disentangle SUSY and non-SUSY Higgs physics scenarios.

To be specific, for the analysis of the $\APelectron\Pelectron \to
2H$ cross sections we have faced the computation of the full set of
$\mathcal{O}(\alpha^3_{ew})$ quantum effects, among them the subset
of $\mathcal{O}(\alpha_{ew}^2\lambda_{3H}^2)$ corrections, and
supplemented them with the leading $\mathcal{O}(\alpha^4_{ew})$
pieces arising from the square of the one-loop diagrams involving
Higgs boson self-couplings. These (finer) corrections are of
$\mathcal{O}(\alpha^2_{ew}\lambda_{3H}^4)$ (see, however, below).
Their inclusion is actually \emph{a must} at large center-of-mass
energies (i.e. once $\sqrt{s}$ has left behind the startup
situation, and of course also in the  $1\,\TeV$ regime). The reason
is that the tree-level $\mathcal{O}(\alpha^2_{ew})$ and the one-loop
$\mathcal{O}(\alpha^3_{ew})$ amplitudes virtually cancel out each
other at high energies, owing to the characteristic large and
negative quantum effects in this regime.  We can convince ourselves
of this fact from the severe depletion exhibited by the one-loop
curves -- see e.g. Figs. \ref{fig:overs_s2}-\ref{fig:overs_s4} -- as
compared to the tree-level ones. Typically, the effect appears after
we increase the center-of-mass energy by $100\,\GeV$ beyond the
initial $\sqrt{s}=500\,\GeV$, and persists till 1 TeV and
afterwards. At large energies, the inclusion of the
$\mathcal{O}(\alpha^2_{ew}\lambda_{3H}^4)$ effects is a consistency
requirement as it prevents the cross-section from becoming negative.
These terms can actually be relevant also at the initial energy
interval $\sqrt{s}=500-600\,\GeV$ at which the linac will first
operate; for example, for the scenarios with maximally enhanced 3H
self-couplings, they may lead to an additional $\sim 20\,\%$
contribution to $\delta_r$, hence pushing the overall quantum
correction even higher. In this respect, we note that our earlier
estimate of these higher order terms, Eq.\,(\ref{eq:estimate2}), was
able to hit the right order of magnitude, although it actually
underestimates the maximum size revealed by the exact numerical
analysis -- not too surprising from such a rough attempt at guessing
their bulk size.

Let us clarify that the two-loop diagrams involving the Higgs
boson-mediated $\vzlA/\vzHA$ vertex corrections, lead of course also
to amplitudes of order $\mathcal{O}(\alpha^2_{ew}\lambda_{3H}^4)$.
These are actually finite, and originate from the interference of
the tree-level amplitude with the two-loop one. An obvious concern
is then the following: can we safely neglect them? Upon inspection
of the Higgs-mediated two-loop diagrams, and taking into account
power counting and dynamical considerations, one can show that they
are actually suppressed, and particularly so for those scenarios
where $\sigma^{(0+1)}(\APelectron\Pelectron \to 2H)$ is optimized.
But, most significantly, there is a deeper observation to be adduced
at this point, which is well in the spirit of the effective field
theory approach to QFT. The (finite) two-loop effects induced by the
enhanced Higgs boson self-couplings can be conveniently reabsorbed
in the form of one-loop corrections to the bare Higgs
self-couplings, which first appear at the one-loop level in the
process under consideration. We can iterate this algorithm at any
order in perturbation theory by further reabsorbing the higher order
Higgs mediated loops into the triple \emph{and} quartic Higgs
self-couplings introduced at one-loop order. In this way we can
define a collection of effective 3H and 4H self-couplings that
encapsulate all these higher order effects.

The above remark is an interesting one, as it tells us that the
application of the stringent tree-level unitarity relations that we
have used can be directly reiterated for these effective couplings.
Therefore, no matter how big are the quantum corrections induced by
the 3H self-couplings in subsequent orders of perturbation theory,
their overall effect is constrained by the same set of formal
unitarity relations. If expressed in terms of the original
couplings, these relations will be equally stringent so that, on
balance, the maximum enhancement capabilities of the 3H
self-couplings are basically the same at any order of perturbation
theory. This is the procedure adopted in the present work, and we
believe it is perfectly consistent, in the sense that we have
retained the contributions to the $\APelectron\Pelectron \to 2H$
amplitudes at leading order in the 3H self-couplings while employing
the unitarity constraints of
Ref.~\cite{Kanemura:1993hm}-\cite{Akeroyd:2000wc}, also derived at
leading order in $\lambda_{3H}$.

How do our results compare with the expectations in the MSSM? As we
have stressed in section II, a lot of work on Higgs boson production
in $\APelectron\Pelectron$ colliders has been reported in the
literature thus far, but mainly within the context of the MSSM. In
particular, supersymmetric radiative corrections to $\sigma(\plA)$
have been considered in detail, although mostly in the context of
LEP and including sometimes a more or less timid incursion into the
TeV-class colliders, cf.
Refs.~\cite{Driesen:1996jd,Djouadi:1999gv,Osland:1998hv,Muhlleitner:2000jj,Heinemeyer:2001iy,Coniavitis:2007me}.
Let us try to see how some of these results compare with ours. For
the sake of definiteness, let us concentrate on the particular case
of Ref.~\cite{Heinemeyer:2001iy}, in which $\sigma(\eeAl)$ is
analyzed in the MSSM for a linear collider operating at LEP 2 energy
and further extended up to $\sqrt{s} = 500\,\GeV$. In the most
favorable regimes, the loop-corrected cross sections may reach up to
barely $30\,\femtobarn$ at $\sqrt{s} = 500\,\GeV$, which falls in
the ballpark of the results that we have obtained here within the
2HDM. This is not surprising because in both cases the leading
amplitude (i.e. the lowest order one) is purely gauge and hence the
order of magnitude of the cross-section is independent of whether we
compute it in the MSSM or in a generic 2HDM.

Quantum effects can be fairly efficient in the MSSM too, and they
may also boost the tree-level predictions substantially ($\sim
20\%$). However, their size is never as big as the maximal 2HDM
contributions; and, no less important for our conceptual
understanding, the fundamental origin of these quantum effects is
completely different from that of the 2HDM. As already mentioned in
the introduction, the leading quantum effects in the MSSM are due to
the genuine Higgs-quark, Higgs-squark and
quark-squark-chargino/neutralino Yukawa-like supersymmetric
couplings, whose contributions carry enhancement factors of the
fashion  $\sim m_b\,\tan\beta/M_W, \sim m_t(A_t -
\mu/\tan\beta)/M^2_{SUSY},\, m_b(A_b - \mu\,\tan\beta)/M^2_{SUSY}$,
none of them related to the structure of the Higgs potential and
hence none of them associated to Higgs boson self-couplings.

In contradistinction to the MSSM, the 3H self-couplings in the 2HDM
embody the full potential for triggering quantum effects in physical
processes, and these effects become thence tied to the very
structure of the Higgs potential. Even if quantitatively similar in
some scenarios, quantum effects in SUSY and non-SUSY two-Higgs
doublet extensions of the SM are, therefore, prompted by
intrinsically different mechanisms. In practice, the task of
distinguishing whether the produced Higgs boson particles at a linac
are supersymmetric or not can start by addressing the decaying
signatures of the Higgs bosons into $b\bar{b}$ or $\tau^{+}\tau^{-}$
pairs, as sketched above. However, in certain regions of parameter
space the number of events can be by itself a direct signature of
the sort of Higgs model that we have behind. For example, in the
MSSM, the tree-level coupling $C_{\vzlA} \sim \cos(\beta-\alpha)$
undergoes a severe suppression with growing values of $\tan\beta$
and/or $M_{\Azero}$ -- as $\alpha$ depends on $M_{\Azero}$ in the
MSSM. Thereby the predicted cross sections, either at the Born or at
the one-loop level, are dramatically weakened as a function of
$M_{\Azero}$. For instance, $\sigma^{(0+1)}(\plA)$ at $\sqrt{s} =
500\,\GeV$ and $M_{\Azero} = 300\, \GeV$ may end up as tiny as $\sim
0.01\,\femtobarn$ (see e.g. Fig.~9 of Ref.~\cite{Heinemeyer:2001iy})
while the 2HDM prediction is of $\mathcal{O}(10\,\femtobarn)$, i.e.
one thousand times bigger!

But there is more to say here, as we have announced above. The
identification strategy can be highly facilitated through the
interplay of further 2H (and eventually 3H) channels, such as
(\ref{3H}) and (\ref{2HX}), which are also highly distinctive in the
2HDM -- see \cite{Ferrera:2007sp} and \cite{Hodgkinson:2009uj}. In
fact, for these channels the difference with the MSSM can be
apparent already at the lowest order of perturbation theory. Let us
be a bit more quantitative here. In Table~\ref{TableEnd}, we compare
the numerical predictions on the cross-sections of these various
processes in truly equitable conditions, i.e. for different sets of
common parameters for all the processes, and for three realistic
values of the center-of-mass energy of the ILC/CLIC (the latter
operating always at the highest edge of the energy band, although it
could nominally reach the 3\,TeV end).

To be more precise, the cross-sections that we are comparing on
equal footing in Table~\ref{TableEnd} are the following: i) the
one-loop corrected $\sigma^{(0+1)}(\eeAl)$; ii) the leading-order
$\mathcal{O}(\alpha^3_{ew})$ cross-sections of the two triple Higgs
production processes $\APelectron\Pelectron
\to\hzero\Hzero\Azero/\Hzero\PHiggs^+\PHiggs^-$; and iii)
$\sigma(\APelectron\Pelectron \to V^* V^* \to \hzero\hzero + X$) at
leading-order $\mathcal{O}(\alpha^4_{ew})$. Notice that the bulk of
the contribution from processes ii) and iii) comes from the
$\mathcal{O}(\alpha^2_{ew}\lambda_{3H}^2)$ and
$\mathcal{O}(\alpha^3_{ew}\lambda_{3H}^2)$ parts, respectively.  In
all cases we take $\tan\beta = 1$, $\alpha = \beta$ and the maximum
allowed value of  $|\lambda_5|$ according to Table~\ref{tab:maxl5},
that is, the most favorable scenario for $\hzero\Azero$ pair
production. We use Sets I-III for the Higgs boson masses. The
resulting predictions for the different channels turn out to be
highly illustrative of the complementary nature of such Higgs-boson
production signatures. For a thorough discussion of the dynamical
features and signatures of the multiparticle processes ii) and iii),
we refer the reader to the original references
\cite{Ferrera:2007sp,Hodgkinson:2009uj}. Here we have recomputed the
cross-sections ii) and iii) under the same kinematical conditions
and in the very same region of the parameter space as the main
processes i) under consideration, which is of course the
indispensable first step that enables us to compare them all
meaningfully. In this way, we are well prepared to infer the final
strategy that can be designed to efficiently search for 2HDM Higgs
bosons at the linear colliders.
\begin{table}
\begin{center}
\begin{tabular}{|c||c|c|c|}  \hline
Process & $\sigma(\sqrt{s}=0.5\,{\rm TeV})$\,\femtobarn &
$\sigma(1.0\,{\rm TeV})$\,\femtobarn & $\sigma(1.5\,{\rm
TeV})$\,\femtobarn
\\ \hline \hline \multicolumn{4}{|c|}{Set I} \\ \hline
$\hzero\Azero$ &  34.13 & 2.89 & 0.70 \\
$\hzero\Hzero\Azero$ & 3.09 & 8.58 & 5.17 \\
$\Hzero\PHiggs^+\PHiggs^-$ & 6.75 & 19.65 & 12.27 \\
$\hzero\hzero + X$ & 13.29 & 79.00 & 146.08 \\ \hline
\multicolumn{4}{|c|}{Set II} \\ \hline
$\hzero\Azero$ &  26.71 & 4.07 & 1.27 \\
$\hzero\Hzero\Azero$ & 0.02 & 5.03 & 3.55 \\
$\Hzero\PHiggs^+\PHiggs^-$ & 0.17 & 11.93 & 8.39 \\
$\hzero\hzero + X$ & 1.47 & 17.36 & 38.01 \\ \hline
\multicolumn{4}{|c|}{Set III} \\ \hline
$\hzero\Azero$ &  11.63 & 6.11 & 2.52 \\
$\hzero\Hzero\Azero$ & below threshold & 1.25 & 1.33 \\
$\Hzero\PHiggs^+\PHiggs^-$ & below threshold & 0.69 & 2.14 \\
$\hzero\hzero + X$ & 0.92 & 9.72 & 23.40 \\ \hline
\end{tabular}
\end{center}
\caption{Comparison of the predictions for the cross sections (in
fb) corresponding to some of the Higgs boson production processes
(\ref{3H}), (\ref{2HX}) and (\ref{2h}). The results are obtained for
the Sets I, II and III of Higgs boson masses (cf.
Table~\ref{tab:masses}) $\tan\beta = 1$, $\alpha=\beta$, and three
different values of the center-of-mass energy. We observe a great
complementarity between the different channels at different
energies: the exclusive 2H channels (\ref{2h}) are dominant at the
ILC startup energy $\sqrt{s}=0.5$\,TeV whereas the others dominate
at higher energies, specially the inclusive $2\PHiggs+X$ one
(\ref{3H})-- which is triggered mainly by weak gauge boson fusion.
\label{TableEnd}}
\end{table}

\jump

The strategy that follows from the results encapsulated in Table
\ref{TableEnd} and the information that we have gathered in the
figures of the last section should be by now crystal-clear, and it
comes down to the following two-step procedure:

\jump\jump

\begin{itemize}

\item \textbf{Step 1}. At the vicinity of the startup energy of the ILC
(approximately in the range $\sqrt{s}=500\pm 100\,\GeV$), the
exclusive neutral double Higgs boson channels $\APelectron\Pelectron
\to \Azero\hzero/\Azero\Hzero$ -- Eq. (\ref{2h}) -- prove to be the
dominant ones as compared to the triple Higgs boson production
processes $\APelectron\Pelectron \to 3\PHiggs$ and the inclusive
double Higgs production ones $\APelectron\Pelectron \to 2
\PHiggs+X$, see equations (\ref{3H}) and (\ref{2HX}). Therefore, in
this lower energy band, the study of potential signatures of new
physics, and in particular the identification of the nature of the
produced Higgs bosons at the ILC, must be conducted through the
careful analysis of the quantum corrections affecting the two-body
channels $\APelectron\Pelectron \to 2\PHiggs$. Here is precisely
where the detailed results of the present work could be most useful,
and indeed should be the first analysis to be implemented when the
linear colliders are set to work in the future;

\item \textbf{Step 2}. At higher energies, however, say for $\sqrt{s}>600$\, GeV and, for that matter,
in the entire \,TeV range (hence at the highest nominal energy
regimes scheduled for the ILC, and specially for CLIC), the
influence of the $\PZ^0$-propagator stifles dramatically the
cross-section of the exclusive neutral double Higgs boson channels
$\APelectron\Pelectron \to 2\PHiggs$. Therefore, in these high
energy domains, they can no longer compete with the mechanisms
providing multiparticle final states, such as (\ref{3H}) and
(\ref{2HX}). The latter process, $\APelectron\Pelectron \to
V^*V^*\to h\,h + X$, becomes indeed the most efficient one at the
highest energies since it is not crippled by the s-channel
propagator; quite on the contrary, its cross-section increases
steadily with the energy (see Ref.\,\cite{Hodgkinson:2009uj} for
details). Therefore, in the upgraded phase of the ILC, and certainly
for the CLIC collider, one must concentrate all the search strategy
power on looking for an anomalously large number of inclusive Higgs
boson pairs of the type $\hzero\hzero$, $\Hzero\Hzero$,
$\hzero\Azero$ and $\Azero\Azero$, which should emerge mostly
acollinear and dynamically thrusted along the beam direction, rather
than appearing in a simple back-to-back geometry characteristic of
the two-body process $\APelectron\Pelectron \to 2\PHiggs$. Notice
furthermore that, in contradistinction to the latter, the fused
pairs consist of identical Higgs bosons. Unmistakeably, the
predicted numbers in Table \ref{TableEnd} could not, by any means,
be accounted for if they were to be ascribed to a supersymmetric
origin. Finally, the cross-correlation of these higher energy
effects with the previous ones at lower energies, which we could
track very well while upgrading the ILC collider from
$\sqrt{s}=0.5$\,TeV up to $\sqrt{s}=1.5$\,TeV -- and eventually till
$\sqrt{s}=3$\,TeV (through CLIC) -- should provide plenty of
unambiguous evidence of non-supersymmetric Higgs
physics\,\footnote{Another production mechanism where the
outstanding enhancing properties of the general 2HDM Higgs-boson
self-couplings can be tested -- and correlated with the processes
under consideration -- is the one-loop production of neutral Higgs
bosons in association with the $Z$ gauge boson:
$\APelectron\Pelectron \to \PZ^0\PHiggslight$\, ($\PHiggslight =
\hzero\, ,\Hzero$), also called ``Higgs-strahlung'' processes; see
\cite{NDJ10} for a thorough study at the quantum level in the
general 2HDM. }.

\end{itemize}

Let us finally introduce a remark that hints at the potentially far
reaching implications of the 2HDM dynamics in different sectors of
Particle Physics. The triple Higgs boson self-couplings could also
have an indirect impact on the best high precision observables at
our disposal, even well before the ILC/CLIC colliders are
commissioned and put effectively to work. For example, they could
have a bearing on the famous electroweak precision parameter $\Delta
r$ -- see Eq.\,(\ref{Deltar}). Since our main aim here has been to
exploit the leading contributions from the Higgs boson
self-couplings in the arena of the direct production processes
(\ref{2h}), the $\Delta r$ part did not enter our quantum
computation; it would only enter at two-loop level. However, we
suspect\,\cite{JSDeltar} that the influence of the 3H self-couplings
on $\Delta r$ might have a real interest \textit{per se}, for it
could induce a correction to $\delta\rho$ and ultimately trigger a
shift in the $W^{\pm}$ mass, and the result might well compete with
the highly accurate calculations that have been performed on this
observable within the alternative framework of the
MSSM\,\cite{Heinemeyer:2006px}.

\jump\jump

To summarize, if the genuine enhancement properties of the 2HDM
effectively hold in the real world, the combined analysis of the
exclusive double Higgs production (2H) processes at the startup
energy  of the ILC, $\sqrt{s}=500\,\GeV$, and subsequently of the
triple (3H) and inclusive double Higgs production processes ($2
\PHiggs +X$) at the upgraded ILC/CLIC ($\sqrt{s}\geqslant 1\,\TeV$),
might reveal strong hints of Higgs boson physics beyond the SM. The
bottom line of our study is that the physics of the linear colliders
can be truly instrumental to unveil the nature of the Higgs bosons.
These bosons may, or may not, have been discovered during the LHC
era while the linear colliders take the floor, but the real issue at
stake here is of highest priority and should strengthen the need for
such machines. High precision Higgs boson physics in a linac can
indeed provide a keen insight into the most sensitive building
blocks of modern gauge theories of weak and electromagnetic
interactions, viz. in the very core architecture of the Electroweak
Symmetry Breaking.

\newpage

\begin{acknowledgments}
This work has been supported in part by the EU project RTN
MRTN-CT-2006-035505 Heptools. DLV thanks an ESR position of this
network, and also the Theory Group at the Max-Planck Institut f\"ur
Physik in Munich for the warm hospitality extended to him while part
of this project was being carried out. The work of DLV has also been
supported partially by the MEC FPU grant Ref. AP2006-00357. JS has
been supported in part by MEC and FEDER under project FPA2007-66665,
by the Spanish Consolider-Ingenio 2010 program CPAN CSD2007-00042
and by DIUE/CUR Generalitat de Catalunya under project 2009SGR502.
JS also thanks the MPI in Munich for the hospitality and financial
support while part of this work was under way. The authors are
grateful to Prof. W. Hollik for useful discussions on this topic and
for his support. Early discussions with N. Bernal and J. Guasch are
also acknowledged.
\end{acknowledgments}


\providecommand{\href}[2]{#2}

\newcommand{\JHEP}[3]{ {JHEP} {#1} (#2)  {#3}}
\newcommand{\NPB}[3]{{ Nucl. Phys. } {\bf B#1} (#2)  {#3}}
\newcommand{\NPPS}[3]{{ Nucl. Phys. Proc. Supp. } {\bf #1} (#2)  {#3}}
\newcommand{\PRD}[3]{{ Phys. Rev. } {\bf D#1} (#2)   {#3}}
\newcommand{\PLB}[3]{{ Phys. Lett. } {\bf B#1} (#2)  {#3}}
\newcommand{\EPJ}[3]{{ Eur. Phys. J } {\bf C#1} (#2)  {#3}}
\newcommand{\PR}[3]{{ Phys. Rept. } {\bf #1} (#2)  {#3}}
\newcommand{\RMP}[3]{{ Rev. Mod. Phys. } {\bf #1} (#2)  {#3}}
\newcommand{\IJMP}[3]{{ Int. J. of Mod. Phys. } {\bf #1} (#2)  {#3}}
\newcommand{\PRL}[3]{{ Phys. Rev. Lett. } {\bf #1} (#2) {#3}}
\newcommand{\ZFP}[3]{{ Zeitsch. f. Physik } {\bf C#1} (#2)  {#3}}
\newcommand{\MPLA}[3]{{ Mod. Phys. Lett. } {\bf A#1} (#2) {#3}}
\newcommand{\JPG}[3]{{ J. Phys.} {\bf G#1} (#2)  {#3}}



\end{document}